\documentstyle[epsfig]{elsart}

\begin{document}
\begin{frontmatter}
\title{Submillimeter Galaxies}
\author[ad1,ad2]{Andrew W. Blain\thanksref{corr}},
\author[ad3]{Ian Smail}, 
\author[ad4]{R. J. Ivison}, 
\author[ad5]{J.-P. Kneib}, 
\author[ad6]{David T. Frayer} 
\thanks[corr]{Corresponding author. E-mail: awb@astro.caltech.edu}
\address[ad1]{Department of Astronomy, Caltech, \cty Pasadena CA 91125, 
\cny USA}
\address[ad2]{Institute of Astronomy, Madingley Road, \cty Cambridge CB3 0HA, 
\cny UK} 
\address[ad3]{Department of Physics, University of Durham, South Road, 
\cty Durham DH1 3LE, \cny UK} 
\address[ad4]{Institute for Astronomy, University of Edinburgh, 
\cty Edinburgh EH9 3HJ, \cny UK}
\address[ad5]{Observatoire Midi-Pyr\'en\'ees, 14 Avenue E. Belin, 
\cty F-31400 Toulouse, \cny France}
\address[ad6]{SIRTF Science Center, Caltech, \cty Pasadena CA 91125, 
\cny USA} 

\begin{abstract}

A cosmologically significant population of 
very luminous high-redshift galaxies has recently been discovered 
at submillimeter (submm) wavelengths. 
Advances in submm detector technologies have opened this new window 
on the distant Universe. 
Here we discuss the 
properties of the high-redshift 
submm galaxies, their significance for our understanding of the 
process of galaxy formation, and the selection effects that apply to deep 
submm surveys. The submm galaxies generate a 
significant fraction of the energy output of all the galaxies in the
early Universe. We emphasize the importance of
studying a complete sample of submm galaxies, and stress that 
because they are 
typically very faint in other wavebands, these 
follow-up observations are very
challenging. Finally, we discuss the surveys that will be 
made using the next generation of submm-wave instruments under 
development. 
\end{abstract}

\begin{keyword}
Dust: extinction \sep Cosmology: observations \sep Galaxies: evolution \sep 
Galaxies: formation \sep gravitational lensing \sep Radio continuum: galaxies 
\end{keyword}
\end{frontmatter}

\section{Introduction}

Discovering the process by which the dense, gravitationally bound 
galaxies 
formed in the Universe from an initially almost uniform gas, and 
understanding the way their constituent populations of stars were born
is a key goal of modern physical cosmology. A wide 
range of well understood physical processes are involved; including 
general relativity, gas 
dynamics and cooling physics, nuclear reactions 
and radiative transfer. However, the range of possible initial 
conditions and the non-linear nature of most of the events, starting with 
the collapse of primordial density perturbations, 
ensure that these intimately connected processes can generate a 
very wide range of possible scenarios and outcomes. Galaxy formation 
can be studied by 
attempting to reproduce the observed Universe via analytical models and 
numerical simulations. 
The information required to constrain these models is provided by both 
forensic 
studies of the current constituents of the Universe, including stellar ages, 
chemical abundances and the sizes and shapes of galaxies, and by 
direct observations 
of the galaxy formation process taking place in the young Universe  
at great distances. 
Direct observations exploit  
both the light emitted by distant galaxies, and the signature of 
absorption due to intervening structures along the line of sight, and began  
almost 50 years ago using sensitive optical and radio telescopes. 
Astronomers must now use all available frequencies 
of radiation to probe the properties of the Universe, 
from the lowest energy radio waves to the highest-energy $\gamma$-rays.  
It is vital to combine the complementary information that can be 
determined about the constituents of the Universe at different 
wavelengths in order to make progress in our understanding. 

This review discusses the results of 
a new type of direct observation of the galaxy formation process, 
made possible by the development of powerful new radiation 
detectors sensitive to wavelengths in the range 200\,$\mu$m 
to about 1\,mm: the 
submillimeter (submm) waveband.   
The detection of submm radiation from distant galaxies  
is one of the most recent developments in observational cosmology, 
and has finally brought this region of the electromagnetic spectrum into 
use for making cosmological observations not directly connected with 
the cosmic microwave background (CMB; Partridge and Peebles,\ 1967). 
With the possible exception of the 
hardest X-ray wavebands, studies of distant galaxies in the 
submm waveband  
remained elusive for the longest period. We will also discuss some observations 
at the mid- and far-infrared(IR) wavebands that 
bound the submm waveband at short wavelengths, usually defined 
as the wavelength ranges from about 5--40 and 40--200\,$\mu$m, respectively.

\begin{figure}[t]
\begin{center}
\epsfig{file=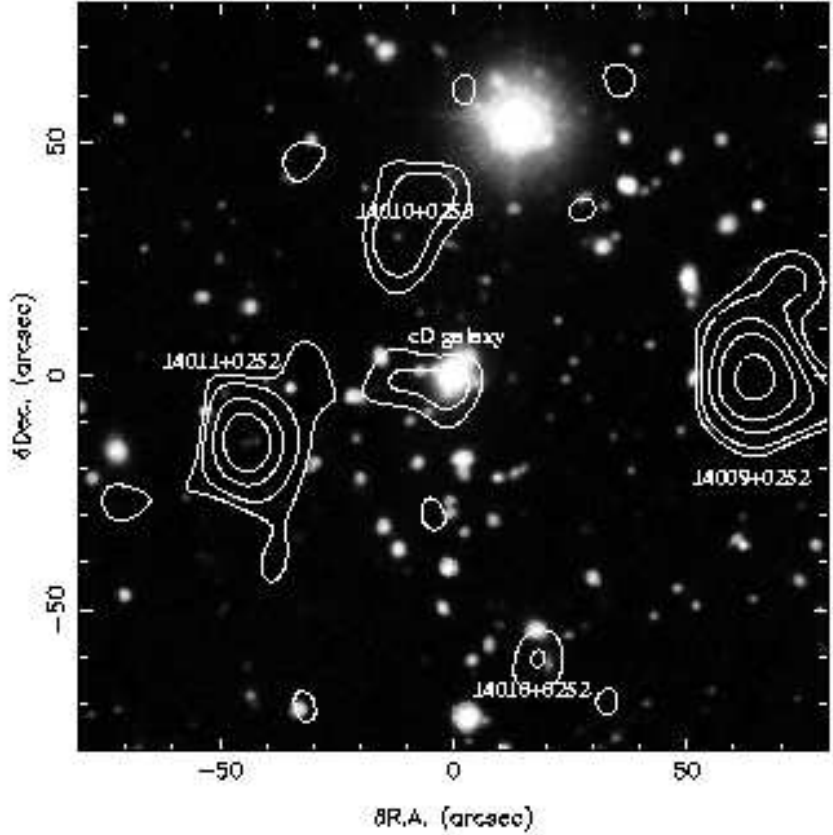, width=12cm} %, angle=-90}
\end{center}
\caption{A comparison of deep optical and submm views of the sky. The 
background image is a 
3-color optical image of the rich cluster of 
galaxies Abell\,1835 at the low/moderate redshift $z=0.25$ 
(Smail et al.,\ 1998b) taken using the 5-m Hale telescope,  
overlaid with the 14-arcsec resolution contours 
of a SCUBA 850-$\mu$m submm-wave 
image of the same field (Ivison et al.,\ 2000a). 
North is up and East to the left.  
The brightest SCUBA galaxies 
at ($-45$,$-15$), (65,0) and (20,$-60$), and the 
central cD galaxy (Edge et al.,\ 1999), all 
have clear radio detections at a frequency of 1.4\,GHz in images with 
higher spatial 
resolution than the SCUBA contours, obtained at 
the Very Large Array (VLA), supporting their reality. 
The bright SCUBA galaxy at ($-$45,
$-$15) is associated with SMM\,J14011+0253, an interacting pair of galaxies 
at redshift $z=2.56$ in the background of the cluster (Frayer et al.,\ 1999). 
Spectacular 
fragmented structure appears in the Easterly red component of this 
galaxy in {\it Hubble 
Space Telescope (HST)} 
images (Fig.\,\ref{fig:blobII}). 
}
\label{fig:A1835}
\end{figure}

The most significant reason for the late flowering of submm 
cosmology is 
the technical challenge of building sensitive receivers 
that work efficiently at the boundary between radio-type  
coherent and optical-like incoherent detection 
techniques. In addition, atmospheric emission and absorption permits sensitive 
submm observations from only high mountain sites, and only in specific 
atmospheric 
windows. The zenith opacity from the best sites in the clearest 
submm atmospheric window at 850\,$\mu$m is 
typically about 0.1. Furthermore, 
the long wavelength of submm radiation limits spatial resolution 
unless very large filled or synthetic 
apertures are available. The largest single apertures 
available 
at present are in the 10--30\,m class, providing spatial 
resolution 
of order 10\,arcsec. This resolution is much coarser than the 
sub-arcsec resolution of optical and near-IR
observations. 
The appearance of the same region of sky at optical and submm wavelengths 
is compared in Fig.\,\ref{fig:A1835} to illustrate this point: 
the multicolor optical image was obtained using the Hale 5-m telescope 
at Mt. Palomar, 
while the 850-$\mu$m submm image was obtained using the 15-m James Clerk Maxwell 
Telescope (JCMT) on Mauna Kea. 
Interferometers can dramatically enhance the resolution of images, but 
so far have only operated at 
longer mm wavelengths. The commissioning of the 8-element Sub-Millimeter 
Array (SMA; Ho, 2000)\footnote{http://sma2.harvard.edu} on 
Mauna Kea in Hawaii with baselines of up to about 500\,m, 
the 
first dedicated submm-wave  
interferometer, will provide images with 
subarcsecond resolution. The much larger 
64-element Atacama Large Millimeter Array 
(ALMA; Wootten, 2001)\footnote{http://www.alma.nrao.edu}  
will be in service at the end of decade. 

A key development was the commissioning of the Submillimetre Common-User 
Bolometer Array (SCUBA) camera at the 
JCMT in 1997 (Holland et al.,\ 1999). 
SCUBA images the sky in the atmospheric 
windows at both 450 and 850\,$\mu$m in a 2.5-arcmin-wide field, using 
hexagonal close-packed arrays of 91 and 37 bolometer detectors 
at the respective wavelengths. SCUBA provided a dramatic 
leap forward from the pre-existing single-pixel or 
one-dimensional array instruments available. The combination 
of field of view and sensitivity was sufficient 
to enable the first searches for 
submm-wave emission from previously unknown distant galaxies. The Max-Planck 
Millimetre Bolometer Array (MAMBO;  
Kreysa et al.,\ 1998) is a 1.25-mm camera with similar capabilities to 
SCUBA, which operates during the winter from the Institut de Radio Astronomie
Millim\'etrique (IRAM) 30-m telescope on Pico 
Veleta in Spain. A similar device---the SEST Imaging Bolometer Array 
(SIMBA)---designed at Onsala in Sweden is 
soon to begin operation on the 15-m Swedish--ESO
Submillimetre Telescope (SEST)
in Chile, providing a sensitive submm imaging capability in the South. 
The capability of mm and submm-wave observatories is not standing 
still: a number of 
larger, more sensitive mm- and 
submm-wave cameras are under construction, including the 
SHARC-II (Dowell et al.,\ 
2001), BOLOCAM
(Glenn et al.,\ 1998) and SCUBA-II instruments.\footnote{Details can 
be found in Table\,\ref{table:instrument}. The next-generation SCUBA-II 
camera for the JCMT is under development at the United Kingdom 
Astronomy Technology Centre (UKATC). See 
http://www.jach.hawaii.edu/JACpublic/JCMT/Continuum\_ 
observing/SCUBA-2/home.html.} 
Bolometer technology continues to advance. The advent of extremely stable 
superconducting 
bolometers that require no bias current and can be  
read out using multiplexed 
cold electronics, should ultimately  
allow the construction of very large submm detector arrays of order 
$10^{4-5}$ elements (for example 
Benford et al.,\ 2000). SCUBA-II is likely to be the 
first instrument to exploit this technology, providing a  
$8 \times 8$-arcmin$^2$ field of view at the resolution limit of the 
JCMT. 

The first extragalactic submm/mm surveys using SCUBA  
and MAMBO 
revealed a population of very luminous high-redshift galaxies, which 
as a population, were responsible for the release of a significant fraction of 
the energy generated by all galaxies over the history of the Universe 
(Blain et al.,\ 1999b). Almost 200 of these galaxies are now known 
(Smail et al.,\ 1997; 
Barger et al.,\ 1998; Hughes et al.,\ 1998; 
Barger et al.,\ 1999;  
Eales et al.,\ 1999, 2000; 
Lilly et al.,\ 1999; Bertoldi et al.,\ 2000; Borys et al.,\ 2002; 
Chapman et al.,\ 2002a; Cowie et al.,\ 2002; 
Dannerbauer et al.,\ 2002; 
Fox et al.,\ 2002; Scott et al.,\ 
2002; Smail et al.,\ 2002; Webb et al.,\ 2002a).
There is strong evidence that almost all of these galaxies are at 
redshifts greater than unity, and that the median redshift of the population 
is likely to be of order 2--3 (Smail et al.,\ 2000, 2002). However, 
only a handful of these objects have certain redshifts and well-determined 
properties at other wavelengths (Frayer et al.,\ 1998, 1999; Ivison et al.,\ 
1998, 2001; Kneib et al.,\ 2002). The results of these mm/submm 
surveys provide complementary information to deep surveys for galaxies 
made in the 
radio (Richards, 2000), far-IR (Puget et al.,\ 1999),  mid-IR (Elbaz et al.,\ 
1999) and optical (Steidel et al.,\ 1999) wavebands. Submm 
observations are a vital component of the search for a coherent picture 
of the formation and evolution of galaxies, which draws on data 
from all wavebands where the distant Universe can be observed. 

In this review, we describe the key features of the submm 
emission processes in galaxies. We 
summarize the current, developing state of submm-wave observations of 
distant galaxies, 
including the results of both blank-field surveys, and
targeted observations of known  
high-redshift galaxies, including radio-galaxies, 
optically-selected quasars/QSOs, X-ray detected active galactic nuclei (AGNs) 
and optically-selected Lyman-break 
galaxies (LBGs). Submm-wave surveys are not immune to selection effects, 
and we discuss their strengths and weaknesses. We describe the 
properties of the class of submm-luminous galaxies, and discuss 
the key results that are required to make 
significant progress in understanding them.  
We consider the relationship between the submm-selected 
galaxies and other populations of high-redshift galaxies, and describe 
models that can account for the properties of 
submm-selected galaxies. We introduce  
the unusually significant effects of the 
magnification of distant submm-selected
galaxies due to gravitational lensing (Schneider et al.,\ 1992). 
Finally, we recap the key developments that are keenly 
awaited in the field, and describe some of the exciting science that will be 
possible in the next decade using future instruments. 

The cosmological parameter values assumed are generally listed where they 
appear. We usually adopt a flat world model with a Hubble constant 
$H_0 = 65$\,km\,s$^{-1}$\,Mpc$^{-1}$, a density parameter in matter 
$\Omega_{\rm m} = 0.3$ 
and a cosmological constant $\Omega_\Lambda = 0.7$. 

\section{Submm-wave emission from galaxies}

There are two major sources of submm radiation from galaxies: 
thermal continuum emission from dust grains, the solid phase of
the interstellar medium (ISM), and line emission from atomic and molecular 
transitions in the interstellar gas. 
The ladder of 
carbon monoxide (CO) rotational transitions, spaced every 115\,GHz, is the 
most important source of molecular line emission, but there is a 
rich zoo of other emitting molecules in the denser phases of the ISM.  
Submm surveys for distant galaxies have so far been made using cameras 
that detect only continuum dust emission, and so this will be the main focus of 
this review. However, the search for line emission is already important, 
and its study will become 
increasingly significant. The spectral resolution provided by line 
observations reveals much more about the 
physical and 
chemical conditions in the ISM, for studies of kinematics, metallicity and 
excitation conditions. Molecular lines can also be used to obtain a
very accurate spectroscopic redshift for the ISM in 
high-redshift galaxies with prior optical redshifts 
(for example Frayer et al.,\ 1998). Searches for redshifts at cm and 
(sub)mm wavelengths using CO lines 
will be possible using future telescopes. 

\begin{figure}[t]
\begin{center}
\epsfig{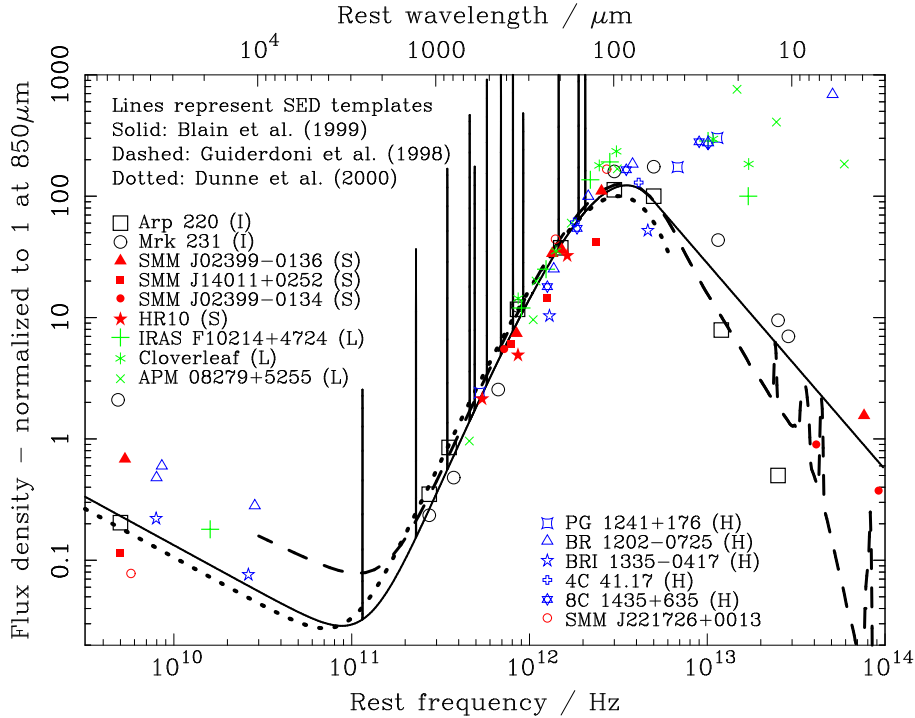}
\end{center}
\caption{Various observed restframe spectral energy distributions (SEDs) 
of galaxies from the radio to 
the near-IR wavebands. Two examples of the most luminous 
low-redshift galaxies detected by {\it IRAS}  
are included (I). Five very luminous high-redshift galaxies that 
have been, or could have been,  
detected directly in deep submm surveys (S), three high-redshift galaxies 
serendipitously magnified and made easier to study by the gravitational lensing 
effect of 
foreground galaxies and also detected by {\it IRAS} (L), 
and five high-redshift AGNs detected in optical or radio surveys (H) 
are also shown.
In addition, three template SEDs are shown. One includes 
the properties of CO and atomic fine-structure emission lines 
in the (sub)mm waveband at wavelengths from 100 to 3000\,$\mu$m 
(Blain et al.,\ 2000b), one includes polycyclic 
aromatic hydrocarbon (PAH) molecular emission 
features at wavelengths $\sim 10$\,$\mu$m 
in the mid-IR waveband (Guiderdoni et al.,\ 1998),  
and one is normalized to the typical SED of a sample of low-redshift {\it IRAS} 
galaxies (Dunne et al.,\ 2000). For further information on 
far-IR SEDs see Dale et al.\ (2001). With the exception of the 
high-redshift AGNs and the lensed galaxies, the 
templates tend to provide a reasonable description of the SED at
wavelengths around and longer than its peak, the regime probed by 
submm surveys. Less luminous galaxies like the Milky Way 
have dust spectra that peak at a wavelength about a factor of 2 
longer than these templates (Reach et al.,\ 1995). 
} 
\label{fig:SED}
\end{figure}

The best studied regions of the Universe in the submm waveband are 
Giant Molecular Clouds (GMCs) in the Milky Way, in which ongoing 
star formation is 
taking place (Hollenbach and Tielens, 1997). 
GMCs are perhaps very low-luminosity archetypes 
for distant dusty galaxies,  
although these galaxies have 
far-IR luminosities that are 
up to 4 orders 
of magnitude greater than that of the whole Milky Way. 

Detailed, resolved submm-wave images and spectra only 
exist for low-redshift galaxies (for example 
Regan et al.,\ 2001; Sakamoto et al.,\ 
1999), and it is often necessary to use them as 
templates to interpret the properties of more distant galaxies. 
A very important 
class of well-studied galaxies similar in luminosity, and perhaps in 
physical properties, to high-redshift submm galaxies are the 
ultraluminous 
IR galaxies (ULIRGs) discovered in the 
{\it InfraRed Astronomy Satellite (IRAS)} all-sky 
survey in the mid 1980's (see the review by 
Sanders 
and Mirabel, 1996). ULIRGs are usually defined as having a 
bolometric luminosity, integrated over all wavelengths at which dust 
emission dominates the SED (from about 1\,mm--8\,$\mu$m), in excess of 
$10^{12}$\,L$_\odot$.\footnote{ 
1\,L$_\odot$ = $3.84 \times 10^{26}$\,W} 
They are amongst the most luminous of all galaxies, but 
number less than 0.1\% of galaxies in the 
local Universe. Due to their selection by {\it IRAS}, they  
are typically at relatively low redshifts, less than 
about 0.3. The first {\it IRAS}-detected high-redshift ULIRG  
was identified by Rowan-Robinson et al.\ (1991) at $z=2.3$. The 
current record redshift for a galaxy detected by 
{\it IRAS} is $z=3.9$ for APM\,08279+5255 (Irwin et al.,\ 1998). Both these 
galaxies appear to be extremely luminous; however, their luminosities are 
boosted by at least a factor of ten due to gravitational lensing 
by foreground galaxies. A compilation of the properties of some 
of the most extreme ULIRGs is given by Rowan-Robinson (2000). 
The IR spectral energy distributions (SEDs) of some 
low-redshift ULIRGs and a 
compilation of results for the more sparsely sampled SEDs of 
high-redshift dusty galaxies are illustrated in Fig.\,\ref{fig:SED}. 

\subsection{The power source for dusty galaxies} 

About 99\% of the energy released by galaxies in the submm and far-IR
wavebands is produced by thermal emission from dust grains; the 
remainder comes from 
fine-structure atomic and molecular rotational line 
emission. However, the source of the energy to power this emission by heating 
dust is often unclear. Any intense
source of optical/ultraviolet (UV) radiation, either young high-mass stars or an 
accretion disk surrounding an AGN, would heat dust grains. Because 
dust emits a featureless modified blackbody spectrum, submm continuum 
observations can reveal 
little information about the physical conditions within the source. Regions 
of intense dust emission are very optically thick, and so little information 
can be obtained by observing optical or UV radiation. 

In typical spiral 
galaxies, with relatively low far-IR 
luminosities of several 10$^{10}$\,L$_\odot$ (for 
example Alton et 
al.,\ 2000,
2001),
the dust emission is known to be significantly 
extended, on the same scale as the 10-kpc stellar  
disk.\footnote{1\,pc = $3.09 \times 10^{16}$\,m} The emission 
is certainly associated 
with molecular gas rich star-forming regions distributed throughout the 
galaxy (Regan et al.,\ 2001), in which dust is 
heated 
by the hot, young OB stars. 

In intermediate luminosity  
galaxies, such as the interacting pair of spiral galaxies NGC\,4038/4039 
`the Antennae'
(Mirabel et al.,\ 1998; Wilson et al.,\ 2000), the most intense knots of 
star-formation activity, from which most 
of the luminosity of the system emerges, are not coincident with either 
nucleus of the merging galaxies, but occur in a deeply dust-enshrouded 
overlap region of the 
ISM of the galaxies. This provides a strong 
argument that almost all of the energy in this system is being generated 
by star formation rather than an AGN.  

In more
luminous ULIRGs that are at sufficiently low redshift for their 
internal structure to be resolved, the great majority of the dust emission 
arises in a much smaller, sub-kpc region (Downes and Solomon, 1998; 
Sakamoto et al.,\ 1999) within a merging system of galaxies. It is 
plausible that a significant fraction of the energy could be derived from an  
AGN surrounded by a very great column density of gas and dust that 
imposes many tens of magnitudes of extinction on the emission from the AGN 
in the optical and UV wavebands, and 
which remains optically thick 
even at near-IR wavelengths.  Alternatively, an ongoing centrally 
condensed burst of star-formation activity, fueled by gas funneled 
into the center of the potential well of a pair of interacting galaxies by a 
bar instability (Mihos, 2000) is an equally plausible power source. 

If the geometry of absorbing and scattering material is known or assumed, 
then radiative transfer models can be used to predict the SED of a galaxy, 
which should differ depending on whether the source of heating is a very small
AGN with a very hard UV SED, or a more extended,
softer-spectrum nuclear star-forming region (for example
Granato et al., 1996). Note that the results are expected to be 
very sensitive to the assumed 
geometry (Witt et al., 1992). In merging galaxies this  
geometry is highly 
unlikely to be spherical or cylindrical, and is 
uncertain for the  
high-redshift galaxies of interest here. In the case of AGN heating, the 
SED would be is expected to peak at 
shorter wavelengths and the mid-IR SED would be expected to be flatter 
as compared with a more extended star-formation power source. Both these
features would correspond to 
a greater fraction of hot dust expected in AGNs  
(see Fig.\,\ref{fig:SED}), and is seen clearly 
in the SEDs of low-redshift {\it IRAS}-detected QSOs  
(Sanders and Mirabel, 1996). 

An alternative route to probing energy sources in these galaxies is provided 
by near- and mid-IR spectroscopy. At 
these longer wavelengths, the optical depth to the nucleus is less than in 
the optical/UV, 
and so the effects of the more 
intense, harder UV radiation field expected in the environs of an AGN 
can be observed directly. These include the 
excitation of 
characteristic highly-ionized lines, and the 
destruction of relatively fragile polycyclic aromatic hydrocarbon (PAH) 
molecules (Rigopoulou et al.,\ 1999; Laurent et al.,\ 2000; 
Tran et al.,\ 2001), leading to the suppression of their  
distinctive emission and absorption features. Mid-IR 
spectroscopic observations with the successor to {\it IRAS}, the 
{\it Infrared Space Observatory (ISO)} 
in the mid 1990's indicated that 
most of the 
energy from low-redshift ULIRGs is likely 
generated by star-formation activity rather than 
AGN accretion. However, the fraction of 
ULIRGs containing AGN 
appears to increase at the highest luminosities (Sanders, 1999). This could 
be important at high redshifts, where the typical luminosity of 
dust-enshrouded galaxies is greater than in the local Universe. In addition, 
there 
may be duty-cycle 
effects present to make an AGN accrete, and perhaps to be visible, 
for only a fraction of the duration of 
a ULIRG phase in the evolution of the galaxy 
(Kormendy and Sanders, 1988; Sanders et al.,\ 1988; 
Archibald et al.,\ 2002). 
X-ray observations also offer a 
way to investigate the power source, as all but the densest, 
most gas-rich galaxies, 
with particle column densities greater than $10^{24}$\,cm$^{-2}$ 
are transparent to hard ($>2$\,keV) X\,rays. 

Ultra-high-resolution radio observations provide a route to probing the 
innermost regions of ULIRGs (Smith et al.,\ 1998; Carilli and Taylor,  
2000). By detected the diffuse emission and multiple point-like radio sources, 
expected from multiple supernova remnants, rather 
than a single point-like core and accompanying 
jet structures expected from an AGN, 
these observations suggest that high-mass star formation contributes at 
least a significant part of the luminosity of the ULIRGs Arp\,220 and 
Mrk\,273.  

It is interesting to note that the observed correlation between the inferred 
mass of the black holes in the centers of galaxies 
and the stellar 
velocity dispersion of the surrounding galactic bulges, in which most 
of the stars in the Universe reside (Fukugita et al., 
1999), might 
inform this discussion  
(Magorrian et al.,\ 1998; Ferrarese and Merritt, 2000; Gebhardt et al.,\ 2000).  
The mass of the bulge appears to exceed that of the 
black hole by a factor of about 200. When hydrogen is processed 
in stellar nucleosynthesis, the mass--energy conversion efficiency is about 
$0.007 \epsilon_*$, where $\epsilon_* (\simeq 0.4)$ is the fraction of 
hydrogen burned in high-mass stars. 
When mass is accreted onto a black hole, the mass--energy conversion efficiency is expected to be about 0.1$\epsilon_{\rm BH}$, with $\epsilon_{\rm BH} \sim
1$ with the definition above. 
If accretion and nucleosynthesis were to generate the same amount of 
energy during the formation of a galaxy, then 
the ratio of mass contained in both processed stars and stellar remnants 
to that 
of a supermassive black hole is expected to be about 0.1$\epsilon_{\rm BH} / 
0.007 \epsilon_*$. For $\epsilon_* = 0.4$ and $\epsilon_{\rm BH}=1$, this 
ratio is about 36. As a mass ratio of about 200 is observed, 
this implies that a greater amount of energy, by a factor of about 6, is 
generated by high-mass star-formation activity than by gravitational accretion. 

If the bulge-to-black-hole mass ratio is in fact greater than 200, then 
either the factor by which star formation dominates 
will exceed 6, or the accretion must have been more 
than 10\% efficient; that is $\epsilon_{\rm BH}>1$. If  
low-efficiency accretion dominates the process of the build up of mass in the 
central black hole, then less than 1 part in 7 of the luminosity 
generated during 
galaxy formation 
will be attributed to accretion as compared with high-mass star formation. 

A greater amount of energy generated by star formation as compared with 
accretion processes appears to be favored by these circumstantial 
arguments. 

\subsection{Continuum emission from dust} 

The dust emission process is thermal, with dust grains emitting a 
modified blackbody spectrum. Grains of interstellar dust, distributed 
throughout the ISM of a galaxy, 
are heated to temperatures between about 20 and 200\,K, depending on 
the spectrum and intensity of the interstellar radiation field (ISRF), 
and the size and 
optical properties of the grains. Higher dust temperatures can be 
produced close to a powerful source of radiation, 
with dust subliming at temperatures 
of order 2000\,K. Very small grains can be heated far above their 
equilibrium temperatures 
by absorbing 
hard-UV photons (see Draine and Li, 2001). Lower dust temperatures, 
always exceeding the CMB temperature, are possible in opaque 
regions of the ISM that are shielded from intense heating, 
in the intergalactic medium or in regions 
with an intrinsically weak ISRF. Unless dust is heated 
by the ISRF in addition to the CMB the galaxy will not be detectable. 
We now consider the properties of the dust emission that are relevant to 
observations of high-redshift galaxies. 

\subsubsection{The emission spectrum, dust mass and temperature} 

The minimum parameters necessary to describe the emission from 
dust grains are a temperature $T_{\rm d}$ and a form of the 
emissivity function $\epsilon_\nu$. In any galaxy there 
will be a distribution of dust temperatures, reflecting the different 
nature and environment of each grain. It is useful to use 
$T_{\rm d}$ to describe the coolest grains that contribute significantly 
to the energy output of a galaxy when discussing submm observations. 
In most cases, spatially and 
spectrally resolved images of galaxies are not available, and so it is 
reasonable to assume a 
volume-averaged description of the 
emissivity function as a function of frequency $\nu$,  
$\epsilon_\nu \propto \nu^\beta$. 
Values of $\beta$ in the range 1--2 are usually assumed. Scattering theory
predicts that $\beta \rightarrow 2$ at low frequencies, while a value
$\beta \simeq 1$ at high frequencies matches the
general trend of the interstellar extinction curve that describes the 
properties of absorption of optical and UV radiation by the ISM 
(see Calzetti et al.,\ 2000 and Section\,2 of the review by 
Franceschini, 2002).

The simplest form of the emission spectrum/SED, $f_\nu$ is 
given by assuming that $f_\nu \propto \epsilon_\nu B_\nu$, in which 
$B_\nu$
is the Planck function ($2 k T_{\rm d} \nu^2 / c^2$ in the Rayleigh--Jeans 
limit,
in units of W\,m$^{-2}$\,Hz$^{-1}$\,sr$^{-1}$).  
This assumes that the emitting source is optically thin. 
For fitting spectra of
galaxies found in deep submm surveys, we assume the simple
$\epsilon_\nu B_\nu$ function to describe the SED. Dunne et al.\ (2000)
and Dunne and Eales (2001) also 
use this functional form to fit the observed submm
spectra of low-redshift galaxies. At the expense of adding another parameter
to describe the SED, there is 
some physical motivation for a SED that includes an optical depth term, 
\begin{equation} 
f_\nu \propto [ 1 - \exp(- \tau_\nu) ] B_\nu, 
\end{equation} 
where $\tau_\nu$ is the frequency-dependent 
optical depth of the cloud, 
and is a multiple of $\epsilon_\nu$. This equation tends to the simpler 
$\epsilon_\nu B_\nu$ function at long wavelengths, and is assumed by,  
for example, 
Benford et al.\ (1999), Omont et al.\ (2001), Priddey and McMahon (2001) 
and Isaak et al.\ (2002), whose submm
data for high-redshift AGNs tends to correspond to 
rest-frame frequencies that are relatively close to the peak of the SED. The 
extra parameter required 
to relate $\tau_\nu$ and $\epsilon_\nu$ can be defined as 
the frequency at which $\tau_\nu=1$ and the cloud becomes optically 
thick. If the opacity near a wavelength 
of 100\,$\mu$m is large, then the 
form of the peak of the SED tends to that of a blackbody spectrum. 
This suppresses 
the emission near to the SED peak relative to the 
emission in the Rayleigh--Jeans
regime, and so this functional form provides a good fit to a set of 
submm and far-IR 
data with a higher 
value of $T_{\rm d}$ as compared with the $\epsilon_\nu B_\nu$ function, 
usually by about 10--20\%. However, because 
most observed SEDs for high-redshift galaxies 
have fewer than four data points (see 
Fig.\,\ref{fig:SED}), the 
difference is unlikely to be very significant. 

It is reasonable to assume that the mid-IR SED can be 
smoothly interpolated from a modified blackbody function at low 
frequencies to a power-law 
$f_\nu \propto \nu^\alpha$ in the mid-IR waveband on the 
high-frequency side of the spectral peak, in order to prevent the  
high-frequency SED from falling exponentially with a Wien spectrum. 
Hotter components of dust, emitting at shorter wavelengths, and ultimately 
stellar emission in the near-IR waveband, are certain to be present 
to reduce the steepness of the SED in the Wien regime.
That an exponential Wien spectrum is inappropriate
can be seen from the well-defined 
power-law mid-IR SEDs of
Arp\,220 and Mrk\,231 shown in Fig.\,\ref{fig:SED}.  
 
It is not always necessary to relate the SED $f_\nu$ and luminosity $L$ of 
a galaxy
to the mass of dust $M_{\rm d}$ that it contains; this can of worms can remain 
closed by normalizing $f_\nu$ in a self-consistent way. However, 
if a dust mass is required, perhaps in order to estimate the metal 
content of the ISM, and so provide information about the integrated 
star-formation activity in the galaxy at earlier times
(Hughes et al., 1997; Omont et al.,\ 2001), then it is  
conventional to define a frequency-dependent 
mass-absorption coefficient $\kappa_\nu$ (Draine and Lee, 1984; 
with units of m$^2$\,kg$^{-1}$), which is proportional to $\epsilon_\nu$. 
$\kappa_\nu$ is the `effective area' for blackbody emission by a certain mass 
of dust,  
\begin{equation} 
L { {f_\nu} \over {4\pi \int f_\nu' \,{\rm d}\nu'}} = 
\kappa_\nu B_\nu M_{\rm d}. 
\end{equation} 
Values of $\kappa_\nu$ at a conventional frequency of around 1\,mm 
are in the range 0.04--0.15\,m$^2$\,kg$^{-1}$ (Hughes, 1996). 
Recent comparisons of optical extinction and submm 
emission from partially resolved edge-on spiral galaxies
have tended to give values of 
0.05--0.4\,m$^2$\,kg$^{-1}$ (see Fig.\, 4 of Alton et al.,\ 2001). 
Domingue et al.\ (1999) derive 0.09\,m$^2$\,kg$^{-1}$ from 
similar far-IR, optical 
and submm data. Dunne et al.\ (2000) adopt a value of 
0.077\,m$^2$\,kg$^{-1}$. Note that there is at least a factor of 3 
uncertainty in these conversion factors. 

An alternative dimensionless function $Q_\nu$ (Hildebrand, 1983) is sometimes
used, which includes information about the mass/volume and surface area 
of a typical grain. If grains are assumed
to be spherical (a big if), with bulk density $\rho$, 
radius $a$, and an emissive cross section $\pi a^2$, then
$\kappa_\nu = 3 Q_\nu / 4 a \rho$. $Q_\nu B_\nu$ is the effective emissivity 
function describing the energy flux from unit area of the
dust grain surface. However, 
dust grains are more likely to be 
irregular in shape, possibly colloidal or in the form of 
whiskers. In that case, the emissivity per unit mass would be increased, and 
the dust mass associated with a fixed luminosity would be overestimated. 

This geometrical uncertainty will inevitably result in 
uncertainty about the mass of dust. Hence, dust masses quoted 
in papers must be treated with caution, and may be best used as a 
comparative measure to distinguish galaxies. 
In general, we will avoid quoting dust masses, as this is
unlikely to provide a reliable physical measure of the properties of galaxies
until detailed resolved images are available, which 
is likely to require observations with the ALMA interferometer. 
This will be a recurring 
theme: observations with 
excellent sensitivity and spatial resolution using 
a large interferometer will resolve many of the questions raised 
throughout the paper. 

Working from submm data, it is also difficult to assess the 
dust mass of a galaxy, even subject to the caveats above, 
without knowing its dust temperature. In the 
Rayleigh--Jeans spectral regime, 
the flux density from a galaxy 
$S_\nu \propto \nu^{2+\beta} M_{\rm d} T_{\rm d}$. If $T_{\rm d}$ is uncertain 
to within a factor, then $M_{\rm d}$ is uncertain to within the same 
factor. The dust mass is 
at least easier to estimate from a single long-wavelength 
observation than the luminosity $L$. As $L \propto M_{\rm d} 
T_{\rm d}^{4+\beta}$, or equivalently $L \propto S_\nu T_{\rm d}^{3+\beta}$,
an uncertainty in $T_{\rm d}$ corresponds to a proportionally much larger 
uncertainty in the inferred value of $L$. 

However, even if the dust mass can be determined reliably at low redshifts, 
it remains unclear whether the same procedure can be applied to 
determine the dust mass in more luminous and more distant 
systems. 
In order to determine the 
dust properties of high-redshift
galaxies, data of the same quality that has been obtained 
for nearby galaxies is required. High-frequency
submm/far-IR observations are necessary 
to provide information about the rest-frame frequency of the peak 
of the SED for a high-redshift galaxy. 

Given the current lack of resolved images of distant galaxies in 
the submm and far-IR wavebands, it is important to neither 
over-parametrize the descriptions nor overinterpret the results 
of observations of their SEDs. When 
spatially-resolved, high-spectral resolution images are 
available, building on existing interferometric 
images of low-redshift dusty  
galaxies (Downes and Solomon, 1998; Sakamoto et al.,\ 1999; 
Wilson et al.,\ 2000), 
it should be possible to study the radiative transfer from sites of 
intense star formation and AGN in these geometrically complex 
opaque galaxies 
(see Ivison et al.,\ 2000a, 2001).  
Models of the SEDs of dust-enshrouded AGN at different viewing 
angles have been developed by Granato et al.\ (1996), while  
star-forming regions embedded in a disk geometry have been analyzed by 
Devriendt et al. (1999). More powerful and efficient 
radiative-transfer 
codes are being developed (for example 
Abel et al., 1999), and it should be 
practical to develop detailed models of 
the appearance of galaxies with realistic geometries to account for 
future, high-resolution  
multi-band submm images. 

At present, we prefer to use a few simple parameters---$\alpha$, 
$\beta$ 
and $T_{\rm d}$---to describe the essential features of the SEDs of 
dusty galaxies. 
Although such a model can encapsulate only a small part of 
the true complexity of the 
astrophysics in a galaxy, it can 
account for the existing SED data for 
a wide variety of dusty galaxies. A simple parametrization is 
preferable to a more baroque, and necessarily at present unconstrained, 
combination of geometry, dust mass and temperature. 
In the following section we list plausible values of our SED parameters 
and discuss the associated degeneracies in fitted values.  

\subsection{The observed SEDs of dusty galaxies} 

Information about the submm SEDs of galaxies has been 
gathered from targeted 
mm and submm observations of samples of low-redshift far-IR-selected 
galaxies from the {\it IRAS} catalog, 
(Andreani and Franceschini, 1996; Dunne et al.,\ 2000; 
Lisenfeld et al., 2000; Dunne and Eales, 2001), and from 
far-IR and submm observations of high-redshift 
galaxies (see Fig.\,\ref{fig:SED}). The 
most extensive local survey (SLUGS; 
Dunne et al.,\ 2000) consists of 850-$\mu$m SCUBA observations 
of 104 galaxies selected from the low-redshift {\it IRAS} 
Bright Galaxy Sample (BGS; 
Soifer et al.,\ 1987).
After fitting single-temperature $\epsilon_\nu B_\nu$ SEDs to the galaxies, 
Dunne et al. found that $\beta = 1.3 \pm 0.2$ and $T_{\rm d} 
= 38 \pm 3$\,K 
described the sample as a whole, 
with a natural dispersion in the properties from galaxy to galaxy. This 
{\it IRAS}-selected sample could be
biased against less dusty galaxies. Dunne et al.\ are  
currently addressing this issue by observing a complementary sample 
of $B$-band selected low-redshift galaxies, which should be representative
of optically luminous low-redshift galaxies as a whole.

\begin{figure}[t]
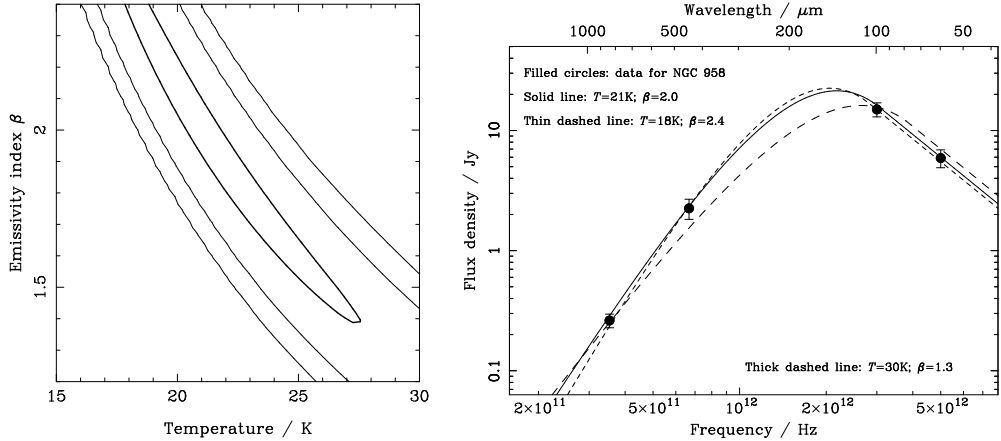

\begin{center}
\epsfig{file=fit_Dunne.ps, width=5.8cm, angle=-90}
\hskip 3mm  
\epsfig{file=SED_Dunne.ps, width=5.8cm, angle=-90}
\end{center}
\caption{An illustration of some of the issues involved in 
describing the SEDs of dusty galaxies. On the left is a probability 
contour plot that shows the 0.5, $5\times10^{-3}$ and 
$5\times10^{-5}$ probability contours
for a fit to an SED model defined by the variable 
parameters $\beta$ and $T_{\rm d}$ 
with a fixed value of $\alpha=-1.95$, 
taking into account four SED datapoints for the 
galaxy NGC\,958 as shown in the right-hand panel 
(Dunne and Eales, 2001). Note that 1\,Jy
= 10$^{-26}$\,W\,m$^{-2}$\,Hz$^{-1}$. 
Note that there is 
a very significant degeneracy in the fitted parameters. 
Adding additional data points with 
small errors close to the peak of the SED at 200\,$\mu$m reduces the extent 
of the probability contours by about 50\%, 
but they remain elongated in the same direction. Note that $\beta > 2$ 
is not expected physically. 
On the right the data are compared with fitted single-temperature SEDs. 
The solid line is 
the best fit to the data. 
The dashed lines correspond to SEDs 
from the ends of the probability `banana' shown 
in the left-hand panel. Note that 
without the 450-$\mu$m point, the thick dashed curve describes the 
best-fit SED, which is defined by 
a significantly greater dust temperature. This SED
is similar to that of a typical luminous 
IR galaxy, whereas
the best fitting model with all four data points is much more 
like the SED of the Milky Way. Note that the shift in the best-fit 
model on adding 450-$\mu$m data is generally less significant than in 
this case.} 
\label{fig:Dunne}
\end{figure}

Note, however, that when fitting only a few datapoints, there is a
significant correlation between values of $\beta$ and $T_{\rm d}$ that can 
account for the data (left panel of Fig.\,\ref{fig:Dunne}). This 
can lead to 
ambiguity in the results, further emphasizing the difficulty in 
associating the dust mass or temperature inferred from a galaxy 
SED with the real physical properties of the galaxy. 

The addition of 450-$\mu$m data for 19 of the 104 
galaxies in the SLUGS 
sample (Dunne and Eales, 2001), tends to split the galaxy SEDs 
into two categories: those that retain a definite 40-K 
spectrum after including the 450-$\mu$m data, 
and those for which cooler single-temperature SEDs, more similar to 
the SEDs of normal spiral galaxies, then provide a better 
fit. The first group are typically 
the more luminous galaxies in the sample, while the second 
includes 3 of the 5 lowest luminosity galaxies from the sample.   
Dunne and Eales (2001) propose a two-temperature model to account 
for the changes in light of the new 450-$\mu$m data; however, a cooler 
single-temperature model with a larger value of $\beta$ provides a fit of
similar quality. 
The results for one of the most significantly 
different fits is shown in the right-hand panel of Fig.\,\ref{fig:Dunne}. 
With the addition of the 450-$\mu$m data, the nature of the SEDs of 
low-redshift, 
low-luminosity galaxies become more 
diverse. However, 
the more luminous galaxies, which are likely to be the most similar to  
typical high-redshift submm galaxies, are still described 
reasonably well by the 
original Dunne et al.\ (2000) 38-K SED. 

An alternative approach is to determine an 
SED that can describe the observed flux density distribution of galaxies in 
the 
far-IR and submm wavebands, which are sensitive 
to galaxies at low, moderate and high redshifts  
(Blain et al.,\ 1999b; Trentham et al., 
1999; Barnard and Blain, 2002). 
Using the $\epsilon_\nu B_\nu$ functional form, 
values of $\beta \simeq 1.5$ and $T_{\rm d} \simeq 40$\,K are required to 
provide a good description of the data, rather similar to the values 
derived for temperatures of individual low-redshift luminous dusty 
galaxies in Dunne et al.\ (2000) and 
Lisenfeld et al.\ (2000), and for both the small number of high-redshift 
submm-selected galaxies with known
redshifts and mid-IR spectral constraints (Ivison et al.,\ 1998a, 2000a)
and typical high-redshift QSOs (for example Benford et al.,\
1999).
These temperatures 
are significantly less than those determined for the most extreme 
high-redshift galaxies (Lewis et al.,\ 1998), 
and significantly 
greater than the $T_{\rm d} = 17$\,K inferred from the maps of 
the Milky Way made using the all-sky survey from the 
FIRAS instrument on the {\it Cosmic Background Explorer (COBE)} 
satellite in the early 1990's 
(Reach et al.,\ 1995). Note that there are examples of moderate-redshift
infrared-selected galaxies with both hotter and colder typical dust 
temperatures than 40\,K: see Deane and Trentham (2001) and 
Chapman et al.\ 
(2002d) respectively. 
At present
it seems likely that a 40-K dust temperature is a reasonable assumption 
for high-redshift submm-selected galaxies. 

Inevitably, however, there will be a population of hotter high-redshift 
galaxies (Wilman et al., 2000; 
Trentham and Blain, 2001). These 
galaxies would be underrepresented in 
existing submm surveys, but 
may make a significant contribution 
to the 240-$\mu$m background radiation intensity (Blain and 
Phillips, 2002). 
Further observational information to test the assumption of a 40-K dust 
temperature is keenly awaited. 
As we discuss below, in Section 2.6, the assumed dust temperature 
has a significant effect on the selection function of submm galaxy 
surveys, and on the properties that are inferred for the galaxies that are 
found in these surveys. 

\subsection{Line emission} 

Emission from  
molecular rotation and atomic fine-structure transition lines 
can be used to diagnose physical conditions within molecular clouds and
photodissociation regions, and to trace out the velocity structure 
within. Some lines, such as those from CS, HCN and HCO$^+$ are excited only in
high-density gas, while others, including the most abundant polar species CO, 
trace more
typical regions in the ISM. 

Studies of many emission lines from 
molecular cloud regions in 
nearby galaxies are possible using existing mm and submm-wave telescopes 
(Wilson et al.,\ 2000; Helfer, 2000). 
However, for more distant galaxies only CO lines have so far been detected 
in significant numbers, almost exclusively from galaxies 
which have been 
subject to strong gravitational lensing by foreground galaxies 
(see the summary in Combes et al., 1999). 
These observations are useful for deriving physical conditions within the 
sources, especially if multiple lines are detected (as in the 
case of APM\,08279+5255; Downes et al.\ 1999b). The improved 
capabilities of the forthcoming mm/submm interferometer arrays---SMA,  
upgrades to the IRAM Plateau du Bure interferometer (PdBI), and the
Combined Array for Research in Millimeter-wave Astronomy
(CARMA)\footnote{http://www.mmarray.org/}---and 
and ultimately the dramatically increased 
sensitivity of ALMA, will make high-redshift lines much easier to 
observe over the next decade (Combes et al.,\ 1999; Blain et al.,\ 2000b).  

One of the most important uses of CO-line 
observations of distant submm galaxies found in continuum surveys 
is their ability to 
confirm an identification absolutely, by tying together an optical and 
submm redshift at the position of the galaxy. 
So far this has been achieved for only three submm 
galaxies (Frayer et al.,\ 1998, 1999; 
Kneib et al.,\ 2002: see Figs. \ref{fig:blob}, \ref{fig:blobII} and 
\ref{fig:ring}). In principle, these observations could be made for 
all continuum-selected galaxies. 
The difficulty is the narrow fractional bandwidth 
available for receivers and correlators. Even at the relatively low 
frequency of 90\,GHz, the redshift of the target must be known to better
than 0.5\% to ensure that a 300\,km\,s$^{-1}$ wide CO line, typical of a 
massive galaxy, with a width equivalent to 0.1\% in redshift 
falls entirely within a 1-GHz band. 
Future cm-, mm-, and submm-wave instruments with  
wider bandwidths will significantly assist the 
the search for redshifts using molecular lines.    
Specially designed low-resolution, ultra-wideband dispersive 
spectrometers covering many 
tens of GHz simultaneously on single-antenna mm-wave 
telescopes also promise 
to provide redshifts for submm galaxies
(Glenn, 2001). 

A complementary search for redshifted cm-wave OH megamaser emission 
to pinpoint the redshifts and positions of 
ultraluminous high-redshift galaxies could be possible using radio 
telescopes 
(Townsend et al.,\ 2001). However, there are very stringent requirements 
on the acceptable level of radio frequency 
interference from terrestrial and satellite 
communications. Observations of low-redshift megamasers are described by 
Darling and Giovanelli (2001).  
Megamaser emission at high redshifts is discussed by Briggs (1999) in 
the context of the proposed Square Kilometer Array (SKA) 
meter/centimeter-wave radio telescope. If it can operate at frequencies 
of several 10's of GHz, then 
the SKA is also likely to be an efficient detector of 
low-excitation high-redshift CO lines (Carilli and Blain, 2002). 

\subsubsection{Line emission contribution to continuum detections} 

An interesting feature of the CO line emission from low-redshift galaxies 
is that lines can lie in the passbands of continuum instruments, and 
could contribute to the continuum flux inferred. For low-redshift galaxies, 
the 345-GHz CO($3\rightarrow2$) line lies within the 
850-$\mu$m atmospheric window, while the 691-GHz CO($6\rightarrow5$) and 
230-GHz CO($2\rightarrow1$) lines lie in the 
450-$\mu$m and 1.25-mm windows, respectively. 

Assuming a 
reasonable template spectrum (Blain et al.,\ 2000b), the equivalent width in 
frequency of 
the CO($3\rightarrow2$) line is 7.4\,GHz. The 
passband of the current SCUBA 850-$\mu$m (353-GHz) filter is about 120\,$\mu$m
(50\,GHz) wide, and so about 15\% 
of the measured continuum flux density of a low-redshift galaxy in the 
850-$\mu$m channel 
is likely to be from the CO line. 
The high-frequency SCUBA passband in the 450-$\mu$m atmospheric window
is 75\,GHz wide, while the equivalent width of the 
CO($6\rightarrow5$) transition is 3.3\,GHz. Hence, a smaller 5\% 
contribution to the continuum flux density from the line is expected at 
450\,$\mu$m.
The CO($2\rightarrow1$) line 
has an expected equivalent
width of 9.2\,GHz, while the wide MAMBO passband 
has half-power points at
210 and 290\,GHz. Contamination of the flux densities detected by MAMBO 
by about 10\% may thus be expected. 

The largest of these correction factors is comparable to the calibration 
uncertainty in 
submm-wave observations, and could be relevant to the detailed 
interpretation of low-redshift observations. 
For example, 
the presence of the CO($3\rightarrow2$) line in the 850-$\mu$m 
window would shift the inferred continuum emissivity spectral index $\beta$ 
in the SLUGS survey from 1.3 to 1.52. 
At high redshifts, any corrections are likely to be less significant, both 
because the 
relatively bright CO($3\rightarrow2$) line redshifts out of the 850-$\mu$m 
passband, and the equivalent width of lines in frequency space decreases as
$(1+z)^{-1}$. 

Although the contribution to measured submm-wave flux densities from 
line emission could be significant at the level of order 10\%, only a 
small fraction of the 
bolometric luminosity from galaxies is detected in the submm waveband. 
More than 99\% of the bolometric luminosity still appears in the 
continuum, predominantly at shorter far-IR wavelengths. 

\begin{figure}[t]
\begin{center}
\epsfig{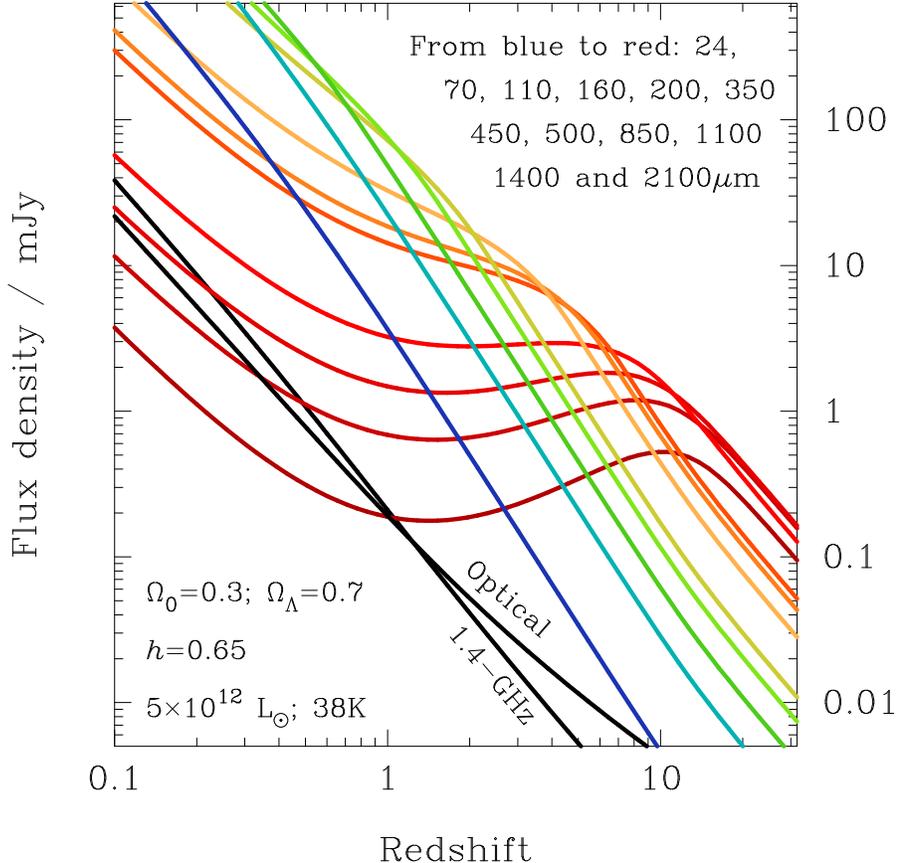}
\end{center}
\caption{The predicted flux density of a dusty galaxy as a function of redshift 
in various submm atmospheric windows, 
and at shorter wavelengths that will be probed by forthcoming space 
missions. Note the powerful K correction in the mm and 
submm wavebands at wavelengths longer than about 250\,$\mu$m, which 
yields a flux density that is almost independent of redshift. 
The template spectrum is chosen to reproduce the 
typical 
properties of distant submm-selected galaxies 
(Fig.\,\ref{fig:SED}). Subtle 
effects due to the additional heating of dust by the CMB, and fine details 
of the radio SED of galaxies are not included; these 
effects are illustrated in
Fig.\,\ref{fig:Svz_highz}.  
} 
\label{fig:Svzcolor}
\end{figure}

\subsection{The observability of high-redshift dusty galaxies} 

The detectable flux density at an observed frequency $\nu$ from a galaxy with 
bolometric luminosity $L$ at redshift 
$z$ with an intrinsic SED $f_\nu$, 
\begin{equation} 
S_\nu = { {1+z} \over {4\pi D_{\rm L}^2}} L { { f_{\nu(1+z)} } \over 
{\int f_{\nu '}\, {\rm d}\nu'} }, 
\end{equation} 
where $D_{\rm L}$ is the luminosity distance to redshift $z$ (for 
example Peebles, 1993).

The key feature that makes submm-wave observations of distant 
galaxies interesting is the ability to sample the SED of a
target galaxy at wavelengths for which the SED is a strongly 
increasing function of frequency (Fig.\,\ref{fig:SED}). 
This ensures that distant 
galaxies are observed at a rest-frame wavelength closer to the peak of 
their SED. There is thus a strong, negative K correction, which 
leads to high-redshift
galaxies being relatively easy to detect at submm wavelengths as 
compared with their low-redshift counterparts. This effect is illustrated 
in Fig.\,\ref{fig:Svzcolor} for 
the template SED from Blain et al.\ (1999b) shown in Fig.\,\ref{fig:SED}. 
The strong K-correction effect 
applies at wavelengths longer 
than about 250\,$\mu$m. At these wavelengths 
the flux density from galaxies at $z > 1$ ceases 
to decline with the inverse square of distance, but instead remains 
approximately constant with increasing redshift. A window is 
thus opened to the detection of 
all galaxies with similar SEDs 
at redshifts up to $z \simeq 10$--20. The effect is 
more pronounced at longer wavelengths: in the mm waveband 
more distant galaxies are expected to produce greater flux densities than 
their more proximate counterparts. 

Note that both the radio and optical flux-density--redshift 
relations decline steeply with increasing redshift, and so 
high-redshift galaxies are not selected 
preferentially in those wavebands. The advantage that faint radio and 
optical galaxy surveys have over submm surveys comes from  
the complementary probe of astrophysical signatures, and the 
combination of greater fields 
of view and finer angular 
resolution. 

A submm telescope that is sufficiently sensitive to detect 
a certain class of galaxy at redshift 
$z \simeq 0.5$, can detect any similar galaxies out to a 
redshift $z \sim 10$ (Blain and Longair, 1993a). Note, however, that surveys 
to exploit this unusual K correction are not 
immune to selection effects. 
The K correction can also only be exploited at 
redshifts for which 
sufficient heavy elements are present in the ISM of the target galaxy to 
form enough dust to reprocess optical radiation. Nor does 
the K correction effect overcome cosmological 
surface brightness dimming for progressively more distant submm 
galaxies: the 
normal $(1+z)^{-4}$ reduction in surface brightness still applies; however, 
it is not expected to become significant 
until redshifts in excess of about 5. 
Because submm-wave telescopes do 
not yet resolve distant galaxies, this effect cannot be observed at 
present. It may provide an opportunity to estimate redshifts for the 
most distant submm-selected galaxies when they can be resolved using ALMA. 

\begin{figure}[t]
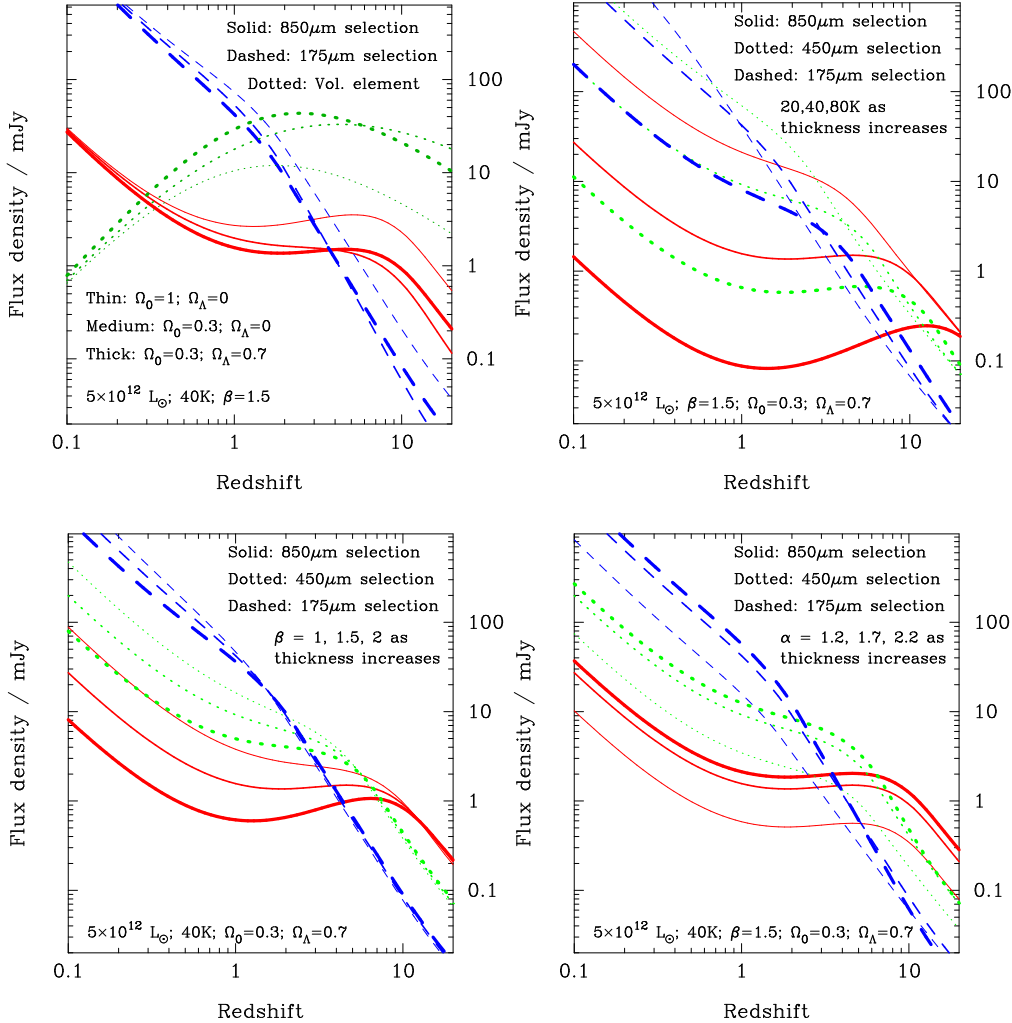

\begin{center}
\epsfig{file=Svz_sel_cos.cps, width=6.5cm, angle=-90} 
\epsfig{file=Svz_sel_T.cps, width=6.5cm, angle=-90}
\end{center} 
\vskip 0.5cm 
\begin{center} 
\epsfig{file=Svz_sel_b.cps, width=6.5cm, angle=-90}
\epsfig{file=Svz_sel_a.cps, width=6.5cm, angle=-90}
\end{center}
\caption{Flux-density--redshift relations illustrating some 
of the uncertainties that apply to the interpretation of 
submm and far-IR surveys. 
In the top-left panel the relatively small 
effects of changing the world model parameters are shown. Note that 
changes to the volume element and flux density received counteract each 
other, and so the chosen cosmology has a small effect on 
the interpretation of surveys. 
The most significant effect---that of 
changing the dust temperature---is shown in the top-right panel.  
At 175\,$\mu$m for galaxies at moderate redshifts, the effect of 
temperature is rather small. However, at 850\,$\mu$m, the 
effects are very significant, and must be remembered when interpreting 
the results of 850-$\mu$m 
observations: doubling the dust temperature 
corresponds to increasing the luminosity associated with a given 
flux density by a factor of about 10. The 
less significant effects of changing the dust emissivity index  
$\beta$ or the mid-IR spectral index $\alpha$ are shown in the bottom-left  
and bottom-right panels respectively.  
}
\label{fig:sel}
\end{figure}

\subsection{Submm-wave selection effects} 

Deep submm-wave observations image the high-redshift 
Universe with very little contamination from low-redshift 
galaxies, and can potentially find a population of galaxies that is quite 
different to those detected in conventional deep optical surveys, and which 
could be undetectable in these surveys. The 
complementarity of submm and optical observations is illustrated by 
the very limited overlap between galaxies detected in the deep submm--optical 
image shown in Fig.\,\ref{fig:A1835}. However, submm surveys 
are certainly subject to 
selection effects. In Fig.\,\ref{fig:sel}
the flux-density--redshift relation for a submm-luminous 
galaxy with a fixed bolometric luminosity is presented as a function of
its SED parameters -- $T_{\rm d}$, $\alpha$ and $\beta$. The relatively minor
effects of different assumed cosmological models are also shown. 
Changing the dust temperature has the greatest effect. The inferred luminosity 
of a dusty galaxy for a fixed observed submm flux density goes up by a 
factor of 10 if the dust temperature is doubled, at all but the very 
highest redshifts.
There is thus a significant potential bias in 
submm surveys against the detection of galaxies with 
hotter dust temperatures for a given bolometric luminosity. 

This effect was noted by Eales et al.\ (1999), when investigating the 
evolution of galaxies in the context of the 
results of deep SCUBA surveys. They suggested that the submm galaxies 
may be cooler than the temperatures of about 60\,K usually assumed, and 
so their significance as 
a population of strongly evolving high-redshift 
galaxies may have been overestimated. 

As discussed in Section\,2.3, a cooler dust temperature of 40\,K is 
compatible 
with observations of the SEDs of individual submm galaxies
with confirmed redshifts 
detected in submm surveys (Ivison et al.,\ 1998a, 2000a) and with the results 
of targeted observations of luminous low-redshift {\it IRAS} galaxies 
and high-redshift QSOs.
If this temperature is assumed, then the 
inferences about galaxy evolution made from the results of 
submm surveys (Blain et al.,\ 1999b, c; Eales et al.,\ 2000; 
Smail et al.,\ 2002) should be reliable. 
However, until a large sample of submm 
galaxies with redshifts and multi-waveband 
SEDs is available, the possibility that a cold or hot 
population of high-redshift dusty galaxies could be missing from or 
misidentified in submm
surveys cannot be ruled out (Eales et al.,\ 1999; 
Blain and Phillips, 2002). The possible 
effects on inferred luminosities of different forms of the 
SED shown in Fig.\,\ref{fig:sel} 
need to be taken seriously, especially when describing the properties 
of individual galaxies selected in submm surveys. 

There is little reliable evidence  
for a systematic relationship between dust temperature and redshift. 
Observations of low-redshift 
{\it IRAS} galaxies (Andreani and Franceschini, 1996; Dunne et al.,\ 2000), 
indicate that any variation of dust temperature with luminosity 
appears to be gradual. However, there is evidence for a significant 
and systematic change in the temperature of dusty galaxies with a 
wider range of luminosities, from about 
20\,K for low-redshift spirals (Reach et al.,\ 1995; Alton et al.,\ 2000; 
Dunne and Eales, 2001) to about 40\,K for more luminous objects typical 
of the galaxies detected in the {\it IRAS} survey. Temperatures 
of up to 110\,K are found for some extremely luminous high-redshift galaxies 
(Lewis et al.,\ 1998). 

We stress that there could be a significant 
selection effect in 
submm surveys that 
depends on the range of dust temperatures in the source population. 
The importance of such an effect can be quantified once a 
complete redshift distribution is available for a submm-selected 
galaxy sample. 

\subsection{Deep submm-wave surveys} 

Images of the redshift $z=0.25$ cluster of galaxies Abell\,1835 
in both the optical and submm wavebands 
were compared in Fig.\,\ref{fig:A1835}. This image provides 
a realistic impression of the appearance of deep optical and 
submm images of the sky. 
Note that the only relatively low-redshift or 
cluster member galaxy that contributes any submm-wave flux is the 
central cD galaxy: 
the other cluster galaxies are either quiescent, neither forming stars nor 
heating dust, or are insufficiently luminous to be detectable at 850\,$\mu$m 
using SCUBA. Background galaxies at much greater redshift, which have 
faint optical counterparts as compared with the cluster member 
galaxies, dominate the 
image.
This is a direct visual demonstration of the strong bias towards the
detection of distant galaxies in submm-wave surveys that was 
illustrated in 
Fig.\,\ref{fig:Svzcolor}. 
These background galaxies are magnified by a 
factor of order 2--3 due to the gravitational lensing 
potential of the foreground cluster over the full extent of the 
image. The effects of gravitational lensing can be determined using 
accurate models of the cluster potential, that are constrained with 
the help of data from {\it HST} images 
and spectroscopic redshifts for multiply-imaged optically-selected 
galaxies. The uncertainty in the results is comparable to the 
uncertainty in the calibration of the submm images. 

As shown in Fig.\,\ref{fig:A1835}, existing deep submm images  
are much less 
visually stimulating than deep optical images, because their angular 
resolution 
is not sufficient 
to image the internal structure in  
distant galaxies. The limited resolution also imposes a 
confusion limit to the depth for submm surveys, at which the  
noise level is dominated not by atmospheric or instrumental noise but 
by the telescope resolution blurring together signals from faint 
unresolved galaxies. It takes about 50\,h of integration using SCUBA to 
reach the practical confusion limit in a single field. Confusion is 
discussed in more detail in Section\,3.1.  

There are a variety of published results from deep submm galaxy surveys. 
Surveys aim to detect high-redshift galaxies exploiting the powerful 
K-correction effect in the submm waveband. 
Over 500\,arcmin$^2$ of blank sky has 
been surveyed using SCUBA by several groups  
(Barger et al.,\ 1998; Hughes et al.,\ 1998; 
Barger et al., 1999a; Eales et al.,\ 1999, 2000; Borys et al.,\ 2002; 
Fox et al.,\ 2002; Scott et al.,\ 2002; Webb et al.,\ 2002a). 
These range from an extremely deep 
survey in the area of the Hubble Deep Field-North (HDF-N; Williams et al.\
1995) by 
Hughes et al.\ (1998) searching for the faintest detectable 
populations of submm galaxies, to wider-field shallower surveys to 
detect brighter sources that might be easier to follow-up and could be 
used to trace large-scale structure (Borys et al.\ 2002; Scott et al.\ 2002). 
About 30 5-arcmin$^2$ lensed cluster fields have been imaged using SCUBA 
(Smail 
et al.,\ 1997, 2002; Chapman et al.,\ 2002a; Cowie et al.,\ 2002;  
Kraiberg Knudsen et al.,\ 2001; van der Werf and Kraiberg 
Knudsen, 2001), to various RMS depths between 0.5 and 8\,mJy.\footnote{1\,Jy 
= 10$^{-26}$\,W\,m$^{-2}$\,Hz$^{-1}$} By exploiting the magnification effect of 
gravitational lensing, which extends over fields several arcminutes across, 
due to rich 
clusters of galaxies at moderate redshifts, the population of distant 
galaxies in the 
source plane behind 
the lensing cluster can be probed to greater depths than is possible in a 
blank field (Blain, 1998). 

The detection 
rate of galaxies using SCUBA based on published papers 
appears to have declined 
over time since 1998. In significant part this is due to the absence of 
the sustained excellent observing conditions on Mauna Kea that were experienced 
during the El Nino winter of 1997--1998, just after SCUBA was 
commissioned. 

A wide variety of 
complementary, and sometimes
overlapping surveys have been made using MAMBO at the IRAM
30-m telescope (Bertoldi et al.,\ 2000, 2001; 
Carilli et al.,\ 2001), during  
several winters. These fields include the cluster Abell\,2125 and the 
ESO-NTT Deep Field (Arnouts et al.,\ 1999). 

A compilation of the results from all the SCUBA and MAMBO surveys 
is presented in Figs.\,\ref{fig:count1} and \ref{fig:count2}. 
A full summary of deep projects that 
have been undertaken or are underway can be found in Ivison (2001). 
In addition, larger shallower submm surveys of the Galaxy 
(Pierce-Price et al.,\ 2000),\footnote{ 
A JCMT project 
(Vicki Barnard et al.) is currently searching for 
candidate high-redshift galaxies detected in wide-field
SCUBA images of star-forming regions in the Milky Way 
(Barnard et al.,\ 2002).} 
and perhaps the CMB images obtained using the 
BOOMERANG balloon-borne experiment 
(Masi et al.,\ 2001), 
can be used to search 
for brighter submm-wave galaxies.

\subsection{Submm observations of known   
high-redshift galaxies and QSOs} 

The advent of SCUBA and MAMBO has also 
provided the opportunity to study the submm properties 
of large samples of interesting high-redshift 
galaxies, including almost all types of previously known distant galaxies. 
Isolated detections of high-redshift AGN-powered radio galaxies and QSOs 
were made in the mid 1990's 
using single-element bolometer detectors (for example Dunlop et al.,\ 1994; 
Isaak et al.,\ 1994); however, the compilation of statistical samples, and 
the secure rejection of
contamination from fluctuating atmospheric noise  
have only been possible more recently, using 
SCUBA and MAMBO, and the 350-$\mu$m one-dimensional
bolometer array SHARC at the 10.4-m aperture 
Caltech Submillimeter Observatory (CSO) on Mauna Kea. 
A key advantage of observing these sources is that both their redshifts
and some of their astrophysical properties are already known, in contrast with 
the submm-selected galaxies discovered in 
blank-field surveys. Some of the targeted galaxies---very faint 
non-AGN radio galaxies, 
mid-IR-selected {\it ISO} galaxies, and X-ray selected AGNs---have only 
been detected very recently. As the relationship between these populations 
of galaxies and 
submm-selected galaxies is still unclear, many of the limits 
will be discussed in 
the context of following up submm surveys in Section\,4. 

Targeted surveys include a search for submm-wave 
continuum emission from 
high-redshift AGN-powered 
radio galaxies (Archibald et al.,\ 2001), and observations 
of various samples of optically-selected QSOs (for example 
Carilli et al.,\ 2001; 
Isaak et al.,\ 2002). In these observations a single bolometer is 
aimed at the position of the target. While this does not 
lead to a fully-sampled 
image of the sky, it provides a more rapid measurement of the flux 
density at a chosen position. The 
results have been the detection and characterization of the  
dust emission spectra for a range of luminous high-redshift galaxies 
and QSOs, 
including APM\,08279+5255 (Lewis et al.,\ 1998), the galaxy with the 
greatest apparent luminosity in the Universe. 
Barvainis and Ivison (2002) have targeted all 
the known galaxies magnified into multiple images 
by the gravitational lensing 
effect of foreground galaxies from the CASTLES gravitational lens 
imaging project,\footnote{http://cfa-www.harvard.edu/castles.} 
significantly expanding the list of high-redshift galaxies magnified 
by a foreground mass concentration with a submm 
detection. The SEDs of several of these galaxies are shown in  
Fig.\,\ref{fig:SED}. 

Archibald et al. (2001) find evidence for
significant evolution in the properties of dust emission with increasing
redshift in a carefully selected sample of AGN radio galaxies, whose
radio properties were chosen to be almost independent of the 
redshift of the observed galaxy. The results perhaps indicate that 
more intense
star-formation activity, as traced by the submm emission, takes place 
alongside the radio source activity at higher
redshifts, and so provide a possible clue to the formation and 
evolution of the massive 
elliptical galaxies thought to host radio galaxies.
Hughes et al.\ (1997), Ivison et al.\ (1998b) and
Omont et al.\ (2001) discuss the consequences of finding large masses of
dust at high redshifts, in terms of the limited cosmic time available
for the formation of the stars required 
to produce the metals and dust required to generate sufficiently intense
submm emission from the host galaxy. 
  
The radio galaxies detected by Archibald et al.\ (2001) in pointed 
single-bolometer SCUBA observations were 
followed up by imaging observations of the surrounding 5-arcmin$^2$ 
fields, 
to search for submm-loud companions.
Ivison et al. (2000b) found that the surface density 
of submm galaxies in some of these fields
is about an order of magnitude greater than 
that in a typical blank field, indicating a significant 
overdensity of sources. This is likely due to some radio galaxies being 
found in high-density regions of biased high-redshift galaxy formation, 
which are possibly `protoclusters'---rich clusters of galaxies in 
the process of formation.    

A similar targeted approach has been taken to try to detect 
submm-wave emission 
from optically-selected LBGs at redshifts 
between 2.5 and 4.5 (Steidel et al.,\ 1999). 
The Lyman-break technique (Steidel et al.,\ 1996) detects the 
restframe 91.2-nm neutral hydrogen absorption 
break in the SED of a galaxy as it passes through 
several broad-band filters. Large 
samples of candidate LBGs can be gathered 
using multi-color optical 
images from 4-m class telescopes. The 
efficiency of the selection method is of order 70\% after spectroscopic 
confirmation of the candidates using 8/10-m class telescopes.
The LBGs are 
the largest sample of spectroscopically 
confirmed high-redshift galaxies, with a well-defined luminosity 
function (Adelberger and Steidel, 2000), a 
surface density of order 10\,arcmin$^{-2}$, and  
inferred star-formation rates between 1 and 10\,M$_\odot$\,yr$^{-1}$. 
They appear to 
be typical of the population of distant galaxies, and their spectra 
provide useful 
astrophysical information. 

Observing the LBGs at submm wavelengths is an important goal, as 
an accurate determination of their submm-wave properties  
will investigate the link (if any) 
between the large well-studied LBG  
sample and the more enigmatic 
submm galaxy population (Blain et al., 1999c; Lilly et al.,\ 1999; 
Adelberger and 
Steidel, 2000; Granato et al.,\ 2001). At present, the
typical very faint limits to the optical counterparts of the subset of 
submm
galaxies with accurate positions 
(Smail et al.,\ 1998a, 2002; Downes et al.,\ 1999; Dannerbauer et 
al., 2002), and 
the detection of Extremely Red Object (ERO) galaxies (with $R-K>6$) 
as counterparts to a significant fraction of submm galaxies 
(Smail et al.,\ 1999, 2002; Gear et al.,\ 2000; Frayer et al.,\ 2002; 
Ivison et al.,\ 2001; 
Lutz et al.,\ 2001), 
argue against a large overlap
between the two populations. 
The direct submm detection of LBGs using SCUBA 
has been largely unsuccessful at the 0.5-mJy RMS 
level: a single galaxy out of 16 was detected by 
Chapman et al.\ (2000; 2002c), while Webb et al.\ (2002b) 
describe a low significance of overlap between LBGs and SCUBA 
galaxies in a wide-field survey.  
The LBG cB58 at $z=2.72$ (Ellingson
et al.,\ 1996; Frayer et al.,\ 1997; 
Seitz et al.,\ 1998; Pettini et al.,\ 2000), which is magnified 
strongly (by a factor of 10--20) by a foreground cluster of 
galaxies at $z=0.37$, and is at 
least ten times brighter than a
typical LBG, was detected in the mm and submm 
by Baker et al.\ (2001) and van der Werf et al.\ (2002). However, after 
correcting for lensing, its 
850-$\mu$m flux density is only about 0.1\,mJy, below the 
level of confusion noise in SCUBA images, 
and similar to the flux density level of the statistical detection 
of high-redshift LBGs in the 850-$\mu$m SCUBA image of 
the HDF-N (Peacock et al.,\ 2000). 
In the field surrounding an overdensity 
of LBGs at $z=3.09$, Chapman et al.\ (2001a) were successful 
in detecting 
bright submm emission that appears to be associated with 
diffuse sources of Lyman-$\alpha$ emission at the redshift of the 
overdensity, but were not 
included in the Lyman-break catalog. A key point to note is that the 
limits on submm-wave emission from LBGs are typically 
lower than expected if the relationship between UV spectral slope and 
far-IR luminosity observed for low-redshift low-luminosity 
starburst galaxies (Meurer et al.,\ 1999) 
continues to high redshifts. Goldader et al.\ (2002) indicate that the 
relationship does not appear to 
hold for the most luminous galaxies. 

The required sensitivity for successful 
submm observations of typical LBGs seems to 
be deeper than can be achieved using existing instruments. Observations 
using future very 
sensitive, high-resolution  
interferometers 
certainly ALMA, and perhaps CARMA and SMA, will shed 
more light on the submm--LBG connection. 

The advent of the current generation of very sensitive X-ray observatories, 
{\it Chandra} and {\it XMM--Newton}, is generating a large sample of faint, 
hard X-ray sources, the luminosity of which is assumed to be dominated by 
high-redshift AGN (Fabian, 2000). Absorption and Compton scattering 
in large column densities of gas 
preferentially depletes soft X-rays, hardening the X-ray SEDs of 
gas-rich AGN. Such a population of hard, absorbed X-ray sources 
is required in order to account for the cosmic X-ray background 
radiation spectrum, which is harder than the typical SEDs of 
individual low-redshift AGN (Fabian and Barcons, 1992; 
Hasinger et al.,\ 
1996). 
Observations of the limited areas of the sky where both 
submm and X-ray data are available (Fabian et al.,\ 1999; 
Hornschemeier et al.,\ 2000; Mushotzky et al.,\ 2001; Almaini et al.,\ 2002) 
have tended to show little direct overlap 
between the X-ray and submm galaxies, although there are examples of 
X-ray-detected submm-wave galaxies (Bautz et al.,\ 2000). 
The combined 
results of Bautz et al. and Fabian et al. reveal that 2 out of 9 
SCUBA galaxies are detected by {\it Chandra}. In the larger-area 
brighter 8-mJy 
survey, Almaini et al.\ (2001) identify only 1 out of 17 SCUBA galaxies 
using 
{\it Chandra}. Page et al.\ (2002) discuss further the submm properties 
of X-ray sources. 
Perhaps of order 10\% of known submm galaxies have faint hard X-ray 
counterparts that would be typical of dust-enshrouded AGN. 
There is 
also a statistical detection of excess submm-wave emission from the 
positions of faint high-redshift hard X-ray sources (Barger et al.,\ 2001) 
and a positive submm--X-ray galaxy correlation function (Almaini et al.,\ 
2002). 
The lack of strong X-ray emission from a majority of submm galaxies 
lends circumstantial support to the idea that much of their luminosity is 
derived from star formation and not from AGN accretion. However, some 
submm galaxies may have hydrogen column densities, and 
thus optical depths to Compton scattering, that are sufficiently 
great to obscure 
soft X-ray radiation entirely ($>10^{24}$\,cm$^{-2}$). 
Even if they contained powerful AGN, these submm galaxies 
would be very faint in {\it Chandra} 
surveys, which reach detection limits of order 
$10^{-17}$\,erg\,cm$^{-2}$\,s$^{-1}$ at soft 0.5--2\,keV wavelengths 
(Giacconi et al.,\ 2002). 
They may be found in very deep observations using the the 
greater collecting area of {\it XMM-Newton} for hard X-ray photons. 
However, note that the 
15-arcsec resolution of {\it XMM-Newton} leads to confusion 
due to unresolved faint sources in the beam that is likely to  
impose a practical limit of order $10^{-15}$\,erg\,cm$^{-2}$\,s$^{-1}$ 
to the 
depth of a survey in the hard X-ray 2--8\,keV band (Barcons et al.,\ 2002). 
Deconvolution of the images from joint {\it XMM-Newton}/{\it Chandra} 
deep fields, exploiting the sub-arcsec positional information from  
{\it Chandra}, will perhaps allow this limit to be exceeded. 

Finally, galaxies detected in far-IR surveys using {\it ISO} 
(for example Puget et al.,\ 1999) out to redshifts $z \sim 1$ have been 
targeted for SCUBA submm observations (Scott et al.,\ 2000). 
The large
arcmin-scale observing 
beam in 170-$\mu$m {\it ISO} surveys makes identification of 
submm counterparts difficult, but progress has been made by 
combining sub-arcsec resolution deep radio images. 
The results 
include some sources with apparently rather cool dust 
temperatures of order 30\,K (Chapman et al.,\ 2002d), and are 
generally consistent with redshifts less than unity and 
dust temperatures of less than about 50\,K for most of the galaxies. 

\subsection{Alternative strategy for deep submm surveys} 
 
An alternative strategy for finding 
submm galaxies has also been tried: 
targeted submm observations of faint radio-selected  
galaxies with accurate positions and radio spectral index information 
detected at a 1.4-GHz flux density level 
brighter than 40-$\mu$Jy 
using the VLA (for example 
Richards, 2000). The radio 
source population at these faint flux density levels is expected to include  
mostly high-redshift star-forming galaxies, and only a minority of 
sources with the more powerful synchrotron-emitting jets and lobes associated 
with particles accelerated by 
AGN. By selecting those faint radio sources with faint K-band 
counterparts (Barger et al., 2000; Chapman et al.,\ 2001b), it should 
be possible to sift out low-redshift galaxies from the target sample, and
so generate a concentrated sample of luminous high-redshift galaxies to study 
in the submm. 
From Fig.\,\ref{fig:Svzcolor}, it is clear that 
a high-redshift faint 20--100-$\mu$Jy non-AGN 1.4-GHz radio source 
is likely to be a very 
luminous galaxy. 
Chapman et al.\ (2001b) claim that high-redshift 
submm galaxies can be detected using SCUBA at a rate of about one every hour 
using this method, which is significantly more rapid than the rate of about 
one every 
10 h achieved in blank-field surveys. The efficiency of submm 
detection of 
optically-faint 1.4-GHz 20-$\mu$Jy radio galaxies 
at an 850-$\mu$m flux density greater than about 5\,mJy 
is of order 30--40\%. Barger et al.\ (2000) detect 5/15 with $I>24$ 
at 6\,mJy, and Chapman et al.\ 
(2001b) detect 20/47 with $I>25$ at 4.5\,mJy.   

The difficulty comes in 
the interpretation of the results. The additional selection conditions of 
requiring first a radio detection, and then a faint optical counterpart 
will inevitably lead
to the omission of a certain fraction of the submm galaxy 
population. Galaxies missing would include both the order of 15\% of distant 
submm-selected 
galaxies that have relatively bright optical counterparts---the `Class-2' 
SCUBA galaxies (Ivison et al.,\ 2000a; Smail et al.,\ 2002)---and 
any very high-redshift
submm galaxies that cannot be detected at the VLA, despite lying on 
the far-IR--radio correlation (Condon, 1992; see Fig.\,\ref{fig:Svz_highz}).
There is also likely to be
a bias (at all redshifts) towards both detecting AGNs, in which 
radio flux densities are 
boosted above the level expected from the standard radio--far-IR
correlation, and the exclusion of a small population of low-redshift 
submm galaxies, 
which would be too bright 
in the radio and $K$\,band to be included in the survey. 
The rare, and perhaps especially interesting, submm galaxies 
at the lowest and highest redshifts are thus likely to be missing 
from radio pre-selected surveys, as are the distant submm galaxies with 
brighter optical
magnitudes that are likely to be easiest to follow-up and investigate. 

The size of these selection 
effects is difficult to quantify at present. However, their existence can be 
inferred from the diverse multiwaveband properties displayed by 
submm-selected galaxies from imaging surveys, a significant fraction 
of which are known not to be 
detected in deep radio observations (Downes et al.,\ 1999; 
Smail et al.,\ 2000). In some cases, the depth of the radio images 
used to make the comparison could be improved by a factor of several; 
it is possible, but we suspect unlikely, 
that all submm sources have radio 
counterparts lurking just below existing detection thresholds. 

It is instructive to compare the results of the blank-field  
and faint radio pre-selected submm surveys. In a true 260-arcmin$^2$ 
blank-field survey, Scott et al.\ (2002) found surface 
densities of $550^{+100}_{-170}$ 
and $180 \pm 60$\,deg$^{-2}$ submm galaxies 
brighter than 6 and 10\,mJy respectively,  
while Borys et al.\ 
(2002) estimate $164 \pm 28$\,deg$^{-2}$ brighter than 12\,mJy in 
a 125-arcmin$^2$ survey, with a 
rather conservative error estimate from 12 detections. 
The surface densities resulting from the faint radio-selected investigations of 
Barger et al.\ (2000) and Chapman et al. (2001b) are 
430\,deg$^{-2}$ brighter than 4\,mJy and 135\,deg$^{-2}$ brighter than 
10\,mJy.\footnote{Barger et al.,\ (2000) also impose a limit of less than
840\,deg$^{-2}$ brighter than 6\,mJy at 850-$\mu$m, based on serendipitous 
detections of sources in their 2.5-arcmin-wide SCUBA images around the 
AGN radio galaxies.} This indicates 
incompleteness in the faint 
radio-selected counts by at least about 25\%. This is 
perfectly acceptable for 
gathering a list of high-redshift galaxies for further 
study; 
however, for statistical analysis of a large sample of many 
tens of submm galaxies, the 
uncertainty introduced in the derived properties of the population 
due to incompleteness is expected to dominate the 
Poisson uncertainty, and so limit the accuracy
of the inferences that can be made.  
Once the bright counts of 
distant 850-$\mu$m galaxies are known from direct, hopefully unbiased,
large-area SCUBA imaging surveys 
(Fox et al.,\ 2002; 
Scott et al.,\ 2002) and matched 
with very deep radio images 
(Ivison et al.,\ 2002), then the value of this shortcut for compiling 
a large sample of submm galaxies can be assessed. The same strategy 
would be very valuable for wide-field MAMBO 1.2-mm surveys 
(Dannerbauer et al.,\ 2002). At present, some  
care needs to be taken in the interpretation of faint radio pre-selected 
surveys. 

\subsection{Determining redshifts of submm galaxies}

By virtue of their uniform 
selection function with redshift, it is impossible to even indicate 
the redshift of submm galaxies from single-wavelength  
submm flux densities alone. The relatively poor positional accuracy 
of submm images also makes it very 
difficult to identify an unambiguous optical counterpart for 
spectroscopic follow-up. This will be discussed in more  
detail when the properties of individual galaxies are 
described in Section 3. Here we discuss some general features of 
the prospects for determining their redshifts from
submm, far-IR and radio observations. 

\subsubsection{Photometric redshifts from far-IR SEDs} 

\begin{figure}[t]
\begin{center}
\epsfig{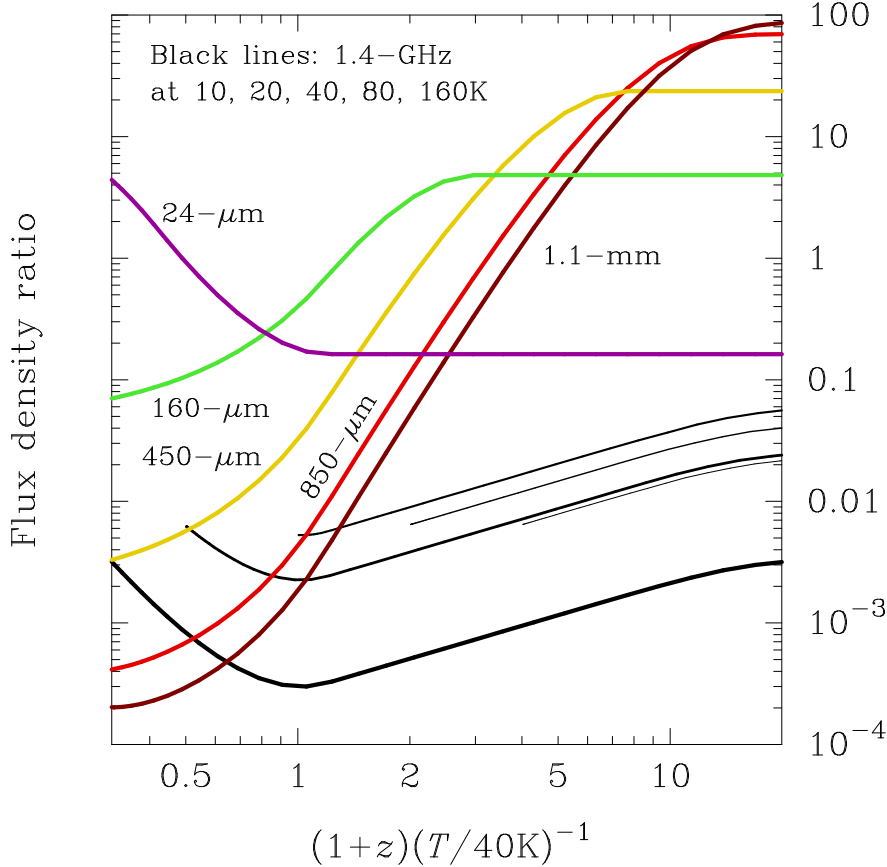}
\end{center}
\caption{The ratio of flux densities expected in different observing bands 
as a function of the degenerate redshift/dust temperature parameter, 
compared with the flux 
density expected in the 70-$\mu$m band at which the {\it SIRTF} satellite 
will be very sensitive. Where the lines have steep gradients, measured 
colors from multi-band 
data locate the 
peak of the dust SED accurately 
in the observer's frame, providing a measurement of temperature--redshift. 
The degeneracy between $T_{\rm d}$ and $z$ can be lifted slightly 
by including radio data (Blain, 1999a; Yun and Carilli, 2002), 
if the dust temperature is 
greater than about 60\,K (see also Fig.\,\ref{fig:CY}). If deep 
near-IR and optical images can be included, and the optical counterpart 
to the galaxy can be 
readily identified, 
then conventional photometric redshifts can be 
determined from stellar synthesis models. 
However, care must be taken as it is unclear whether 
the SEDs of very dusty 
galaxies have familiar restframe-optical spectral breaks. 
}
\label{fig:photz}
\end{figure}

The submm--far-IR SED of dusty galaxies is 
thermal, and so redshifting a fixed template SED affects the observed colors 
in exactly the same way as changing the dust temperature. 
This means that even when multi-frequency far-IR 
data is available, the redshift
will be uncertain, unless there is information about the intrinsic dust 
temperature of the source. This effect is related to the selection 
effect in favor of cold galaxies illustrated in Fig.\,\ref{fig:sel}. 
The expected colors of dusty galaxies, as a function of $(1+z)/T_{\rm d}$, 
the parameter that can be constrained in light of this degeneracy, is shown in 
Fig.\,\ref{fig:photz}. Colors in a variety of observing bands in both  
submm atmospheric windows and for 
the observing bands of the MIPS instrument on the 
{\it Space InfraRed Telescope Facility 
(SIRTF)}\footnote{sirtf.caltech.edu} 
satellite are included. It is vital to stress that without 
knowledge of the dust temperature, it is
impossible to determine a redshift from any combination of 
multicolor broadband 
far-IR/submm data. The degeneracy between 
$T_{\rm d}$ and $z$ is lifted partially by combining information derived 
from radio observations, if $T_{\rm d} > 60$\,K. It may also become clear 
that there is 
a Universal temperature--luminosity relation that extends to high redshifts, 
and can be exploited to 
determine redshift information
using a far-IR--submm color--magnitude relation.  
In the absence of such a Universal relation, far-IR and submm colors  
can only be used to fix the parameter 
$(1+z)/T_{\rm d}$ reliably. 

\subsubsection{Radio--submm photometric redshifts} 

In Fig.\,\ref{fig:Svzcolor}, an estimate of the flux density of 
radio emission of a template 
dusty galaxy was shown as a function of redshift. 
We now investigate the radio properties of 
submm galaxies, which provide useful information about 
their SED and redshift.  
There is an excellent observed correlation between the radio and far-IR 
(60- and 100-$\mu$m) flux densities of low-redshift 
galaxies over 4 orders of magnitude in 
luminosity, reviewed by Condon (1992) and recently 
investigated out to $z \simeq 0.3$ by Yun et al.\ (2001). If 
this correlation is assumed to hold to high redshifts, then  
submm-selected galaxies should be 
detectable in the deepest 10-$\mu$Jy-RMS 1.4-GHz VLA radio images out to 
redshifts of order 3.
Note that as a result of this correlation, optical spectroscopy of 
faint non-AGN radio galaxies 
alone can be used to trace the evolution of star-formation activity 
to redshifts $z \simeq 1.2$, beyond which spectroscopic 
redshifts are hard to determine (Haarsma et al.,\ 2000).
 
The far-IR--radio correlation is thought to be due to a match between 
the rate at which the optical/UV radiation from young stars is absorbed by 
dust on a local scale in  
star-forming regions of galaxies, and 
re-emitted as thermal far-IR 
radiation, and the radio luminosity  
from the same regions (Harwit and Pacini, 1975). The radio luminosity is 
due to  
both free--free 
emission in H{\sc ii} regions, and more importantly at frequencies 
less than about 10\,GHz, to the 
synchrotron emission from relativistic electrons accelerated in supernova 
shocks. If an AGN is present, then it is likely that its accretion disk will  
provide 
an additional source of UV photons to heat dust, and both the disk and 
outflows will generate 
shocks to accelerate relativistic electrons.
There is little reason to expect these effects to be proportional, unlike 
UV heating by massive stars and particle acceleration by 
supernova shocks in star-forming regions. 
Radio-quiet QSOs tend to lie on the radio-loud side 
of the far-IR--radio correlation, while radio-loud AGN can lie 
up to three orders of magnitude away. 

\begin{figure}[t]
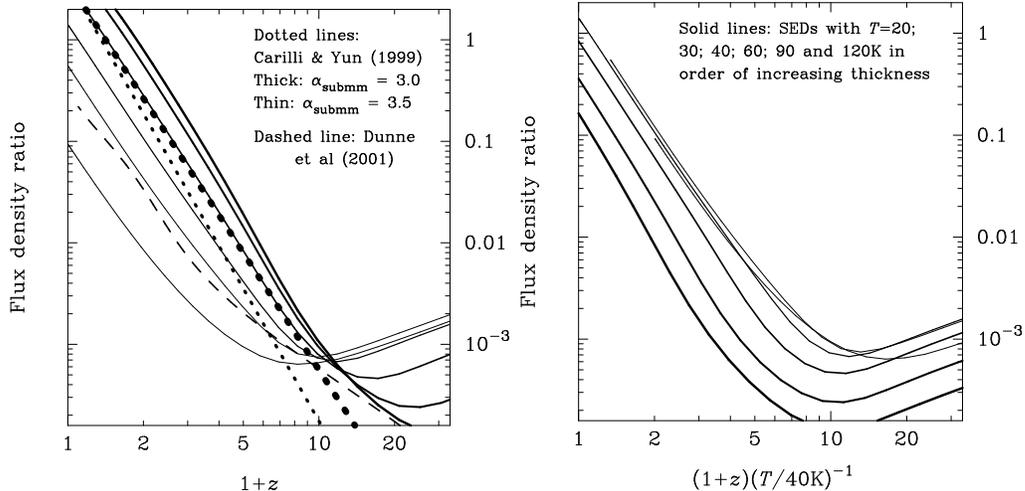

\begin{center}
\epsfig{file=CY_z.ps, width=6.5cm, angle=-90}
\epsfig{file=CY_Tz.ps, width=6.5cm, angle=-90}
\end{center} 
\caption{The behavior of the Carilli--Yun 1.4-GHz to 
850-$\mu$m radio--submm redshift indicator. 
The left panel shows the ratios of 1.4-GHz:850-$\mu$m 
flux density predicted from empirical SEDs by Carilli and 
Yun (1999; dotted lines) 
and Dunne et al.\ (2000; dashed line; see 
also Carilli and Yun, 2000). Predictions for the ratio based 
on the results of Blain (1999a) are also shown assuming radio--far-IR 
SEDs 
with various dust 
temperatures, but which all lie on the far-IR--radio 
correlation (Yun et al.,\ 2001; solid lines). The flux ratio is a 
good indicator of redshift, clearly separating high- and low-redshift 
galaxies.  
Both synchrotron and free--free 
radio emission are included, and the dust temperature and radio properties 
evolve with redshift self-consistently, as modified by the CMB.  
In the right panel, the solid curves are replotted 
as a combined function of temperature and redshift, emphasizing 
that for $T_{\rm d} < 60$\,K, the inferred temperature and 
redshift are degenerate, just as for a thermal spectrum 
(Fig.\,\ref{fig:photz}). For $T_{\rm d} > 60$\,K 
the flux ratio becomes a non-degenerate 
redshift indicator. 
}
\label{fig:CY}
\end{figure}

Carilli and Yun (1999, 2000) demonstrated that the radio--submm color 
is a useful redshift indicator, assuming that dusty galaxies have simple
synchrotron SEDs in the radio waveband and thermal dust spectra in the 
submm and far-IR wavebands. The radio--submm color was also considered 
in the interpretation of the redshifts of galaxies detected in the 
submm surveys by Hughes et al.\ (1998), 
Lilly et al.\ (1999) and Eales et al.\ (2000). 
The Carilli--Yun redshift indicator is subject to a degeneracy between dust 
temperature and redshift, at dust temperatures less than about 60\,K: 
see Fig.\,\ref{fig:CY} in which its form is shown for 
a variety of SEDs. 
Hot, distant galaxies are difficult to 
distinguish from cool, low-redshift ones (Blain, 1999a, b). 
Despite this degeneracy, the Carilli--Yun redshift indicator
is very useful, especially for investigating 
optically faint submm galaxies for which almost no other
information is available (Smail et al.,\ 2000). AGN, 
which are expected to be radio-loud as compared with the 
standard far-IR--radio correlation, lead to conservatively low 
Carilli--Yun estimated redshifts. 

\begin{figure}[t]
\begin{center}
\epsfig{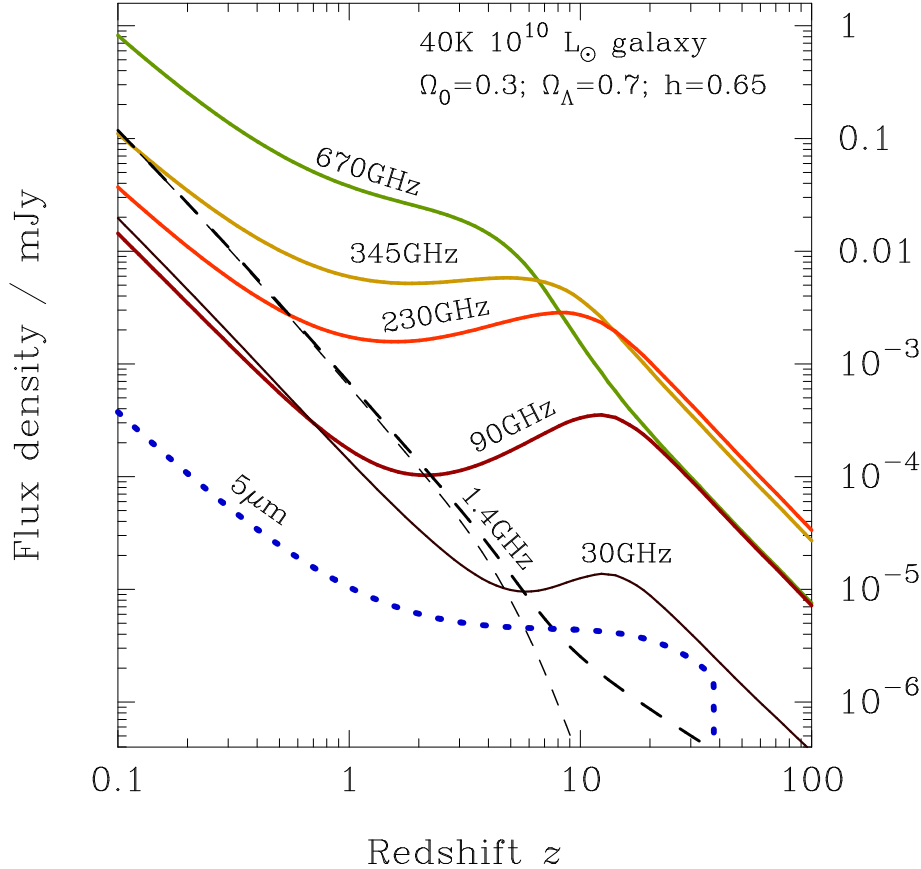}
\end{center}
\caption{Some key features of flux density--redshift relations 
expected at a range of wavelengths, extending to very high redshifts. 
CMB heating of dust at $z > 10$ prevents the mm-wave 
K correction from assisting the detection of very high redshift galaxies:  
the flux density--redshift relation has the same redshift dependence 
beyond $z \simeq 15$ at 230, 90 and 30\,GHz. CMB cooling of relativistic  
electrons suppresses synchrotron radio emission beyond $z \sim 5$, as 
shown by the thin dashed line. Realistic free--free emission is also 
included in the model represented by the thick dashed line 
(Condon, 1992; Yun et al., 2001), 
significantly increasing the radio 
emission expected from very high redshifts. 
An estimate of the flux density from a $3\times10^4$-K stellar photosphere 
at 5\,$\mu$m is also shown, cutoff at the redshift beyond which Lyman-$\alpha$ 
absorption is redshifted through the band.  Note that 
there is probably a 
maximum redshift above which dust does not exist, and so 
beyond which thermal emission from dust can never be detected; 
this effect is not included here. 
} 
\label{fig:Svz_highz}
\end{figure}

At the highest redshifts, two additional factors arise to modify the 
relation expected. 
First, synchrotron emission from relativistic 
electrons is likely to be suppressed. 
The intensity of 
synchrotron emission depends on only the energy density in the interstellar 
magnetic field, while the total cooling rate of the electrons depends on the 
sum of the energy densities in the interstellar magnetic field, for 
synchrotron losses, and in the 
ISRF, for inverse Compton scattering losses. 
The tight low-redshift far-IR--radio 
correlation implies that these energy densities are 
proportional over a very wide range of galaxy properties (V\"olk, 1988). 
However, above some critical redshift, the energy density in the CMB 
will always rise to dominate the ISRF, 
upsetting this balance. Inverse-Compton electron cooling will then  
dominate and the amount of synchrotron emission will decline. 
Note, however, that free--free emission does not suffer this  
suppression at high redshifts, 
and so should remain detectable out to any practical redshift. The almost 
flat SED of free--free emission ensures a more favorable K correction 
for high-redshift galaxies than expected for a pure   
synchrotron emission spectrum. The free--free emission spectrum cuts off 
only at photon energies greater 
than the thermal energy of emitting electrons in H{\sc ii} regions. In 
the absence of free--free optical depth effects, for 
$10^{4-5}$\,K this corresponds to optical frequencies (Yun and Carilli, 
2002). In 
Fig.\,\ref{fig:Svz_highz} we show the effects of CMB suppression of 
synchrotron emission, for a ratio of energy densities in the 
magnetic field and the ISRF of 0.33, which is reasonable for M82 and 
the Milky Way (Hummell, 1986). We assume a galaxy SED template that 
lies on the standard far-IR--radio correlation at $z=0$ 
(Condon, 1992). 

Secondly, again most significant at high redshifts, a minimum dust temperature 
is imposed by the rising CMB temperature: dust 
must be hotter than the CMB. Given that observed dust temperatures in 
ultraluminous galaxies seem to lie in the range 40--100\,K, then this 
may become an important factor at redshifts $10 < z < 30$, if 
an early generation of stars 
generates the heavy elements required to form 
dust prior to 
these redshifts. For cooler Milky-Way-like SEDs, this effect would be 
important at $z \simeq 5$. 
An increase in dust temperature due to CMB heating at 
high redshifts shifts the peak of the dust SED to 
higher frequencies, counteracting the beneficial K correction illustrated 
in Fig.\,\ref{fig:Svzcolor}. As a result, there is 
a firm upper limit to the redshift at which submm continuum 
radiation can be exploited to image the most distant galaxies efficiently, 
even if these  
galaxies do contain dust. This is illustrated by
the flux density--redshift relations at 230, 90 and 30\,GHz shown in 
Fig.\,\ref{fig:Svz_highz}: at redshifts greater than about 15 all three 
curves have the same form. 

Both of these factors were included in the derivation of the curves in 
Fig.\,\ref{fig:CY}, which should thus provide 
an accurate guide to the usefulness of the Carilli--Yun 
redshift indicator out to the 
highest redshifts. Note that the indicator becomes of little use for 
the most distant galaxies, for which an almost constant 
radio:submm flux ratio of 
about $10^{-3}$ is expected. In addition, the hottest dusty galaxies 
may be more likely to contain AGN, and thus to lie on the radio-loud
side of the far-IR--radio correlation. This effect could make 
the interpretation of Fig.\,\ref{fig:CY} for determining 
redshifts ambiguous, even if $T_{\rm d}>60$\,K. However, it should 
always guarantee a 
conservative estimate of the redshift for any observed galaxy. 

Foreground absorption is not a problem for  
very high-redshift sources of any radio, submm or far-IR radiation. 
It is thus  
likely that the most sensitive future instruments at submm and radio 
wavelengths, ALMA and the SKA will both be able to 
detect `first light' galaxies. 
Note, however, that a mask of foreground structure may be significant for 
radio observations (Waxman and Loeb, 2000). A practical 
limit to the capability of determining the history of very early star
formation from an `SKA deep field' could also be set by the low, 
cosmologically-dimmed 
surface brightness of galaxies at the highest redshifts, and their 
potentially overlapping emission regions. The importance of both of these 
factors is expected to depend critically on the unknown physical sizes of 
the first galaxies. ALMA will probably be limited to observe 
galaxies out to a maximum redshift set by the requirement 
that sufficient metals have been generated to form obscuring dust at that 
early epoch. 
To compare what 
might be possible in the near-IR waveband, the emission from a 
$3\times10^4$-K blackbody stellar photosphere
at 5\,$\mu$m is shown in Fig.\,\ref{fig:Svz_highz}, 
cutoff at the redshift beyond which absorption by redshifted Lyman-$\alpha$
becomes important in the band. This indicates the potential for 
probing the earliest galaxies using a near-IR camera on the 
{\it Next Generation Space Telescope (NGST)}.  
Free-free emission and redshifted near-IR 
stellar emission may thus be the best routes to the detection of 
the first galaxies 
at redshifts greater than 10, if they exist. 

\section{The observed properties of submm-selected galaxies} 

\begin{figure}[t]
\begin{center}
\epsfig{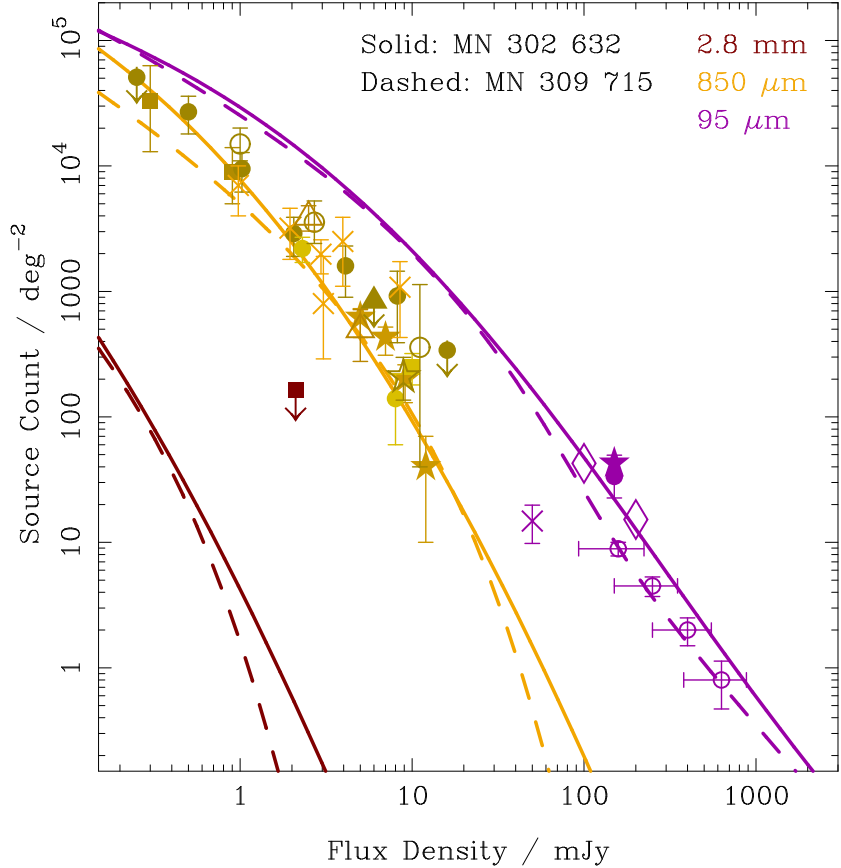}
\end{center}
\caption{A summary of count data from several mm, 
submm and far-IR surveys. The overplotted curves are 
derived in models that provide good fits to the compilation of data, 
and are updated from the results in the listed MNRAS papers 
(Blain et al.,\ 1999b, c).
Identical symbols represent post-1999 data from the same source. The errors 
are shown as $1\sigma$ values unless stated.   
The 2.8\,mm data (square) is from Wilner and Wright (1997). 
At 850\,$\mu$m, in order of increasing flux (less than 15\,mJy), data is 
from Blain et al.\ (1999a), Cowie et al.\ (2002) with 90\% confidence limits; 
Hughes et al.\ (1998),
Chapman et al.\ (2001b), Barger et al.\ (1999a, 1998), 
Smail et al.\ (1997), Eales et al.\ (1999, 2000) consistent with the 
increased area reported by Webb et al.\ (2002a), Borys et al.\ (2001), 
Barger et al.\ (2000) and Scott et al.\ (2002). The data points between 
about 2 and 10\,mJy are consistent with a steep integral source count  
$N(>S) \propto S^\alpha$, 
with a power-law index $\alpha \simeq -1.6$. 
The counts at brighter flux densities are likely to steepen considerably; 
note that the counts must turn over at fainter flux densities to have
$\alpha < -1$ to avoid the background radiation intensity diverging. 
At 95\,$\mu$m, the 
data is from Juvela et al.\ (2001), Kawara et al.\ (1998), 
Matsuhara et al.\ (2000), Serjeant et al.\ 
(2001) and Linden-Vornle et al.\ (2000). 
}  
\label{fig:count1}
\end{figure}

\begin{figure}[t]
\begin{center}
\epsfig{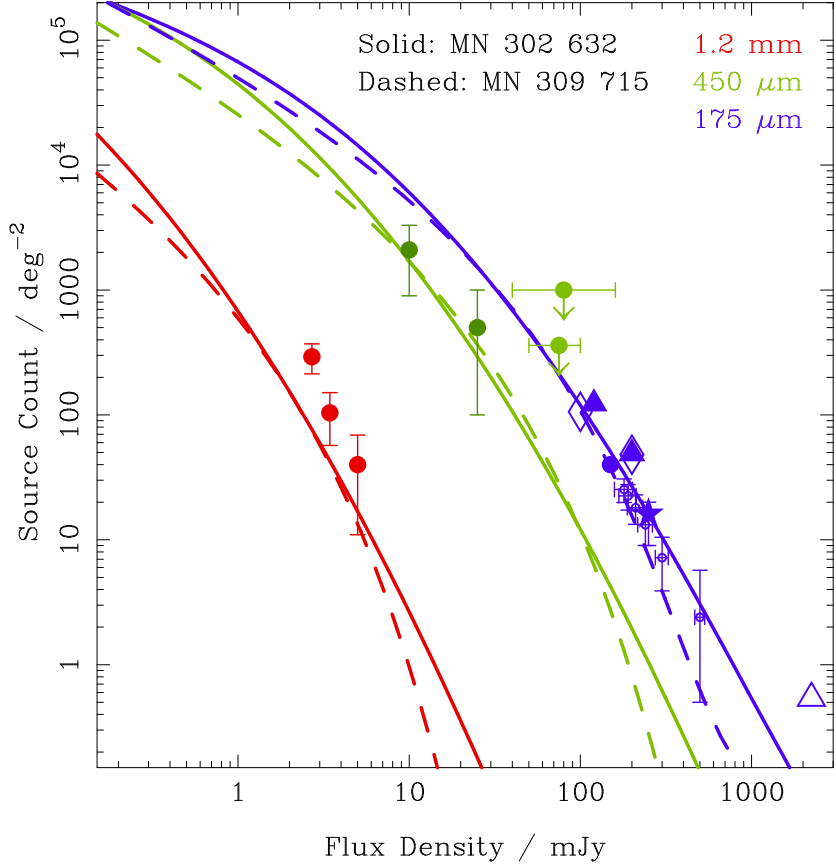}
\end{center}
\caption{Counterpart to Fig.\,\ref{fig:count1} for three other 
observing bands. The data at 1.2\,mm (circles at flux densities less 
than 10\,mJy) are from Bertoldi et al.\ (2001), 
Carilli et al.\ (2000) and Carilli (2001). 
The data at 450\,$\mu$m (circles at 
10--50\,mJy) are from 
Smail et al.\ (2002), with limits from Smail et al.\ (1997) 
and Barger et al.\ (1998). The data at 
175\,$\mu$m ($\> 100$\,mJy) 
are from Kawara et al.\ (1998), Puget et al.\ (1999), Matsuhara 
et al.\ (2001), 
Juvela et al.\ (2001), Dole et al.\ (2001) and Stickel et al.\ (1998). 
} 
\label{fig:count2}
\end{figure}

Well over 100 submm-selected galaxies are now known 
(see Fig.\,\ref{fig:count1}), 
although their redshifts and detailed astrophysical properties are 
very largely 
uncertain. The key information available about their properties comes from 
observations of discrete galaxies made using the SCUBA and MAMBO bolometer 
array cameras at wavelengths of 450, 850 and 1200\,$\mu$m. Counts of 
distant galaxies at far-IR wavelengths of 95 and 175\,$\mu$m have 
also been measured using  
the PHOT instrument aboard {\it ISO}. Limits to 
the counts at 2.8\,mm have been obtained using the Berkeley--Illinois--Maryland
Association (BIMA) mm-wave 
interferometer. The results of all the relevant observations are compiled 
in Figs.\,\ref{fig:count1} and \ref{fig:count2}.
Information is also 
available about the population of mid-IR 15-$\mu$m sources 
using the CAM 
instrument aboard {\it ISO} (Altieri et al.,\ 1999; Elbaz et al.,\ 1999): 
see Fig.\,\ref{fig:count15}. 

\begin{figure}[t]
\begin{center}
\epsfig{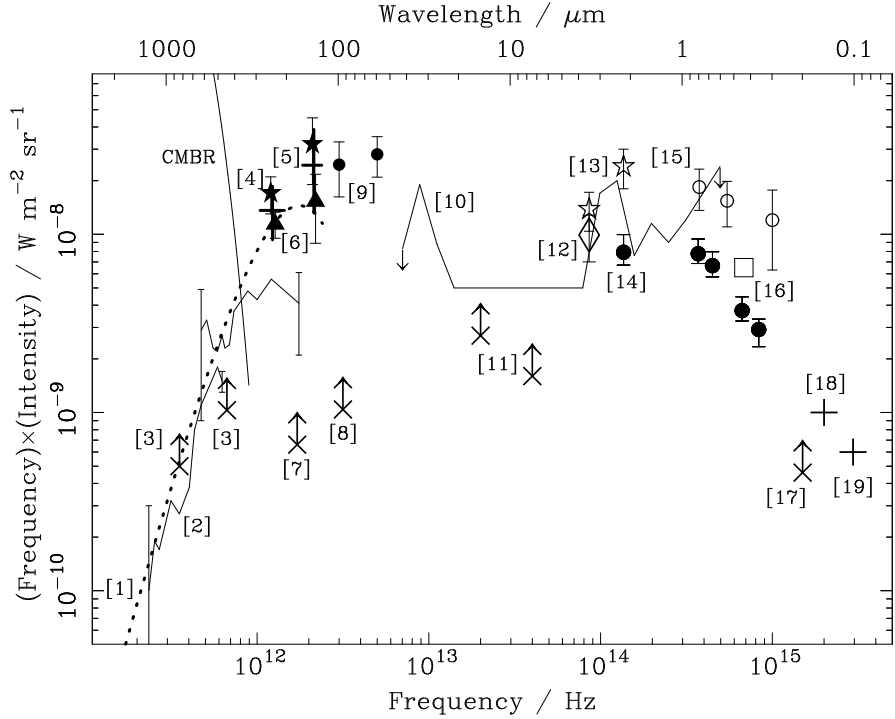}
\end{center}
\caption{The observed intensity of cosmic background radiation between 
the radio and far-UV wavebands. The great majority of the 
background energy density in the Universe derived from sources other than 
the CMB 
is represented in this figure. Almost all of the rest 
appears in the X-ray waveband. Some significant uncertainty remains, 
but the combination 
of measurements and limits indicates that a comparable amount of energy is 
incorporated in the far-IR background, which peaks at a wavelength of 
about 200\,$\mu$m, and in the near-IR/optical background, which peaks 
at a wavelength between 1 and 2\,$\mu$m. 
The data originates from a wide range 
of sources: 1. Fixsen et al.\ (1998); 2. Puget et al.\ (1996); 3. 
Blain et al.\ (1999a); 4. Schlegel et al.\ (1998); 
5. Hauser et al.\ (1998); 6. Lagache et al. (2000a) see also Kiss et al.\
(2001); 
7. Puget et al.\ (1999); 
8. Kawara et al.\ (1998); 9. Finkbeiner et al. (2001); 
10. Stanev and Franceschini (1998); 11. Altieri et al. (1999); 
12. Dwek and Arendt (1998); 13. Wright and Johnson (2002); 14. 
Pozzetti et al. (1998); 15. Bernstein (1999) and 
Bernstein et al.\ (2002); 
16. Toller et al.
(1987); 17. Armand et al. (1994); 18. 
Lampton et al. (1990); and 19. Murthy et al.\ (1999).  
For a detailed review of cosmic IR backgrounds see Hauser and Dwek (2001). 
Note that Lagache et al.\ (2000a,b) claim that the Finkbeiner et al.\ points (9) 
could 
be affected by diffuse zodiacal emission. Where multiple results are  
available in the literature the most sensitive result is quoted. 
} 
\label{fig:back}
\end{figure} 

Measurements of the observed intensity of background radiation from the radio 
to far-UV wavebands are shown in Fig.\,\ref{fig:back}. The 
background measurements include lower limits obtained by summing over 
the observed counts plotted in Figs.\,\ref{fig:count1} and \ref{fig:count2}. 
the deepest counts at 850 and 15\,$\mu$m come from surveys made in fields
magnified by gravitational lensing clusters. Surface brightness 
conservation in the lensing process 
ensures that the mean background intensity in the direction 
of a lens should be the same as that in a blank field.   

From the properties of the counts and backgrounds alone, without any 
details of the individual galaxies involved, 
it is possible to 
infer important details about the population of distant 
dust-enshrouded galaxies. 

The significant surface density of the faint SCUBA and MAMBO galaxies, 
when coupled to plausible SEDs (Blain et al.,\ 1999b; Trentham et al.,\ 1999; 
Dunne et al.,\ 2000), clearly indicates that the luminosity function of 
distant dusty submm galaxies is much greater than that 
of low-redshift {\it IRAS} galaxies (Saunders et al.,\ 
1990; Soifer and Neugebauer, 1991), and  
undergoes very strong evolution. 
An extrapolation of the low-redshift luminosity function 
without evolution predicts a surface density of galaxies brighter 
than 5\,mJy at 850\,$\mu$m of only about 0.25\,deg$^{-2}$, 
as compared with the observed density of 
several 100\,deg$^{-2}$ (Fig.\,\ref{fig:count1}). Because of 
the flat flux density--redshift  
relation in the submm shown in Fig.\,\ref{fig:Svzcolor}, a 5-mJy SCUBA 
galaxy at 
any moderate or high redshift ($z>0.5$) has a 
luminosity greater than about 
$8\times10^{12}$\,L$_\odot$. Immediately, this tells us that the comoving 
density of high-redshift galaxies with
luminosities in excess of about $10^{13}$\,L$_\odot$ is 400
times greater than at $z=0$. We stress that the submm K correction 
ensures that the redshift has little effect on the results: the 
count would be approximately the same whether the population 
is concentrated at $z \simeq 1$ or extends from $z \simeq 2$ to 10. 

This estimate is subject only to an uncertainty in the dust
temperature, which is assumed to be about 40\,K. Even if the dust temperature 
of some of the galaxies is as low as the 20\,K found for low-redshift 
spiral galaxies, then their luminosity is still 
about $8\times10^{11}$\,L$_\odot$, considerably greater than the several 
$10^{10}$\,L$_\odot$ expected 
for typical spiral galaxies. This issue can be addressed by 
taking into account both the observed 
background spectrum and the counts at different 
wavelengths. 

The submm-wave 
background radiation spectrum can also 
be exploited to provide information about 
the form of evolution of the luminosity function. 
The submm-wave 
background, measured directly using {\it COBE}-FIRAS (Puget et al.,\ 1996; 
Hauser et al.,\ 1999; Schlegel et al.,\ 1999), 
reasonably exceeds the sum of the measured flux densities of 
discrete galaxies detected in 
SCUBA surveys (Smail et al.,\ 1997, 2002; Blain et al.,\ 1999a). However, 
the submm background makes  
up only a small fraction of the total energy density in the far-IR 
background, which peaks at a wavelength of 
about 200\,$\mu$m and is generated by galaxies at 
redshift $z \sim 1$. The relatively flat source SEDs and the rate of change
of the 
cosmic volume element at this redshift 
conspire to generate most of the background light, just 
as in the radio, X-ray, optical and 
near-IR wavebands. The mm and submm 
background radiation is unique in originating at a higher redshift. Very little 
of the background is expected to be generated at 
$z < 1$, and so it is an important signature of 
high-redshift galaxy formation. Despite representing only a small 
fraction of the total energy density in the cosmic background radiation, 
the mm-wave background 
is one of the cleanest measures of activity in the distant Universe.  

There are significant consequences for the evolution of galaxies at high 
redshifts due to 
the observed smooth power-law form of the 
background spectrum, $I_\nu \propto \nu^{2.64}$, 
for $\nu < 500$\,GHz (Fixsen et al.,\ 1998), which 
originates at moderate to 
high redshifts, on account of the submm-wave K correction. 
The shape of the background radiation spectrum at frequencies greater than 
about 100\,GHz can be approximated quite accurately by
associating an evolving comoving volume emissivity ($\rho_{\rm L}$
in units of W\,m$^{-3}$) with an SED  
that peaks at a single frequency $\nu_0$, 
so that $\epsilon_\nu \propto \rho_{\rm L}(z) 
\delta(\nu - \nu_0)$ (Blain and Longair, 1993b), and then integrating over 
cosmic volume over a fixed angle on the sky. 
If the SED, via $\nu_0$, is assumed not to evolve strongly with 
redshift---there is no clear evidence that it does---then in order to 
reproduce the observed slope of the mm/submm background 
spectrum, $\rho_{\rm L}(z) \propto (1+z)^{\simeq -1.1}$ is required for 
$z \gg 1$, and so the comoving luminosity density of 
dust-enshrouded galaxies must decline at 
large redshifts. If it did not decline, 
then the background spectrum measured by 
{\it COBE} would be too flat, with too much energy appearing at long 
wavelengths. This argument has been made using Monte-Carlo 
simulations of 
$\rho_{\rm L}(z)$ by 
Gispert et al. (2000). A similar set of simulations have been carried out 
by Eales et al.\ (2000), taking into account the observed background 
radiation spectrum, counts and inferred redshift distribution of 
submm-selected galaxies. 

An approximately equal fraction of the cosmic background radiation energy 
density emerges in 
the near-IR/optical and far-IR wavebands 
(Fig.\,\ref{fig:back}). Because dusty galaxies do not dominate the total 
volume emissivity at low redshifts (Sanders, 1999;
Yun et al.,\ 2001), then the volume emissivity of dusty galaxies must 
increase by a factor of at least 10, matching the significant 
evolution of the population of galaxies observed in the optical waveband at 
$z < 1$ (Lilly et al.,\ 1996), to avoid the intensity of the
far-IR background radiation being significantly less than observed. 
Only a very small fraction of the 
total far-IR luminosity from all low-redshift galaxies comes from 
galaxies more luminous than 
10$^{12}$\,L$_\odot$, yet as 
discussed above in the context of the submm-wave counts, these 
luminous galaxies are much more numerous at high redshifts, by a factor 
of several hundred. These twin constraints demand that the 
form of evolution of the luminosity function of dusty galaxies cannot 
be pure density evolution, a simple increase in the comoving 
space density of all far-IR-luminous galaxies. If the counts were to be 
reproduced correctly in such a model, 
then the associated background radiation spectrum would be 
much greater than observed. 
A form of evolution similar to pure 
luminosity evolution, in which the comoving space density of galaxies 
remains constant, but the value of $L^*$, the  
luminosity that corresponds to the knee in the luminosity function, 
increases---in this case by a factor 
of order 20---is  
consistent with both the submm-wave counts and  
background intensity. 

By a more rigorous process, taking into account all available information, 
including the need to normalize the results to the observed low-redshift 
population of dust-enshrouded galaxies from the {\it IRAS}
luminosity function and the 
populations of galaxies observed by {\it ISO} at $z \simeq 1$, the 
evolution of the luminosity density 
$\rho_{\rm L}$ can be constrained. The results have been 
discussed by Blain et al. (1999b, c), as updated in 
Smail et al.\ (2002), and by Eales et al.\ (2000). 
They are discussed further in Section\,5  
below. 

\subsection{Confusion}  

Source confusion, the contribution to noise in an image due to the 
superimposed signals from faint unresolved sources clustering on the 
scale of the observing beam (Condon, 1974; Scheuer, 1974), is a 
significant problem for observations in 
the submm waveband (Blain et al.,\ 1998; Eales et al.,\ 2000; 
Hogg, 2001). This is due to 
the relatively coarse ($\simeq 10$\,arcsec) spatial resolution currently 
available. 
In fact, a significant fraction of the noise in the deepest 850-$\mu$m SCUBA 
image of the HDF-N (Hughes et al.,\ 1998) 
can be attributed to confusion  
(Peacock et al.,\ 2000). At present, the
practical confusion limit for galaxy detection in SCUBA 
observations at the atmospherically 
favored 850-$\mu$m wavelength is about 2\,mJy. 
This limit makes it difficult 
to determine accurate sub-arcsec positions for the centroids of 
the submm emission from 
faint SCUBA-selected galaxies, 
rendering follow-up observations more challenging. 
Unfortunately, experience has shown that many known high-redshift 
galaxies, especially optically-selected LBGs, are typically 
fainter than the confusion 
limit, and so are difficult to study using SCUBA.

The variety of count data for dusty galaxies shown in 
Figs.\,\ref{fig:count1} and \ref{fig:count2} can be used to estimate 
the effect of source confusion in observations made at 
a wide range of frequencies and 
angular scales. 

\begin{figure}[t]
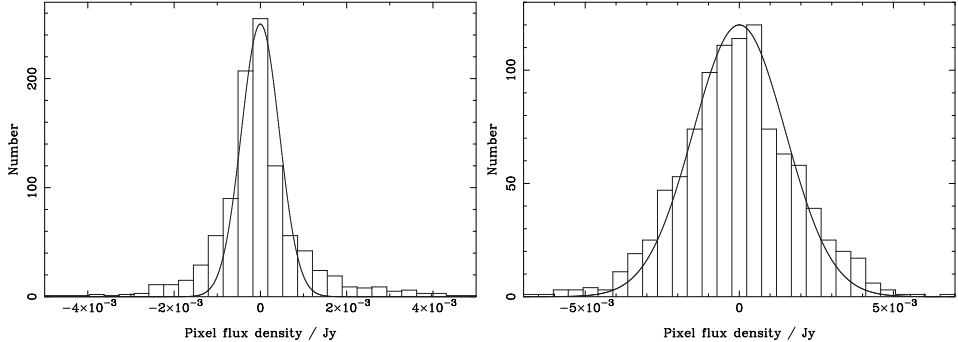

\begin{center}
\epsfig{file=confusion_nonoise.ps, width=4.5cm, angle=-90}
\epsfig{file=confusion_noise.ps, width=4.5cm, angle=-90}
\end{center}
\caption{Histograms showing the simulated effects of confusion noise in 
deep SCUBA integrations at 850\,$\mu$m. Left: the 
expected distribution of pixel flux densities when the telescope 
samples the sky in a standard ($-0.5$,1,$-0.5$) chopping scheme, with no 
additional noise terms present. The flux distribution is non-Gaussian, 
with enhanced high- and low-flux tails as compared with the 
overplotted Gaussian, which has the width predicted by simple 
calculations (Fig.\,\ref{fig:confcont}). Right: the same confusion 
noise distribution is shown in the right-hand panel, but convolved with 
Gaussian instrument and sky noise 
with an RMS value of 1.7\,mJy, which is typical of the noise level in the 
SCUBA Lens Survey (Smail et al.,\ 2002).  
At this noise level, the additional effect of confusion 
noise is small.} 
\label{fig:confhist}
\end{figure}

The distribution of flux density values from 
pixel to pixel in an image due to confusion noise depends on the 
underlying counts of detected galaxies. Confusion noise is always 
an important factor when the surface
density of sources exceeds about 0.03\,beam$^{-1}$
(Condon, 1974). 
The results of a 
confusion simulation for the 14-arcsec SCUBA 850-$\mu$m beam are shown in
Fig.\,\ref{fig:confhist}: 
see also Hogg (2001), Eales et al.\ 
(2000) and Scott et al.\ (2002). Note that Eales et al. assume a 
very steep count and obtain larger values of confusion noise
than those shown here.  
Observations made with finer beams at the 
same frequency suffer
reduced confusion noise, while for those made in coarser beams the
effects are more severe: compare the results for the much larger 
5-arcmin beam in the 
three highest frequency submm-wave observing bands of the planned {\it Planck
Surveyor} space mission all-sky survey shown in Blain (2001a). 

\begin{figure}[t]
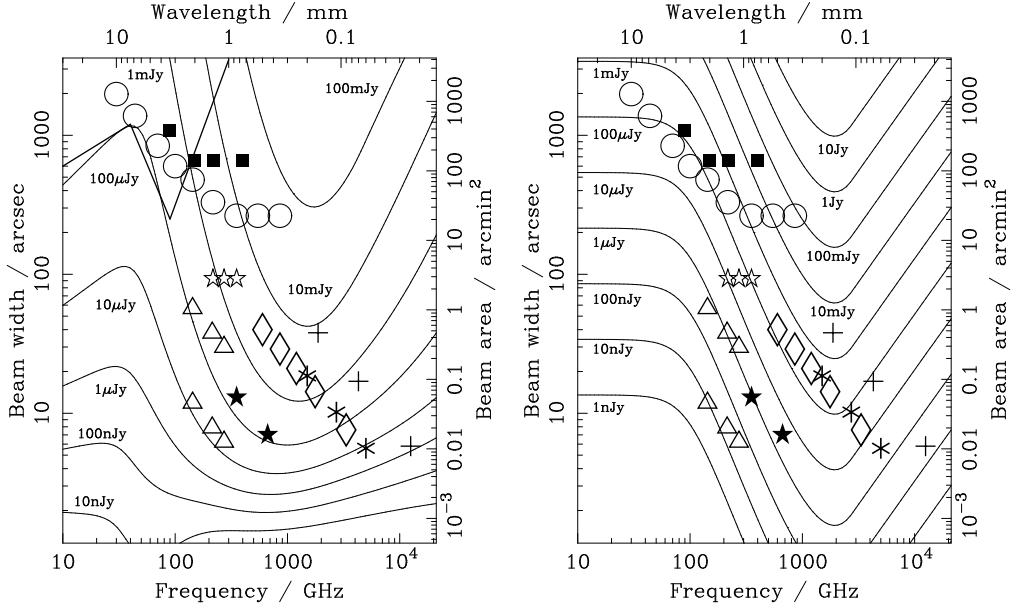

\begin{center}
\epsfig{file=conf_ex.ps, width=8.0cm, angle=-90} 
\hskip 2mm
\epsfig{file=conf_gal.ps, width=8.0cm, angle=-90}
\end{center} 
\caption{An approximate measure of the 1-$\sigma$ confusion noise expected 
as a function of both observing frequency and 
angular scale from the 
mm to mid-IR waveband (updated from Blain et al.,\ 1998). The 
contributions from extragalactic and Galactic sources are shown in the left 
and right panels respectively. 
Radio-loud AGN may make a significant contribution to the top left of the
jagged solid line (Toffolatti et al.,\ 1998).
A Galactic cirrus surface 
brightness of $B_0=1$\,MJy\,sr$^{-1}$ at 100\,$\mu$m is assumed. 
The ISM confusion noise is expected to scale as $B_0^{1.5}$
(Helou and Beichman, 1990; Kiss et al.,\ 2001).
The bands and beamsizes of existing and future experiments 
(see Tables\,\ref{table:ground} and \ref{table:space}) are shown by: 
circles --- {\it Planck Surveyor}; squares ---
BOOMERANG; empty stars --- the SuZIE mm-wave Sunyaev--Zeldovich instrument; 
triangles --- BOLOCAM, as fitted to CSO (upper 3 points) and the 50-m 
Large Millimeter 
Telescope
(LMT; lower 3 points);
filled stars --- SCUBA (and SCUBA-II); 
diamonds --- {\it Herschel}; asterixes --- Stratospheric Observatory for 
Infrared Astronomy (SOFIA);
crosses --- {\it SIRTF}. The resolution limits of the 
interferometric experiments ALMA and {\it SPECS} lie far 
below
the bottom of the panels. The confusion performance of the 
2.5-m aperture BLAST balloon-borne instrument 
is similar to that of 
SOFIA. 
Confusion from extragalactic sources is expected
to dominate over that from the Milky Way ISM for 
almost all of these instruments. 
}
\label{fig:confcont}
\end{figure}

The simulated confusion noise distribution
is non-Gaussian (Fig.\,\ref{fig:confhist}), but can be represented quite 
accurately by a log-normal
distribution, leading to many more high-flux-density peaks in an image
than expected assuming a Gaussian distribution of the same width. 
The width of the central peak of the distribution in flux density is 
approximately the same as the flux density at which the count 
of sources exceeds 1\,beam$^{-1}$. This provides a useful indication of 
the angular scales and frequencies for which confusion noise is likely to 
be significant, and of the limit imposed to the effective 
depth of surveys by confusion for specific instruments: 
see Fig.\,\ref{fig:confcont}. 
  
\subsubsection{Confusion and follow-up observations of submm galaxies}

The real problem of confusion for identifying and conducting multiwaveband
studies of submm-selected galaxies is illustrated by the results 
of the first generation of surveys. The very deepest optical image that 
matches a submm-wave survey is the HDF-N, in which  
there are several tens of faint optical galaxies (at $R>26$) that could be the 
counterpart to each SCUBA detection (Hughes et al.,\ 1998; Downes et 
al.,\ 
1999a). It is thus impossible to be certain that a correct
identification has been made from the submm detection image 
and optical data alone: compare the identifications in Smail et al.\ 
(1998a, 2002). In some cases, 
EROs and faint non-AGN   
radio galaxies (Smail et al.,\ 1999, 2000;  
Gear et al.,\ 2000; Lutz et al.,\ 2001) 
can be associated with submm galaxies, especially after 
higher-resolution mm-wave 
interferometry observations have provided more accurate astrometry for 
the submm 
detection (Downes et al.,\ 1999; Frayer et al.,\ 2000; Gear et al.,\ 
2000; Lutz et al.,\ 2001), to reduce the effects of submm confusion, with 
the investment of significant amounts of observing time. 
The surface 
density of 
both EROs and faint non-AGN 
radio galaxies is less than that of the faintest optical 
galaxies, and so the probability of a chance coincidence between one 
and a submm galaxy is reduced.
A very red color and detectable radio emission from a
high-redshift galaxy are both 
likely to indicate significant star-formation/AGN
activity and/or dust extinction, making such galaxies better candidate 
counterparts even in the presence of confusion-induced 
positional uncertainties. 

\subsection{Multi-waveband follow-up studies} 

A great deal of telescope time has been spent so far to detect and 
study submm-selected galaxies in other wavebands. In 
many cases, rich archival data predated the submm observations:
most notably in HDF-N (Hughes et al.,\ 1998). Considerable data was 
also available in the fields of rich clusters (Smail et al.,\ 1997, 
1998a, 2002; Cowie et al.,\ 2002), in the region of the Eales et al.\ 
(1999, 2000) surveys in Canada--France Redshift Survey (CFRS) fields, 
which include the Groth Strip, and in the deep Hawaii survey fields (Barger 
et al.,\ 1999a).  
The results of follow-up 
deep optical and near-IR imaging (Frayer et al.,\ 2000) and 
spectroscopy (Barger et al.,\ 1999b), mm-wave continuum imaging
(Downes et al.,\ 1999a; Bertoldi et al.,\ (2000); 
Frayer et al.\ 2000; Gear et al.\ 2000; 
Lutz et al.,\ 2001; Dannerbauer et al.,\ 2002) and molecular line 
spectroscopy (Frayer et al.,\ 1998, 1999; 
Kneib et al.,\ 2002) have been published, and many additional studies are 
under way. The time spent following up the SCUBA Lens
Survey (Smail et al.,\ 2002) exceeds by almost an order of magnitude the 
time required to make the submm discovery observations (Smail et al.,\ 1997). 
The difficulty of the task is highlighted by the identification of plausible 
counterparts to these galaxies being only about 60\% complete over 4\,y 
later 
at the start of 2002. 
The follow-up results from a well-studied subset of 
galaxies in the SCUBA Lens Survey 
are shown in Figs.\,\ref{fig:blob}--\ref{fig:n4}, 
in order of 
decreasing 850-$\mu$m flux density. 
These are chosen neither to be a 
representative sample 
of submm galaxies, nor to be a sufficiently large sample for statistical 
studies, but rather to present a flavor of the range of 
galaxies that can be detected in submm-wave surveys for which high-quality 
multi-waveband data is available. The galaxies that 
are presented are chosen to have good positional information, and redshifts 
where possible. Other 
detections for which excellent multi-wavelength follow-up 
data are available include the brightest source in the HDF-N 
SCUBA image (Hughes et al.,\ 1998; Downes et al.,\ 1999a), an ERO detected 
in the CUDSS survey by Eales et al.\ (1999) (Gear et al.,\ 2000), a 
$z=2.8$ QSO in a cluster field (Kraiberg-Knudsen et al.,\
2001), and the substantially overlapping catalogs of galaxies detected 
by Cowie et al.\ (2002) in deeper images of 3 of the 7 clusters 
in the Smail et al. lens survey.  

\subsubsection{Optical/near-IR} 

The properties of submm galaxies in the optical waveband, corresponding to 
the rest-frame UV waveband, appear to be very diverse (Ivison et al.,\ 
2000a). This may be due in part to their expected broad redshift distribution. 
However, given that two submm galaxies at the reasonably high redshifts 
$z = 2.5$ and 2.8 are 
known to be readily detectable at $B \simeq 23$ (Ivison et al.,\ 
1998, 2000, 2001)---before correcting each for 
the magnitude (factor of about 2.5) of amplification due to 
the foreground cluster lens---while
most others are very much 
fainter (Smail et al.,\ 2002; Dannerbauer et al.,\ 2002), 
it is likely that much of the spread 
in their observed 
properties is intrinsic. As most counterparts are extremely faint,  
confirmation of their nature requires a large, completely identified 
sample 
of submm galaxies with known redshifts, which is likely to be some 
time away. It is possible that optically faint submm galaxies have 
similar properties, an issue that can be addressed when deep near-IR 
observations are available.  

Ivison et al.\ (2000a) and Smail et al.\ (2002) proposed a 
3-tier classification system to 
stress the varied nature of 
submm galaxies (three being an eminently sensible number of classes for 
15 galaxies!). Class-0 galaxies 
are extremely faint in both the observed 
optical and near-IR wavebands. Class-1 galaxies are EROs, very faint in the 
optical but detectable in the near-IR, while Class-2 galaxies are relatively 
bright in both bands. It is unclear how closely 
this scheme reflects the underlying 
astrophysics of the submm galaxies; however, 
the classification separates the optically bright galaxies (Class 2s), for 
which the acquisition
of optical redshifts and confirming 
CO redshifts are likely to be practical, and the fainter galaxies, for 
which this will be a great challenge (Class 0s). A similar approach for 
MAMBO sources has been discussed by Dannerbauer et al.\ (2002). 
Note that submm galaxies could change classification 
by having different redshifts, despite identical intrinsic SEDs.

\subsubsection{Ultradeep radio images} 

The surface density of the faintest radio sources that can be detected using 
the VLA (Richards, 2000) is
significantly less than that of optical galaxies, and so an incorrect  
radio counterpart to a submm-selected galaxy is relatively 
unlikely to be assigned 
by chance. If 1.4-GHz VLA images are available at a flux limit approximately 
1000 times deeper than 850-$\mu$m images, then the radio counterparts to 
non-AGN submm galaxies 
should be detectable to any redshift (see Fig.\,\ref{fig:CY}).  
Deep radio follow-up observations of submm-selected galaxies yielding 
cross identifications have been 
discussed by Smail et al.\ (2000), and further information about 
very faint radio sources in the field of the UK 8-mJy SCUBA survey 
(Fox et al.,\ 2002; 
Scott et al.,\ 2002) should soon be 
available in Ivison et al.\ (2002). Despite an extremely deep radio 
image (Richards,\ 2000), the brightest submm galaxy detected in 
HDF-N (Hughes et al.,\ 1998) does not have a radio detection, probably 
indicating a very 
high redshift. The survey results reported by Eales et al.\ (2000) and 
Webb et al.\ (2002a) were discussed in the context of radio data 
covering the same fields; however, the radio survey is not deep 
enough to detect a significant fraction of the relatively faint submm 
sources.  
As can be seen for the specific submm galaxies shown in 
Figs.\,\ref{fig:blob}--\ref{fig:n4}, when they are available, 
deep high-resolution radio images are 
very useful for determining accurate positions and even astrophysical 
properties of submm 
galaxies: see Ivison et al.\ (2001).

\subsubsection{CO rotation line emission and continuum mm-wave interferometry} 

The detection of CO line emission from submm-selected galaxies is a
crucial step in the confirmation of their identification. 
It has been demonstrated in only three cases so far 
(see Figs.\,\ref{fig:blob}, 
\ref{fig:blobII} and \ref{fig:ring}), using the OVRO Millimeter 
Array (MMA; Frayer et al.,\ 1998, 1999), in one case in combination with the 
BIMA array (Ivison et al.,\ 2001), and the IRAM PdBI 
(Kneib et al.,\ 2002). These observations are very 
time-consuming,  
typically requiring tens of hours of observing time. 
The $\simeq 1$\,GHz instantaneous bandwidth of existing line-detection 
systems also means that a redshift accurate to at least 0.5\% must be known 
before attempting a CO detection. In other cases, continuum emission 
is detected using the interferometers, confirming the reality of the 
initial submm 
detection and providing a better 
position 
(Downes et al.,\ 
1999a; Bertoldi et al.,\ 2000; 
Frayer et al.,\ 2000; Gear et al.,\ 2000; Lutz et al.,\ 2001; 
Dannerbauer et al.,\ 2002), 
but no absolute confirmation of a correct optical/near-IR 
identification or a crucial redshift. 

The ALMA interferometer array will have the collecting area and 
bandwidth to make rapid searches for CO line emission in the direction of 
known submm continuum sources from about 2010. Specialized 
wide-band mm-wave spectrographs to search for multiple high-redshift 
CO lines separated by $115\,{\rm GHz}/(1+z)$ that are currently under 
development (Glenn, 2001). Wide-band cm-wave receivers 
at the 
100-m clear-aperture Green Bank Telescope (GBT) 
could detect 
highly redshifted 115-GHz CO($1\rightarrow0$) line emission. 

\subsubsection{X-ray observations} 

Based on synthesis models of the X-ray background radiation 
intensity (Fabian and 
Barcons, 1992, Hasinger et al.,\ 1996), Almaini 
et al.\ 
(1999) and Gunn and Shanks (2002) suggested that 10--20\% of the 
submm galaxy population could be associated with the hard X-ray 
sources that contribute this background. Observations of fields with 
common deep {\it Chandra} and SCUBA images were discussed in 
Section\,2.8. 

The small degree of observed 
overlap between the submm and X-ray sources, implies that 
if a significant fraction of submm galaxies are powered by 
accretion in AGN, then the accretion must occur behind an extremely thick 
absorbing 
column of gas, and less than 1\% of the X-ray emission from the AGN can be  
scattered into the line of sight   
(Fabian et al.,\ 2000; Barger et al.,\ 2001; Almaini et al.,\ 2002). 
In order to avoid detection using SCUBA, high-redshift 
hard X-ray {\it Chandra} 
sources must either contain a very small amount of gas and dust, and thus have 
only a small fraction of their energy reprocessed into the far-IR
waveband, which seems unlikely given their hard spectra; or they must  
contain dust at temperatures much higher than appears to be typical for 
submm-selected galaxies. 
The detection of {\it Chandra} sources in 
Abell\,2390 using {\it ISO} at 15\,$\mu$m, but not using SCUBA at 850\,$\mu$m,
argues in favor of at least some {\it Chandra} sources having very hot dust 
temperatures (Wilman et al.,\ 2000). 
Comparison of larger, deep {\it ISO} 15-$\mu$m images 
with {\it Chandra} images shows that most of the faint, red AGN detected 
by {\it Chandra} are detected in the mid-IR (Franceschini et al.,\ 2002).

This should be readily confirmed 
using wide-field sensitive mid-IR observations of {\it Chandra} fields 
using {\it SIRTF}, images which will also yield 
well-determined SEDs for the dust emission from the detected galaxies.  

\subsubsection{Mid- and far-IR observations} 

Distant submm galaxies are too faint at far- and 
mid-IR wavelengths to have been 
detected in the all-sky {\it IRAS} survey. However, the 
first submm-selected galaxies were detected while the next-generation 
{\it ISO} space observatory was 
still operating, and there were both late-time {\it ISO} 
observations of submm fields, and some serendipitous overlap of 
fields. In general, the small aperture and small-format detector arrays 
of {\it ISO} still led to relatively little overlap between SCUBA 
and {\it ISO} galaxies, for example in the HDF-N 
(Hughes et al.,\ 1998; Elbaz et al.,\ 
1998) and Abell\,2390 (Fabian et al.,\ 2000; Wilman et al.,\ 
2000). The brightest SCUBA galaxies in the cluster Abell\,370 (Ivison 
et al.,\ 1998a) have fluxes measured by {\it ISO} at 
15\,$\mu$m (Metcalfe, 2001), 
providing 
valuable, and otherwise difficult to obtain, 
constraints on their short-wavelength SEDs (Fig.\,\ref{fig:SED}), 
and thus their 
dust temperatures. 
The much larger detector arrays aboard {\it SIRTF}  
should allow many more  
constraints to be imposed on the mid-IR SEDs of submm galaxies. 
Existing deep submm fields are included for imaging within 
the {\it SIRTF} 
guaranteed time programs, and individual submm galaxies have been 
targeted for mid-IR spectroscopy.\footnote{Details of {\it SIRTF} 
observing programs can be found at  
sirtf.caltech.edu.}

\subsection{A gallery of follow-up results} 

\begin{figure}
\begin{center}
\epsfig{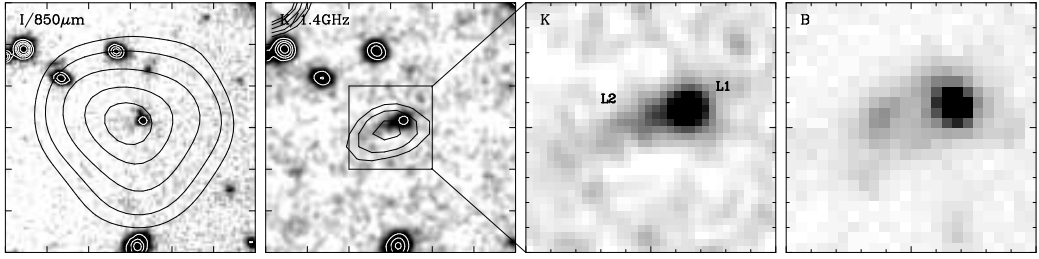}
\end{center}
\caption{Multi-waveband images of SMM\,J02399$-$0136 (23\,mJy; 
Ivison et al.,\
1998; Frayer et al.,\ 1998). The format of this figure is the 
template for those that follow. 
Note that these multiwaveband figures are presented 
in order of reducing 850-$\mu$m flux density, without correcting 
for gravitational lensing amplification.  
The leftmost panel shows black 
contours of 850-$\mu$m emission superimposed on a grayscale $I$-band 
image. The second panel shows black contours of faint 1.4-GHz 
radio emission superimposed on a $K$-band image. These two left-hand 
images are both 30\,arcsec on a side. The third panel shows a 
10-arcsec zoom of the $K$-band image (from UKIRT unless otherwise 
stated; Smail et al.,\ 2002). 
The rightmost panel shows 
a $B$-band CFHT image in this figure; in the figures that follow this 
panel shows an {\it HST} image. Here and in the figures that follow,  
white contours are added to 
show contrast in saturated regions of the grayscale. North is up 
and East is to the left. 
SMM\,J02399$-$0136 is a merging 
galaxy with a confirmed optical/radio 
counterpart, and a CO redshift $z=2.808$: see Vernet and 
Cimatti (2001) 
for a new high-quality spectrum showing Lyman-$\alpha$ emission 
from this galaxy extended over 12\,arcsec.} 
\label{fig:blob}
\end{figure}

\begin{figure}
\begin{center}
\epsfig{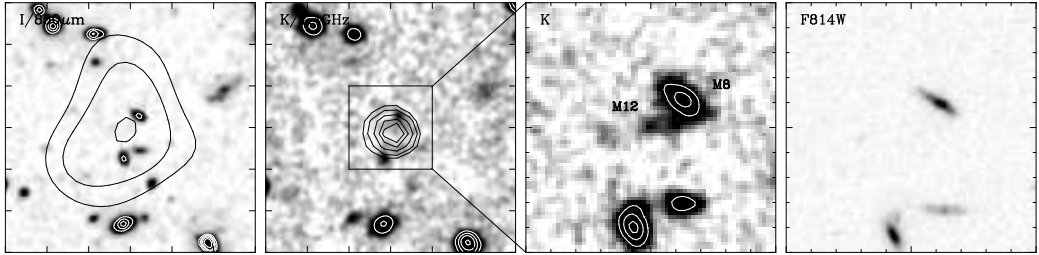}
\end{center}
\caption{Images of SMM\,J00266+1708 (18.6\,mJy;
Frayer et al.,\
 2000). 
The left-hand $K$-band image is from UKIRT; 
the right-hand $K$-band image is from Keck-NIRC. 
The $K$-band detection is located at the position of the very red 
galaxy M12 
in a 1.1-mm continuum image obtained using the OVRO 
MMA.  
}
\label{fig:m12} 
\end{figure}

\begin{figure}
\begin{center}
\epsfig{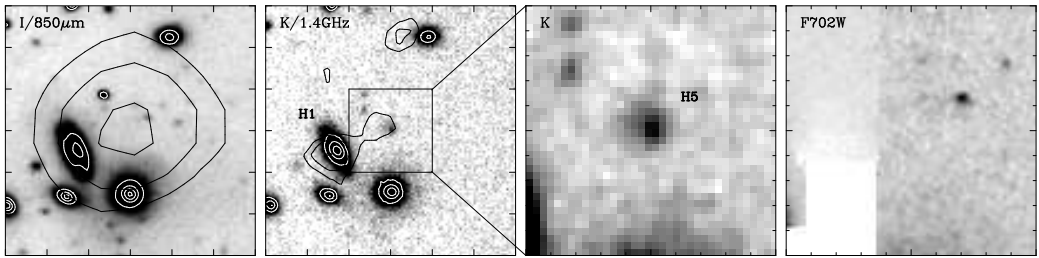}
\end{center}
\caption{Images of SMM\,J0942+4658 (17.2\,mJy), an 
ERO counterpart (Smail et al.,\ 
1999). Faint radio emission and extended, rather bright $K$-band 
emission make this a good candidate for the source of the submm 
emission. H1 is a low-redshift spiral galaxy in the foreground 
of Abell\,851.} 
\label{fig:h5} 
\end{figure}

\begin{figure}
\begin{center}
\epsfig{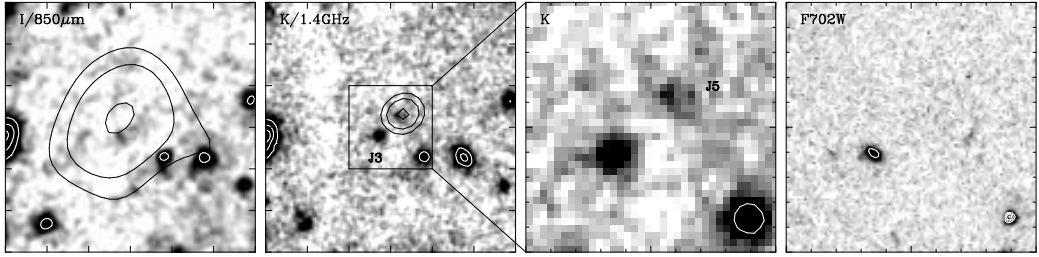}
\end{center}
\caption{Multi-waveband images of SMM\,J14009+0252 (14.5\,mJy), 
the bright 
radio-detected submm galaxy  
Abell\,1835 (Fig.\,\ref{fig:A1835}; Ivison et al.,\ 2000). 
Two faint near-IR 
counterparts can be seen in the $K$-band image. Of these, J5 is 
extremely red,  has no counterpart in the {\it HST}-F702W image, 
and is aligned accurately  
with the centroid of the radio emission.  
\label{fig:j5} 
}
\end{figure}

\begin{figure}
\begin{center}
\epsfig{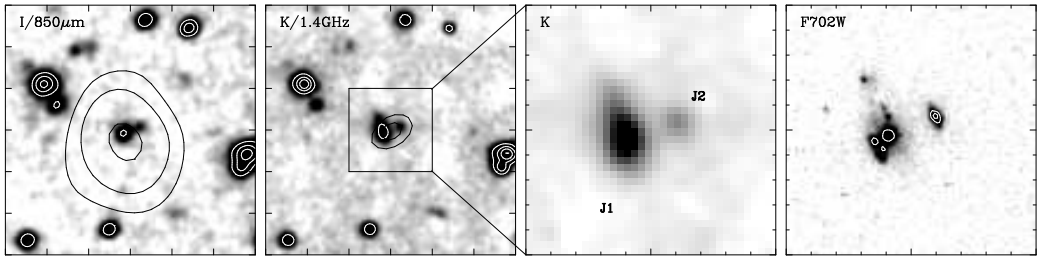}
\end{center}
\caption{Multi-waveband images of SMM\,J14011+0252
(12.3\,mJy; 
Ivison et al.,\ 2000, 2001). This complex merging system 
has a confirmed optical/radio 
counterpart, and a CO redshift $z=2.565$ (Frayer et al.,\ 1999). 
High-resolution CO 
and radio images are presented in Ivison et al.\ (2001). Note that 
the Northern extension of J1 is extremely red,  and is close to 
the centroid of the radio emission. J2 is blue, while 
J1 is red. The complexity of this system is a caution against 
simple treatment of extinction as a uniform screen in submm 
galaxies: for a detailed discussion see Goldader et al.\ (2002) and 
references therein.}
\label{fig:blobII} 
\end{figure}

\begin{figure}
\begin{center}
\epsfig{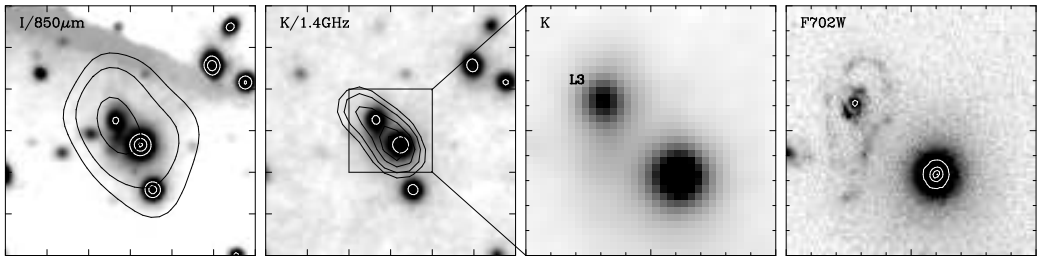}
\end{center}
\caption{Multi-waveband images of SMM\,J02399$-$0134
(11.0\,mJy; 
Kneib et al.,\ 2002). This ring galaxy has a confirmed optical/radio 
counterpart, and a CO redshift $z=1.06$. Its low redshift accounts 
for its very bright $K$-band image and mid-IR {\it ISO} 
detection at 15\,$\mu$m. The other galaxy in the $K$-band
image is a member of Abell\,370.}
\label{fig:ring} 
\end{figure}

\begin{figure}
\begin{center}
\epsfig{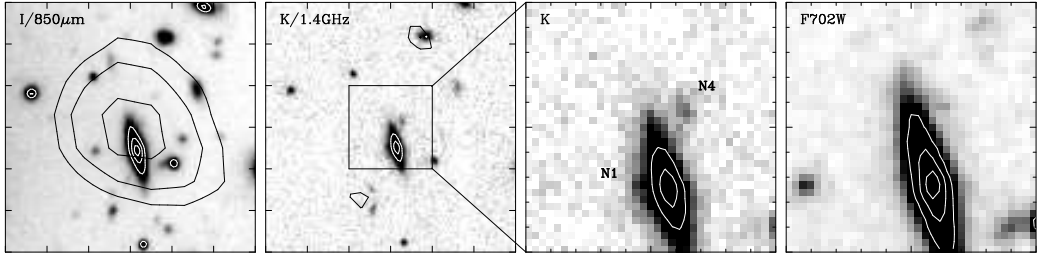}
\end{center}
\caption{Images of SMM\,J04431+0210 (7.2\,mJy), a very red 
counterpart (Smail et al.,\ 1999). 
A tentative H$\alpha$ redshift of $z=2.5$ is determined from a 
near-IR Keck-NIRSPEC observation. Unlike SMM\,J09429+4658 this 
galaxy has no radio emission.}  
\label{fig:n4} 
\end{figure}

In this section we show some of the submm-selected galaxies with the 
most complete and comprehensive follow-up information, including all three  
with confirmed redshifts (Figs.\,\ref{fig:blob}, \ref{fig:blobII} and 
\ref{fig:ring}). To reveal more of the diversity of 
counterparts to the SCUBA galaxies, we also show a 
relatively strong radio source with a faint red $K$-band counterpart 
(Fig.\,\ref{fig:j5}), two $K \simeq 19.5$ galaxies, one a formal ERO 
and the other with very red colors 
(Smail et al.,\ 1999; Figs.\,\ref{fig:h5} and \ref{fig:n4}), which are 
likely to be 
the correct counterpart 
on the grounds of the relatively low surface density of 
EROs; and a mm-continuum source located by the 
OVRO MMA at the position of a SCUBA-selected galaxy, 
with a very faint K-band counterpart observed using the 
NIRC instrument on the Keck telescope (Frayer et al.,\ 2000; 
Fig.\,\ref{fig:m12}). Similar enigmatic faint red galaxies have been 
reported as counterparts to well-located submm-selected 
galaxies from other submm 
surveys by 
Gear et al.\ (2000) and Lutz et al.\ (2001), while Dannerbauer et al.\ 
(2002) do not find counterparts to three well-located MAMBO galaxies 
to a 3-$\sigma$ limit of $K_{\rm s} = 21.9$. 

The galaxies shown in Figs.\,\ref{fig:blob}--\ref{fig:n4} are certainly an 
unrepresentative sample of submm-selected 
galaxies, missing galaxies that are either intrinsically very 
faint at other 
wavelengths or lie at the highest redshifts. 
It is to be hoped that within the next few years, deep follow-up 
observations, especially near-IR ground-based observations\footnote{
Near-IR imaging to 2$\sigma$ $K \simeq 24$ of 
all of the SCUBA Lens Survey 
submm galaxies is underway using the NIRC camera at the Keck telescopes: 
see Fig.\,\ref{fig:m12}.} 
and mid-IR observations using {\it SIRTF} 
will reveal the nature of
the majority of submm-selected galaxies.\footnote{More  
information about this sample 
can be found in the catalog paper of the SCUBA Lens Survey (Smail et al.,\ 
2002 and references therein).}  

\subsection{Clustering properties} 

The deep submm-wave surveys made to date provide relatively little 
information about the spatial distribution of the detected 
galaxies. There is 
some indication from the UK 8-mJy SCUBA survey 
(Almaini et al., 2002; Fox et al.,\ 2002; Ivison et al.,\ 2002; Scott et al.,\ 
2002) and from the widest-field MAMBO surveys (Carilli et al.,\ 2001) 
that the clustering strength of 
SCUBA-selected galaxies is greater than that of faint optically-selected 
galaxies, yet less than that of $K \simeq 20$ 
ERO samples (Daddi et al.,\ 2000). Webb et al.\ (2002b) point out that the 
angular clustering signal expected in submm surveys is likely to be 
suppressed by smearing in redshift, as the submm galaxies  
should have a 
wide range of redshifts, and so the spatial correlation 
function might in fact be stronger than that of the EROs. 
As most $K \simeq 20$ EROs are evolved elliptical galaxies at $z \simeq 1$, 
which are likely to be amongst the first galaxies to form in the most 
overdense regions of the Universe, their strong clustering is 
easily explained. 
However, a definitive result on the clustering of submm galaxies 
awaits a much larger sample of galaxies than considered by Scott et al.\ 
(2002) and Webb et al.\ (2002b).  
Characteristic brightness fluctuations on the angular scales expected 
from faint unresolved dusty galaxies have been 
found in both the 850-$\mu$m SCUBA image of the HDF-N 
(Peacock et al.,\ 2000) and in deep, confused 175-$\mu$m {\it ISO}  
images (Lagache and Puget, 2000; Kiss et al.,\ 2001).  

Haiman and Knox (2000) have discussed the details of measurements of the 
correlation function of unresolved submm galaxies on arcminute angular  
scales in the context of CMB experiments, finding 
that the correlated signal 
can carry important information about the nature and evolution of the 
submm galaxy population. A simpler 
investigation by Scott and White (1999) drew similar conclusions, while there is 
further discussion by Magliocchetti et al.\ (2001). 

\section{Submm galaxy luminosity functions and their relationship 
with other populations} 

Submm-selected galaxies are an important component of the Universe, 
but are typically very faint in other wavebands, and so difficult to study. 
This immediately 
implies that the submm population does not overlap significantly
with other types of high-redshift galaxies, although it may be possible 
to infer their properties if detailed information about  
these other classes is available (Adelberger and Steidel, 2000). 
The likely lack of overlap is reinforced by the relatively low surface 
density of submm galaxies as compared with the faintest 
optically-selected galaxies. 
Nevertheless, a 
detailed understanding of the process of galaxy formation demands that the 
relationship of the submm galaxies 
to other populations of high-redshift galaxies is determined. 

\subsection{Optically-selected Lyman-break galaxies (LBGs)} 

The LBGs (Steidel et al.,\ 1999) are sufficiently 
numerous to have a well-defined luminosity 
function (Adelberger and Steidel, 2000). 
The effects of dust 
extinction on the inferred luminosity of a small subset of LBGs have 
been estimated reliably from near-IR observations of H$\alpha$ emission: 
corrections by factors of 4--7 are indicated 
(Pettini et al.,\ 1998, 2001; Goldader et al.\ 2002). 
At present, it is difficult to confirm this degree of extinction directly, as 
attempts to detect LBGs using submm-wave 
instruments have not so far been successful: see Section\,2.8. 
Observations  
suggest that a typical LBG has a 
850-$\mu$m flux density of order 0.1\,mJy, 
well below the confusion limit at the 
resolution of existing submm-wave images. 

Adelberger and Steidel (2000) have discussed the various selection effects 
associated 
with submm, optical and faint radio selection of high-redshift galaxy samples. 
They assumed that the relation between the slope of the UV SED of a 
galaxy and the 
fraction of its 
luminosity emitted in the far-IR waveband that is observed for low-redshift 
{\it IUE} starbursts with luminosities less than about 
$10^{11}$\,L$_\odot$ (Meurer 
et al.,\ 
1999) holds at greater redshifts and luminosities.
A common, 
smooth luminosity function can then account for the properties of 
LBGs and submm galaxies. A priori, there must be an underlying 
multi-waveband luminosity function 
of all high-redshift galaxies from which both classes of galaxies 
are drawn. However, 
while observations of some submm galaxies 
(a key example being SMM\,J14011+0252; Ivison et al.,\ 2000a, 
2001; Fig.\,\ref{fig:blobII})
seem to support this interpretation at first sight, it is clear that 
only the J2 
region of this galaxy would be identified as a LBG, while 
the submm emission is concentrated nearer to J1. Further discussion 
can be found in Goldader et al.\ (2002). 
Because of the apparent diversity of optical--submm properties of submm 
galaxies (Ivison et al.,\ 2000a; Smail et al.,\ 2002), 
this simple transformation is unlikely to hold. Hence, 
a fraction of submm galaxies will probably never be detected  
in rest-frame UV continuum surveys because of 
their extreme faintness. The confirmed ERO submm galaxies (Smail et al.,\ 
1999; Gear et al.,\ 2000; Lutz et al.,\ 2001) are clear 
examples of such a population. 

\subsection{Extremely Red Objects (EROs)} 

The development of large format near-IR detectors has enabled relatively deep, 
wide-field IR surveys, and lead to the discovery of a class 
of faint EROs (galaxies with colors in the range 
$R-K > 5.5-6$), supplementing traditional 
low-mass stellar EROs (Lockwood, 1970). At first EROs were found one by one, 
(Hu and Ridgeway, 1994; Graham and Dey, 1996), but  
statistical samples are now detected in $K < 20$ near-IR surveys 
(Thompson et al.,\ 1999; Yan et al.,\ 2000; 
Daddi et al.,\ 2000; Totani et al.,\ 2001), 
in parallel to their identification in multiwaveband 
surveys (Smail et al.,\ 1999; 
Pierre et al.,\ 2001; Smith et al.,\ 2001; Gear et al.,\ 2000; 
Lutz et al., 2001).
The clustering of relatively bright $K < 19.2$ 
EROs (Daddi et al.,\ 2000) is observed to be very strong, 
fueling speculation that EROs are associated 
with the deepest 
potential wells that have the greatest density contrast at any epoch. 
This has 
been used as an argument in favor of their association with submm 
galaxies, which could be good candidates for massive elliptical galaxies 
in formation (Eales et al.,\ 1999; Lilly et al.,\ 1999; Dunlop, 2001).  

There are two obvious categories of extragalactic EROs: very evolved galaxies, 
containing only cool low-mass stars, and 
strongly reddened galaxies, with large amounts 
of dust absorption, but which potentially have a very blue underlying 
SED. Only the second are 
good candidates for identification with submm-luminous galaxies.  
Detections of faint radio emission associated with young supernova 
remnants in EROs, and 
determinations of signatures of 
ongoing star-formation in their rest-frame UV colors should allow these cases 
to be distinguished. 
Radio and submm follow-up observations of EROs have tended to show that 
most are passively evolved non-star-forming galaxies without detectable  
radio emission (Mohan et al.,\ 2002), 
with at most about 10--20\% being candidates for 
dust-enshrouded starbursts/AGN. 
It is important to remember that few EROs selected 
from wide-field near-IR surveys, which reach limits of 
$K \simeq 20$, are actively star-forming submm galaxies (see the 
summary of results in Smith et al.,\ 2001). 

A small but significant fraction of 
submm galaxies appear to be associated with EROs at bright  
magnitudes $K<20$ (Smail et al.,\ 1999, 2002). Many more 
submm galaxies probably fit the ERO color criterion, but at much fainter
magnitudes; 
see for example the $K=22.5$ SCUBA galaxy shown in Fig.\,\ref{fig:m12} 
(Frayer et al.,\ 2000). It is certainly possible that future fainter 
ERO samples with $K > 22$ could contain a greater fraction of 
submm-luminous galaxies and fewer passive ellipticals than the 
$K \simeq 20$ samples. Note also that EROs have 
unfavorable K corrections for detection at high redshifts (Dey et al.,\ 1999; 
Gear et al.,\ 2000): beyond $z \simeq 2.5$ any ERO would be 
extremely faint at even near-IR wavelengths. 
This could account, in part or in whole, for the extreme 
faintness of counterparts to submm galaxies, if many do have 
extremely red intrinsic colors. 

\subsection{Faint radio galaxies} 

As discussed above, the faintest radio galaxies should be detectable in 
the submm waveband if the far-IR--radio correlation 
remains valid at high redshifts. The narrow dispersion of this 
correlation suggests that submm galaxies and faint 
radio galaxies are perhaps the most likely populations of high-redshift 
galaxies to overlap substantially.
Surveys made using SCUBA to search for optically-faint, and thus 
presumably high-redshift, galaxies with radio flux densities close to 
the detection threshold 
of the deepest radio surveys (Barger et al.,\ 2000; Chapman et al.,\ 2001b)
have been used to detect many tens of high-redshift 
dusty galaxies much more rapidly than blank-field surveys. 
The selection effects at work when making a radio-detected, 
optically-faint cut from a radio survey are not yet sufficiently well 
quantified to be sure that these catalogs are 
representative of all 
submm galaxies. The typical optical magnitudes of the radio-selected 
objects with submm detections are clustered around $I \simeq 24$ and 
greater. Hence, bright optical 
counterparts to 
mJy-level submm galaxies are rare (Chapman et al.,\ 2002b). 
Because relatively accurate positions are 
available from the radio observations, it should be possible to 
determine spectroscopic redshifts for a significant number of these 
galaxies, providing a valuable contribution to our knowledge of the 
distances to at least a subset of submm galaxies. 

\subsection{Active galaxies and X-ray sources} 

An important category of objects that could be associated with 
submm galaxies are accreting AGN. As discussed in Sections\,2.8 and 3.2.4, 
the overlap between {\it Chandra} and SCUBA galaxies does not appear 
to be very great---at about the 10\% level. However, the 
gravitational energy released when forming the 
supermassive black holes 
in the centers of galactic bulges (Magorrian et al.,\ 1998) 
is likely to be only a few times less than the 
energy released by stellar nucleosynthesis over the lifetime of the 
stars in the bulge, and so a case remains for 
the existence of both a 
significant population of Compton-thick AGN submm sources with no 
detectable X-ray 
emission at energies less than 10\,keV, and for hot dusty AGN detectable  
in the mid- and far-IR but not the submm wavebands (Wilman et al.,\ 2000; 
Blain and Phillips, 2002). 

\subsection{Gamma-ray burst (GRB) host galaxies}

An interesting new development is the idea that if  
GRBs are likely to be associated 
with the deaths of massive stars, then the rate of GRBs and the global 
high-mass star-formation rate should be linked (Krumholz et al.,\ 
1998). By searching for submm emission 
from the directions of GRBs, it may be possible to test whether either submm 
or UV-bright galaxies are the dominant population to host high-mass stars, 
and what fraction of the submm galaxies are powered by non-GRB-generating 
AGN
(Blain and Natarajan, 2000). About 10\% of GRBs are expected to be in hosts with 
850-$\mu$m flux densities greater than 5\,mJy (Ramirez-Ruiz et al.,\ 
2002),  
if submm galaxies dominate the cosmic star-formation rate and are not 
typically powered by AGN. 
Two excellent candidates for submm-loud host galaxies of GRBs are now known 
(Berger et al.,\ 
2001; Frail et al.,\ 2002), the first based on deep VLA radio images, 
the second on direct SCUBA and MAMBO mm/submm observations. 
Most GRB hosts appear to be
associated with $R \simeq 25$ optical galaxies (Bloom et al.,\ 
2002), which could also be typical of the submm galaxy 
population. It is 
difficult to detect GRB host galaxies without hitting the confusion limit 
using SCUBA, but attempts are underway. As a byproduct of surveys for 
submm afterglow emission, Smith et al.\ (1999, 2002) imposed limits to 
the submm host galaxy emission from the direction of 12 GRBs. An ongoing JCMT 
program (Nial Tanvir et al.) is searching directly for submm emission from 
the host galaxies of accurately located 
GRBs (Barnard et al.,\ 2002). 
Detecting and resolving the submm emission from GRB host galaxies 
should ultimately 
be very simple using ALMA, requiring observations of only 
a few minutes minutes per burst. 

\subsection{Prospects for the follow-up observations in the future} 

In order to make detailed submm-wave studies of the astrophysics of 
high-redshift
galaxies, 
high-resolution images will be required. 
Existing mm-wave interferometers can provide 
high-quality images of the brightest submm galaxies (for example 
Frayer et al.,\ 2000; 
Gear et al.,\ 2000; Lutz et al.,\ 2001); 
however, in order to detect rapidly 
typical LBGs, EROs and hard X-ray sources, the additional sensitivity 
and resolution that should be provided 
by ALMA is required. Improved information on the SEDs of 
dusty galaxies will be obtained using the {\it SIRTF}, SOFIA, {\it ASTRO-F} and 
{\it Herschel} space- and air-borne observatories: see 
Tables\,\ref{table:ground} and \ref{table:space}. 

\section{Modeling the evolution of submm galaxies} 

As soon as the first submm galaxies were detected in 1997, it was clear  
that they made a significant contribution to the luminosity 
density in the high-redshift Universe, subject to the plausible 
assumptions that their 
SEDs were similar to those of luminous low-redshift 
dusty galaxies, and that their redshifts were not 
typically less than about $z=0.5$. These assumptions remain plausible and have 
been confirmed to an acceptable level by subsequent observations 
(Smail et al.,\ 
2000, 2002). Despite an initial suggestion that 30\% of the SCUBA galaxies 
could be at $z<1$ (Eales et al.,\ 1999), it now seems that a median 
redshift of submm-selected galaxies is of order 2--3 (Eales et al.,\ 
2000; Smail et al.,\ 2002). 

We have already discussed that evolution by a factor of 
about 20 in the 
value of $L^*$ is required to account for the properties of the submm 
source counts and background radiation intensity. That is 
the key result from submm surveys, but how can it be explained in models 
of galaxy evolution? In this section we will not describe the modeling 
process 
in great detail, but we highlight the key features 
of such analyses, and the most important future tests 
of our current understanding. 

It is important is to be aware that there is still considerable uncertainty in 
the exact form of evolution required to explain the submm observations. While 
strong luminosity evolution of the dusty galaxy population is 
required out to $z \simeq 1$ and 
beyond, the detailed form of that evolution is rather loosely constrained 
by count 
and background data. Redshift distributions are an essential requirement 
in order to determine the form of evolution accurately. 

\subsection{An array of possible treatments} 

A variety of approaches have been taken to making predictions for and 
interpreting the results of submm surveys. 
This work began after the deep 60-$\mu$m counts of {\it IRAS} galaxies 
were derived, and it was clear that strong evolution was being observed 
out to $z \sim 0.1$ 
(Hacking and Houck, 1987; Saunders et al.,\ 1990; 
Bertin et al.,\ 1997). Observations of more distant galaxies at 
longer wavelengths could probe the extrapolated form of evolution, and 
disentangle the degenerate signatures of density and luminosity 
evolution (Franceschini et al.,\ 1988; Oliver et al.\ 
1992). 

Before the first deep submm survey 
observations in 1997, Franceschini et al.\ (1991),  
Blain and Longair (1993a, b, 1996) and Pearson and Rowan-Robinson (1996) 
made a variety of predictions of what might be detected in submm
surveys. Guiderdoni et al.\ (1998) and Toffolatti et al.\ (1998) did 
similarly as the first observational results become available. 
Generally, the observed surface density of submm galaxies was 
underpredicted. Prior to the recognition that  
an isotropic signal in the far-IR {\it COBE}-FIRAS 
data was an extragalactic background (Puget et al.,\ 1996) and not 
Zodiacal emission
(Mather et al.,\ 1994), this could be accounted for by the use of this 
unduly 
restrictive limit on the intensity of the submm background spectrum. Once 
submm count data became available after SCUBA was commissioned, 
it was possible to take either a more empirical or a more 
theoretical view of the 
consequences for dusty galaxy evolution.  

\begin{figure}[t]
\begin{center}
\epsfig{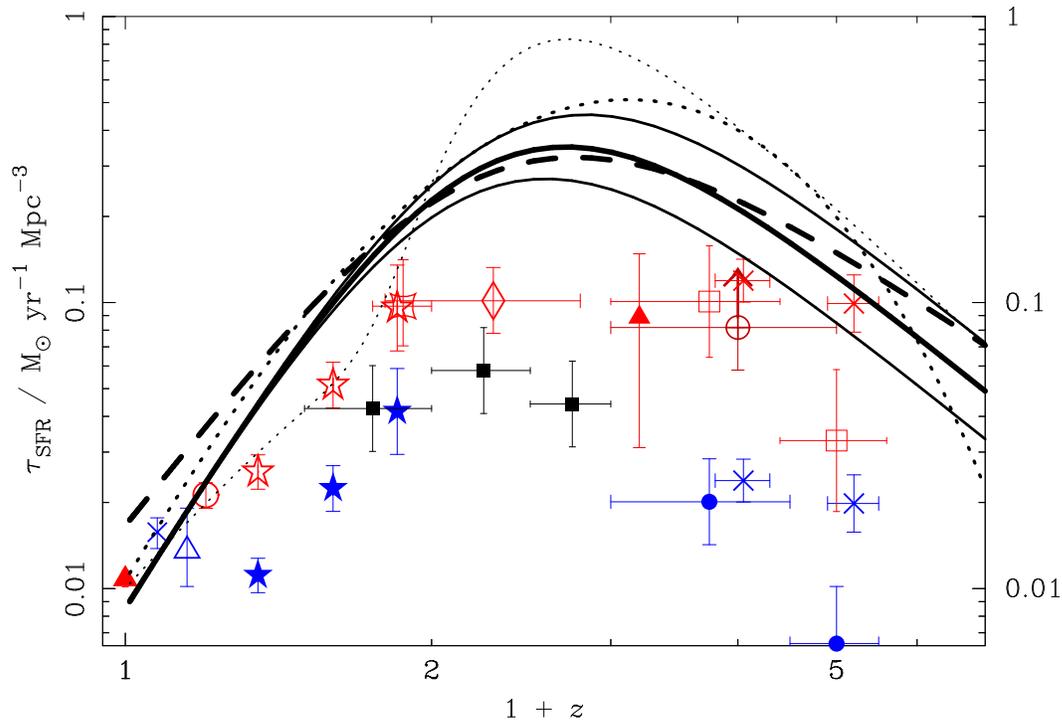}
\end{center}
\caption{The history of energy generation in the Universe, parameterized as
a star-formation rate per unit comoving volume. The absolute
normalization of the curves depends on the assumed stellar IMF
and the fraction of the dust-enshrouded luminosity 
of galaxies that is generated by AGN. 
The points show results derived from a large number of optical 
and near-IR studies, for which detailed references can be found in 
Blain et al.\ (1999b) and  
Smail et al.\ (2002). The most important results are from Lilly et al.\ (1996; 
filled stars) and Steidel et al.\ (1999; high-redshift diagonal crosses). 
The up-pointing arrow comes from the submm-based estimate of 
Hughes et al.\ (1998). 
An important new measurement of the extinction-free 
low-redshift star-formation rate from radio data, that is not 
plotted,
has been obtained by 
Yun et al.\ (2001): $0.015 \pm 0.005$\,M$_\odot$\,yr$^{-1}$.  
The thick solid and dashed lines represent the 
current best fits to far-IR and submm data in a simple
luminosity evolution model and a hierarchical 
model of luminous merging galaxies respectively, 
as updated to reflect additional data and a currently favored
non-zero-$\Lambda$ cosmology. The thinner
solid lines show the approximate envelope of 68\% uncertainty in the results 
of the luminosity evolution model. The thin and thick dotted lines represent 
the best-fitting results obtained in the original derivations 
(Blain et al.,\ 1999b, c). 
}
\label{fig:madau}
\end{figure}

On the empirical side, forms of evolution of the 
low-redshift luminosity function 
of dusty galaxies (Saunders et al.,\ 1990; Soifer and Neugebauer, 1991) that 
were required to fit the count and background data could be determined 
(Malkan and Stecker, 1998, 2001; Blain et al.,\ 1999b; Tan et al.,
1999; Pearson, 2001; 
Rowan-Robinson, 2001).
At redshifts less than about unity, this is done by 
requiring that the predicted counts of low-redshift {\it IRAS} and  
moderate-redshift
{\it ISO} galaxies are 
in reasonable agreement with observations. 
At greater redshifts, where the form of evolution is constrained by submm 
count and background data, 
there is significant degeneracy in the models: strong evolution could 
proceed all the way out to a relatively low cutoff redshift ($z \sim 2$--3), 
or the strong evolution could terminate at a lower
redshift $z \sim 1$, followed by a tail of either non-evolving or 
declining luminosity density out to greater redshifts ($z \geq 5$) (see Fig.\,9 
of Blain et al.,\ 1999b). This 
degeneracy occurs because the far-IR background radiation (like almost
all backgrounds) is generated predominantly at $z \sim 1$, while submm 
galaxies can contribute to the counts equally 
at almost any redshift $1 < z < 10$.  
It can be broken by determining a redshift distribution of submm galaxies, 
which would be 
very different in the two cases. The form of evolution that is 
consistent with the latest observational constraints and radio-derived redshift 
information for submm-selected galaxies (Smail et al.,\ 2002) is 
shown by the thick solid and dashed lines in Fig.\,\ref{fig:madau}. 

Note that the assumptions that underlie these derivations are not yet all 
verified by observations. It is unclear whether all high-redshift
dusty galaxies detected in submm surveys 
have similar SEDs. 
It is possible that the 
properties of the dust grains in galaxies evolve 
with redshift, leading to 
a systematic modification to the temperature or 
emissivity index. It 
is reasonable to expect 
the dust-to-gas ratio in the highest redshift 
galaxies to be lower than in low-redshift galaxies, as less enrichment has 
taken place. However, note that enrichment proceeds very rapidly once 
intense star formation activity is underway. Even the very first regions of 
intense star formation could thus be readily visible in the submm, 
despite the global metallicity being extremely low.
While it seems unlikely, based on a handful of observations 
(Fig.\,\ref{fig:SED}), it is certainly possible that a population of dusty 
galaxies with a significantly different SED is 
missing from current calculations (Blain and Phillips, 2002). 

A more theoretically-motivated approach, based on making assumptions about the 
astrophysical processes at work in galaxy evolution and then  
predicting the observational consequences, has rightly become popular
in recent years. These `semi-analytic' models, which were 
generally developed to explain 
optical and near-IR observations, take a representative set 
of dark-matter halos
that evolve and merge over cosmic time, from the results of 
N-body simulations, and determine  
their star-formation histories and appearance using a set of recipes for 
star-formation and feedback 
(White and Frenk, 1991; Kauffmann and White, 1993; 
Cole et al.,\ 1994, 2000; Guiderdoni et al.,\ 1998; 
Granato et al.,\ 2000, 2001; Baugh et al.,\ 2001; Benson et al.,\ 2001; 
Somerville et al.,\ 2001). 
Unfortunately, at present there is
insufficient information from submm observations to justify a model 
that contains more than a handful of uncertain parameters, and so it is 
difficult to exploit the full machinery of semi-analytic models to explain the 
submm observations. Despite the free parameters available, 
semi-analytic models have had 
limited success in accounting for the observed 
population of high-redshift submm galaxies, 
without adding in an extra population of more 
luminous galaxies to the standard prescription (Guiderdoni et al., 1998), or 
breaking away from their traditional reliance on a universal initial 
mass function (IMF). 
As more information becomes available, then the full capabilities of 
the semi-analytic models can hopefully be applied to address dusty galaxy 
evolution. 

In a blend of these approaches, assuming that the SCUBA galaxies are 
all associated with merging galaxies (Ivison et al.,\ 1998a, 2001; 
Figs.\,\ref{fig:blob} and \ref{fig:blobII}), and yet  
not being sure of the physical processes by which the mergers 
generate the luminosity we observe, Blain et al.\ (1999c) 
used a minimally-parametrized semi-analytic model to investigate 
the change in the properties of merging galaxies required to reproduce the 
submm and far-IR counts and backgrounds: see also 
Jameson (2000) and Longair (2000). 
The observed 
background spectrum can only be reproduced for strong evolution of
the total luminosity density out to redshift 
$z \simeq 1$, by a factor of about 20 
(see Fig.\,\ref{fig:madau}). In addition, the 
lifetime of the luminous phase associated with mergers, and thus the 
mass-to-light ratio also had to be 
reduced by a large factor at high redshift in order to reproduce the 
observed submm counts. The physical reason for this change must be 
an increased efficiency of star formation during starburst activity or 
of AGN fueling at 
increasing redshifts; both make sense in light of the greater gas 
densities expected at high redshifts. The model has the advantage of being 
able 
to reproduce the faint optical counts, if blue LBGs 
are also associated with merging galaxies. The submm 
galaxies release about four times more energy in total than the LBGs,  
and do so over a period of time during a merger that is about 
ten times shorter. It is likely that the total baryonic mass and geometry of 
the merging galaxies
also play important roles in determining the details of star-formation 
activity or AGN fueling during a merger. 

When reviewing the predictions and 
results of any model, note that it is easy to produce a 
model that can account for the far-IR--submm background radiation 
intensity; more difficult to account for the submm counts; and 
more difficult again to reproduce a plausible redshift distribution. 

\subsection{Observational tests of models} 

The key observational test of models of submm-wave galaxy formation 
is the redshift distribution, which is know in outline from 
radio--submm observations (Smail et al.,\ 2000). Determining the 
redshift distribution is a key 
goal of extensive ongoing follow-up observations, but the process 
has proved to 
be difficult and time-consuming, as documented extensively by 
Smail et al.\ (2002). 
The crucial 
problems are the faintness of the counterparts, combined 
with the relatively poor positional accuracy of the centroids of the 
submm galaxies, which are unresolved due to the coarse spatial resolution of
existing submm images. 

Detailed measurements of the counts of galaxies at both brighter 
and fainter flux densities than those shown in Fig.\,\ref{fig:count1} 
would also constrain models. 
However, determining 
the bright counts requires a large-area survey, which is likely to be 
relatively inefficient 
(Fig.\,\ref{fig:rate}), while determining the faint counts requires 
greater angular resolution than can be provided by the telescopes 
used to make existing surveys, to avoid
source confusion. The very bright counts will certainly be probed 
directly towards the end of the decade by the {\it Planck Surveyor} 
all-sky survey at a resolution of 5\,arcmin, 
and sooner by large-area surveys using forthcoming large-format 
mm/submm-wave 
bolometer arrays on ground-based  
telescopes, including BOLOCAM (Glenn et al.,\ 1998) and SCUBA-II. 
Limits on the bright submm-wave counts can 
be imposed from the number of candidate point sources that can be 
found in large-area submm maps of Galactic fields (Pierce-Price et al.,\ 2001; 
Barnard et al.,\ 2002). 
The faint counts will ultimately be determined directly 
using the SMA, CARMA and 
ALMA interferometers. 

The results of deep {\it ISO} surveys have been regularly cited as a 
useful constraint on galaxy evolution (Rowan-Robinson et al.,\ 1997; Xu, 2000; 
Chary and Elbaz, 2001). This is certainly true out to 
$z \simeq 1$. However, when 
estimating a total luminosity density from 15-$\mu$m data, 
it is vital that the 
correct SED is used to extrapolate 
to longer wavelengths, as it is easy to overestimate the 
amount of luminosity associated with a 15-$\mu$m source by assuming a
mid-IR SED that is too steep. 
For example, compare the inferred luminosity density results 
at redshifts $z \simeq 0.7$ quoted by Rowan-Robinson et al.\ (1997) 
and Flores et al.\ (1999). The results differ 
by a factor of 5; 
Flores et al.\ (1999) obtain the lower result by  
using radio observations to constrain the total luminosity 
of the galaxies detected at 15\,$\mu$m. 
Extrapolating mid-IR data towards the peak 
of the SED at longer wavelengths is more difficult than  
extrapolating submm observations to fix the position of 
the peak of the SED that lies at 
shorter wavelengths. This is both because the form of the SED is 
intrinsically simpler on the long-wavelength side of 
the peak, and because the well-determined spectrum of the far-IR background 
radiation can be used to constrain 
the luminosity-averaged dust 
temperature of the submm galaxies. Mid-IR observations with 
{\it SIRTF} after 2002 will provide much more information about the 
SEDs and evolution of dusty galaxies to redshifts $z \simeq 2$. 

\begin{figure}[t]
\begin{center}
\epsfig{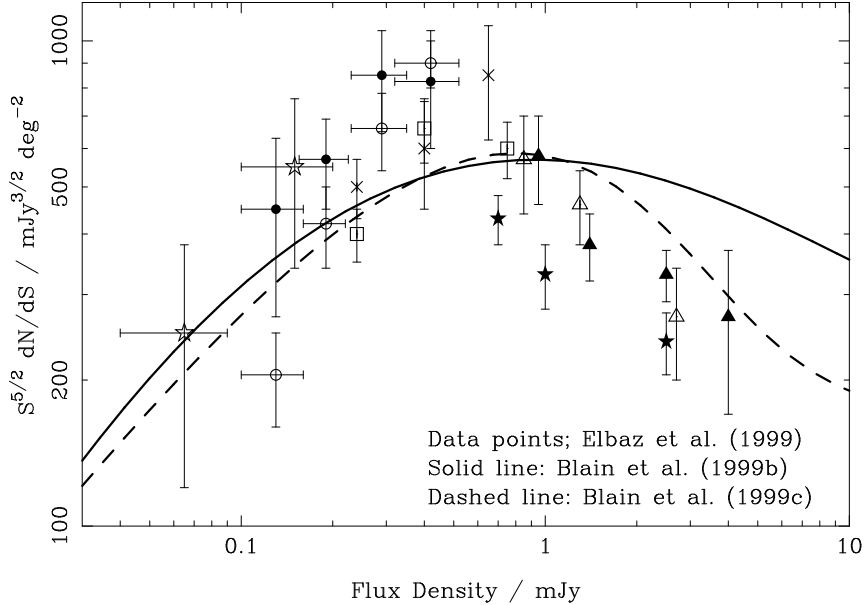}
\end{center}
\caption{A summary of observed 
(Elbaz et al.,\ 1999) and predicted (Blain et al.,\ 
1999b, c)
differential counts of galaxies in the 15-$\mu$m 
{\it ISO} band.  
The model predictions assume only a power-law SED in the mid-IR, 
with $f_\nu \propto \nu^{-1.95}$: no   
fine-tuning with PAH features in the SED is included. The hierarchical 
model (Blain et al.,\ 1999c) 
provides a better fit to the data, but both agree reasonably well 
with the observations. Both models asymptote to the same count 
at greater flux densities. Note that the relative form of the 
model counts reflects that seen in Fig.\,\ref{fig:count1}, with the 
hierarchical model having the steepest rise. 
}
\label{fig:count15}
\end{figure}

In Fig.\,\ref{fig:count15} we show the deep 15-$\mu$m 
counts predicted by models designed to account for the submm data 
(Blain et al.,\ 1999b, c), updated to the current data and cosmology. 
If the mid-IR 
SED is chosen appropriately, then the fit is quite acceptable. Including PAH 
emission features or varying the mid-IR SED index $\alpha$ has 
relatively little effect on the result. The same approach can be used to 
estimate the deep cm-wave radio counts. If we assume just the  
form of the radio--far-IR correlation (Condon, 1992), without 
any fine tuning, and a radio SED of the form $f_\nu \propto \nu^{-0.6}$, then 
the predicted 8.4-GHz counts brighter than 10\,$\mu$Jy, 
based on the submm-based models are 
1.05 and 0.98\,arcmin$^{-2}$ 
respectively; the corresponding power-law indices of the 
count function $N(>S) \propto S^\alpha$ are 
$\alpha = -1.4$ and $-1.3$ respectively. 
The results in both models match 
the observed 8.4-GHz 10-$\mu$Jy count of $1.01 \pm 0.14$\,arcmin$^{-2}$ 
with  
$\alpha = -1.25 \pm 0.2$ (Partridge et al.,\ 1997). 
The reasonable agreement between the predictions of the 
models, which are constrained only by 
observations in the submm and far-IR, and the observed 
deep mid-IR and radio counts confirms that the models are reliable. 
The source confusion estimates 
shown in Fig.\,\ref{fig:confcont}, which are based on the same 
models, should thus 
be reliable over a wide wavelength range from about 10\,cm to 10\,$\mu$m.

\subsection{Modeling the detailed astrophysics of the submm galaxies}  

It is not possible to separate the modeling of the evolution of the 
population of submm galaxies fully from studies of the nature of the 
galaxies themselves. Their luminosities and masses  
(see Frayer et al.,\ 1999) demand that the submm-luminous 
phase be short-lived as compared with the age of the Universe. 
The observational information for most 
of the submm galaxies is insufficient to be confident 
that their nature is understood at present. Of the galaxies with 
reliable counterparts, there are three 
bright Class-2 galaxies 
(Soucail et al.,\ 1999;
Ivison et al.,\ 1998a, 2000a, 2001; Vernet and Cimatti, 2001) with 
optical redshifts and 
CO detections (Frayer et al.,\ 1998, 1999; Kneib et al.,\ 2002), 
and a total of 
four Class-1 galaxies
(Smail et al.,\ 1999; Bertoldi et al.,\ 2000; Gear et al.,\ 2000; 
Lutz et al.\ 2001), 
which are known to be either very red galaxies or formal EROs. Other  
submm-selected galaxies with accurate positions 
from radio observations 
(Smail et al.,\ 2000) or mm-wave interferometry 
(Downes et al.,\ 1999b; Bertoldi et al.,\ 2000; 
Frayer et al.,\ 2000; Dannerbauer et al.,\ 
2002) remain enigmatic. 
All that can be said about these 
galaxies is that they all 
appear to have thermal dust spectra, are all very faint at optical 
wavelengths, and most also appear to be very faint at 
near-IR wavelengths. 

The Class-2 galaxies are all clearly undergoing mergers or interactions. 
Much less is known about the morphology of the faint, but typically 
extended Class-1 galaxies (Figs.\,\ref{fig:h5}, \ref{fig:j5} and \ref{fig:n4}).
It is certainly 
possible that they too are involved in interactions, which appear to trigger 
the dramatic luminosity of almost all 
the low-redshift ULIRGs (Sanders and Mirabel, 1996). 
Programs of ultradeep near-IR imaging on 10-m-class telescopes should 
soon test this idea. 

Hydrodynamical simulations of gas-rich mergers by Mihos and Hernquist (1996), 
Bekki et al.\ 
(1999) and Mihos (2000) show the formation of very dense 
concentrations of 
gas, which could be associated with short-lived, very-intense bursts of 
star formation. However, at present it is not possible to simulate a 
representative sample of mergers with the range of geometries likely to 
be encountered, the necessary time resolution, and a sufficiently 
accurate treatment of the 
detailed astrophysics of star-formation to make a reliable 
connection between the limited observations and the underlying
galaxy properties. The spatial extent of the three bright Class-2 galaxies 
in the 
optical waveband appears to be considerably greater than that of most 
low-redshift ULIRGs. It is thus difficult to be sure that  
simulations
of well-studied low-redshift ULIRGs
adequately represent the 
properties of the high-redshift submm galaxies. Note, however, that the 
precise spatial relationship between the optical and submm emission in these
objects is still unclear (Ivison et al.,\ 2001); the submm emission could 
be more compact than the optical galaxy. 

Larger submm galaxy samples  
will be available over the next few years, boosting 
the likelihood that examples of the full range of submm 
galaxies will be available to be studied in detail. More sensitive 
observations of the properties of the known galaxies will also improve our 
knowledge of their astrophysics. One key question is the relationship of 
the submm galaxies to the formation of elliptical 
galaxies (Lilly et al.,\ 
1999). Whether the bulk of submm galaxies are high-redshift 
low-angular-momentum gas clouds, forming elliptical 
galaxies in a single episode by a `monolithic collapse' (Eggen et al.,
1962), as advocated by  
for example Archibald et al.\ 
(2002), or galaxies observed during one of a series of 
repeated mergers of gas-rich, but pre-existing galaxy sub-units,  
likely to take place at relatively lower redshifts, as discussed by  
Sanders (2001), and which might ultimately yield elliptical merger 
remnants, is an important question that future follow-up observations will 
address.     
Existing observations of extended and disturbed counterparts to submm
galaxies (Ivison et al.,\ 1998a, 
2001; Lutz et al.,\ 2001) tend to favor the second 
explanation, in which well-defined pre-existing stellar systems merge. 
However, in either scenario it is likely that the bulk of the stellar 
population in the resulting galaxies form during the submm-luminous phase. 

\subsection{The global evolution of dust-enshrouded galaxies} 

Fig.\,\ref{fig:madau} summarizes the current state of knowledge of 
the strong evolution of the comoving luminosity density contributed by luminous 
far-IR 
galaxies, whose emission is redshifted into the 
submm (Blain et al.,\ 1999b, c; 
Smail et al.,\ 2002). The derivation of these 
results was discussed briefly above, and is explained in 
much more detail in these 
papers.
The results of both the luminosity evolution and hierarchical models, 
which both include strong luminosity evolution 
of the population of low-redshift {\it IRAS} galaxies out to 
high redshifts, are fully compatible with the redshift 
information 
available for submm galaxy samples. This would not be the case 
if the redshift evolution of the luminosity density was markedly different 
from the forms shown in Fig.\,\ref{fig:madau}. 
For example, if the luminosity density of submm galaxies were to match rather 
than exceed the value denoted by the datapoints at $z \simeq 1$ in 
Fig.\,\ref{fig:madau}, and remain at the same high level out to $z \simeq 10$, 
then the submm-wave spectrum of the cosmic background radiation 
(Fig.\,\ref{fig:back}) 
would tend to be too flat, the intensity of the far-IR 
background radiation would fall short of the observed level, and the  
predicted redshift distribution 
of submm galaxies would be biased strongly to the highest redshifts, 
which at present seems not to be the case (Smail et al.,\ 2002). The 
determination of complete 
redshift distributions for existing samples of submm galaxies, and of more 
redshifts for individual very luminous galaxies drawn from large, future  
submm-selected galaxy catalogs, will ultimately allow the evolution of distant 
dusty galaxies to be traced in detail. 

\section{Gravitational lensing in the submm waveband} 

The first submm-wave surveys for distant galaxies exploited both 
the weak to moderate gravitational lensing magnification, by 
about a factor of 2--3, 
experienced throughout the inner few square arcminutes of rich 
foreground cluster of galaxies 
at a moderate redshift in the range $z \simeq 0.2$--0.4 
(Blain, 1997) 
and the greater magnification produced along 
critical lines for much smaller areas of the background sky. 
A 5-arcmin$^2$ 
SCUBA field centered on a moderate redshift cluster includes 
both these regions, enhancing 
the flux density from all high-redshift background 
galaxies (Smail et al.,\ 1997). 
More, and in some cases deeper,  
SCUBA images of clusters have been taken (Smail et al.,\ 2002; 
Chapman et al.,\ 2002a; Cowie et al.,\ 2002), especially in 
Abell\,2218, where a multiply-imaged source has been 
detected (van der Werf and Kraiberg Knudsen, 2001). 
Whether the magnification acts to increase the surface density of 
background galaxies on the sky depends on the form of their counts.  
Lensing by both galaxies and clusters 
could have significant applications in future 
submm surveys, especially those sampling the steep counts of 
bright submm galaxies in wide fields, 
including the all-sky survey from {\it Planck Surveyor}  
(Blain, 1998), and surveys using BLAST, {\it Herschel}-SPIRE and SCUBA-II 
(Tables\, \ref{table:ground} and \ref{table:space}) covering many tens of 
square degrees. 

\subsection{Magnification bias}

Because surface brightness is 
conserved by all gravitational lenses, 
the net effect of magnifying a population of 
background galaxies depends on the slope of the counts ${\rm d}N(>S)/{\rm d}S$, 
where $N(>S)$ is the number of galaxies per unit area on the sky brighter than
$S$ (Schneider et al.,\ 1992). 
Subject to a magnification $\mu$, 
the count becomes $[1/\mu^2]{\rm d}N[>(S/\mu)]/
{\rm d}S$. For a power-law count with ${\rm d}N(>S)/{\rm d}S \propto S^\alpha$, 
a value of $\alpha < -2$ corresponds to an increase in surface density if 
$\mu > 1$. This 
threshold value corresponds to a slope of $-1$ for the integral counts 
$N(>S$)
shown in 
Fig.\,\ref{fig:count1}. Note that for a uniform non-evolving population of 
galaxies $\alpha = -2.5$. 

As shown in Fig.\,\ref{fig:count1}, submm-wave counts  
are expected to be steep, and to change slope sharply 
at mJy flux density levels, as compared with deep 
optical or radio counts. 
The significant changes in the count slope are particularly unusual, 
and not found in any other waveband. 
As a result, the magnification bias can be large, 
increasing the number of detectable bright galaxies 
(Blain, 1996, 1997), and providing a way to investigate   
very faint counts by comparing lensed and unlensed fields; 
for example in the innermost regions of clusters of galaxies 
(Blain, 2002). 

That a significant magnification bias can be exploited using 
relatively weak lensing by clusters of galaxies can 
be seen by comparing the number of 10--20-mJy 850-$\mu$m galaxies detected 
in the SCUBA Lens Survey 
(Smail et al.,\ 1997, 2002), 
and in the larger field of the unlensed 8-mJy survey 
(Scott et al.,\ 2002). The ratio is about 
3:1, showing a clear positive magnification bias, and  
indicating that if the 850-$\mu$m counts at flux densities greater 
than 10\,mJy are represented by a power-law, then the index 
$\alpha < -2$.

\subsection{Conditions for exploiting submm lensing by galaxies} 

The key advantage of observing background galaxies that are 
gravitationally lensed 
by foreground mass concentrations in the submm waveband is that the 
K correction (Fig.\,\ref{fig:Svzcolor}) acts to brighten the distant 
background lensed galaxy as compared with the lens.   
This is already very familiar from surveys of lensed radio AGN 
(Rusin, 2001), and is illustrated clearly in Fig.\,\ref{fig:A1835},
in which only the central cD galaxy in the lensing cluster  
shows any significant submm emission.  

In SCUBA cluster lens surveys, both 
the image separations, and the extent of the 
high-magnification 
regions are of order 1\,arcmin, a scale which is well matched both to 
the 15-arcsec resolution of the JCMT and to the 2.5-arcmin field of view of 
SCUBA. The magnification ensures that a significantly greater fraction of 
the submm-wave 
background radiation intensity is thus 
resolved into detectable galaxies in surveys in the fields of gravitational 
lensing clusters
than in even the deepest blank-field surveys (Blain et al.,\ 1999a). 
However, for background 
sources lensed by galaxies rather than clusters, 
the relevant image separations and 
the extent of the high-magnification region are only 
of order 1\,arcsec, and so cannot be resolved using any single-antenna 
telescope. 
High-resolution submm 
observations are required to disentangle lensed and unlensed galaxies; this 
capability will be provided by ALMA (Blain, 2002), while pilot studies of 
should be possible using the CARMA, SMA and IRAM PdBI interferometers. 
The most luminous lensed sources can already be resolved into multiple 
images using the IRAM mm-wave interferometer 
(Alloin et al.,\ 1997). 

The only caveat for exploiting galaxy-scale lensing is that the 
source size must be 
small as compared with the area of sky behind the lens that is 
strongly magnified. The intense far-IR and submm emission from 
low-redshift ULIRGs is typically very compact (several hundred pc 
across; Downes and Solomon, 1998), 
and would easily 
meet this condition; however, there are indications that the dust 
emission from at least some luminous 
high-redshift submm galaxies could extend over 
scales greater than 10-kpc (Papadopoulos et al.,\ 2001; 
Chapman et al.,\ 2001a; 
Lutz et al.,\ 2001; Isaak et al.,\ 2002; 
Ivison et al.,\ 2001). The whole area of sky covered by these galaxies would 
not then 
be lensed efficiently by an intervening galaxy, although bright knots 
of emission within them could still be magnified by large factors. 
This concern about lensing efficiency and 
the angular size of distant submm galaxies 
does not apply to lensing by much larger 
clusters of galaxies, which will always be 
effective. 

\subsection{Prospects for the lensing studies in the future} 

Larger area surveys for brighter samples of luminous dusty galaxies 
using the array of forthcoming ground-based, air- and space-borne 
instruments, including BOLOCAM, BLAST, SOFIA, 
SCUBA-II, {\it SIRTF}, 
{\it Herschel} and {\it Planck Surveyor} (see 
Table\,\ref{table:space}), should 
be subject to an enhanced magnification bias. High-resolution follow-up 
observations using ALMA should then yield a large 
sample of strongly magnified 
high-redshift lensed systems to complement the systematically selected 
CLASS sample of gravitational lensed AGN identified at radio 
wavelengths 
(Rusin, 2001).  
These surveys will not be subject to any 
extinction bias due to absorption by dust in the lensing galaxies, 
and should yield a very complete and 
reliable catalog of up to several thousand lenses (Blain, 1998). 

\section{Future developments in submm cosmology} 

During the last 4 years, the first steps have been taken towards investigating 
the Universe using 
direct submm-wave surveys. 
The technologies of the class of detectors that made these initial surveys 
possible are still developing 
rapidly. Many instrumentation projects are underway, which will allow 
us to increase the sizes 
of samples of distant submm galaxies, and to study known 
examples in more detail; some of their key 
features are outlined in 
Tables\,\ref{table:ground}--\ref{table:instrument}. 

\subsection{New technologies for instrumentation} 

A key technology under development is for bolometers with 
superconducting temperature-sensitive elements, 
including transition-edge sensors (TESs). These are much more stable than the 
semiconducting 
thermistors used in existing systems, and so can be read out using 
multiplexed, and therefore much simpler, cold electronics. 
Another advantage of TES devices is 
that they require no bias current, and so need fewer heat-conducting, 
difficult-to-assemble connections to each device.  
The prototype Fabry--Perot
spectroscopic device FIBRE, which uses 
TES bolometers (Benford et al.,\ 2001), 
was tested successfully at the CSO in May 2001.

TES devices offer the 
prospect of increasing the size of the arrays of detector elements  
in mm/submm-wave cameras from of order 100 to 
of order 10$^{4-5}$, providing much larger fields of view. Filled-array 
detector devices using conventional semiconducting bolometers are being 
demonstrated in the SHARC-II and HAWC cameras 
for the CSO and SOFIA, while the SCUBA-II camera has a goal of at least a 
$8 \times 8$\,arcmin field of view---about 25 times greater 
than the field of view of SCUBA---is under 
development in Edinburgh and is expected to integrate large arrays and 
superconducting 
bolometers. SCUBA-II will supplement its much  
wider field of view with an enhanced point source sensitivity: 
the same galaxies should be detectable about 8 and 4 
times faster using SCUBA-II as compared with SCUBA at wavelengths of 
850 and 450\,$\mu$m 
respectively: see Table\,\ref{table:ground}. 
A 10-m telescope operating at 850-$\mu$m with a $10^5$-element 
detector would have a square field of view 
about 1\,deg on a side,  
at a Nyquist-sampled plate scale: much larger than the 5-arcmin$^2$ fields 
of view of SCUBA and MAMBO. With such large 
fields of view, it is not unreasonable to survey most 
of the sky down to the confusion limit of a 10--30-m telescope in an 
observing campaign lasting for several  
years. 

By combining large numbers of bolometer detectors with dispersive 
mm-wave optics (Glenn, 2001),
it should be possible 
to obtain low-resolution mm-wave 
spectra of galaxies over a very wide band, perhaps 100\,GHz, to search for 
CO and atomic fine-structure 
line emission from high-redshift submm galaxies detected in 
continuum surveys, while simultaneously carrying out unbiased surveys for 
line-emitting galaxies within the field of view (Blain et al.,\ 
2000b). Development of several such systems is underway. 

Phase-sensitive heterodyne submm detectors are already 
very efficient; however, only small numbers of these detectors can 
currently be fabricated into 
an array. Their strength is in  
very high resolution submm-wave spectroscopy, 
and as sensitive coherent 
detectors in  
existing mm-wave interferometers. They will be 
fitted to the forthcoming SMA and CARMA, and will be 
be exploited to the full with the 
large collecting area of the ALMA array. 

Sensitive arrays of mid- and far-IR detectors should soon be flying, 
both in space aboard {\it SIRTF} and {\it ASTRO-F}, and
in the upper atmosphere, on balloons such as BLAST, and aboard  
SOFIA. Limits to the continuum flux
densities of the most luminous 
high-redshift galaxies derived using these facilities, measured close 
to the peak of their SED (see Fig.\,\ref{fig:SED}), will provide valuable 
information about their properties. Spectrographs, both aboard these 
facilities and on ground-based telescopes, 
will provide detections of and sensitive limits to the line 
radiation from the same objects, providing redshift information and 
astrophysical diagnostics.  

\subsection{New telescopes}  

At present, a wide range of submm-wave
telescopes are available. Single-antenna telescopes include the 10.4-m 
CSO on Mauna Kea, the 10-m 
Heinrich Hertz Telescope (HHT) on Mount Graham, 
the 15-m JCMT on Mauna Kea, the 15-m SEST 
at La Silla in Chile, 
the 30-m IRAM telescope in Spain, 
and the 45-m antenna at Nobeyama in Japan. The Large 
Millimeter Telescope (LMT), 
a 50-m mm-wave telescope is under construction on a 5000-m peak 
near Puebla in Mexico, and it is hoped that the 100-m Green 
Bank Telescope (GBT)  in West Virginia can operate at 90\,GHz/3\,mm 
during the winter. New 
single-antenna telescopes with large 
survey cameras have been proposed for the excellent submm observing sites at 
the South Pole and  
the ALMA site in Chile. The 
{\it Planck Surveyor} CMB imaging mission will generate an all-sky map in 
the submm at a resolution of 5\,arcmin, and the 3.5-m {\it Herschel} 
space telescope will carry out pointed submm imaging and 
spectroscopic observations of known galaxies, and carry out  
deep confusion-limited cosmological surveys over fields several hundred square 
degrees in size. 
Cameras exploiting the 2.5-m telescope aboard SOFIA and BLAST and other 
dedicated 
ultra-long-duration balloon instruments will allow far-IR and submm-wave 
observations from the upper atmosphere. 

Existing mm-wave 
interferometers include the $6\times15$-m IRAM PdBI, 
the $6\times10.4$-m OVRO MMA, the $10\times6$-m BIMA array at 
Hat Creek in California and the $6\times10$-m Nobeyama Millimeter Array. 
The 8 $\times$ 6-m SMA is under construction on Mauna Kea, the first imaging 
submm-wave interferometer, while it is planned to combine 9 of the BIMA 
antennas with the OVRO MMA  
at a high site in the Inyo Mountains east of Owens Valley in California 
to form CARMA. The 
international 64 $\times$ 12-m ALMA submm interferometer array in 
Chile will provide a tremendous increase in the capability of submm-wave 
spectral line and continuum imaging, providing 
10- to 30-$\mu$arcsec resolution, and detailed 
images of even the most distant galaxies. The most luminous submm 
galaxies so far 
discovered, with 850-$\mu$m flux densities of about 
25\,mJy, could be detected at a 10-$\sigma$ significance by ALMA in about 
a second. Its 
excellent sensitivity and wide 8-GHz instantaneous bandwidth will 
allow a significant fraction of the galaxies detected in deep  
surveys 
to be  
detected simultaneously in the continuum and CO rotation lines, 
providing direct and 
exact redshifts. 
As the redshifted ladder of CO lines are separated by $115/(1+z)$\,GHz, 
about 25\% of galaxies at $z\simeq2.5$
will have a CO line 
lying within the 
8-GHz-wide ALMA band (Blain et al.,\ 2000b). 

\begin{figure}[t]
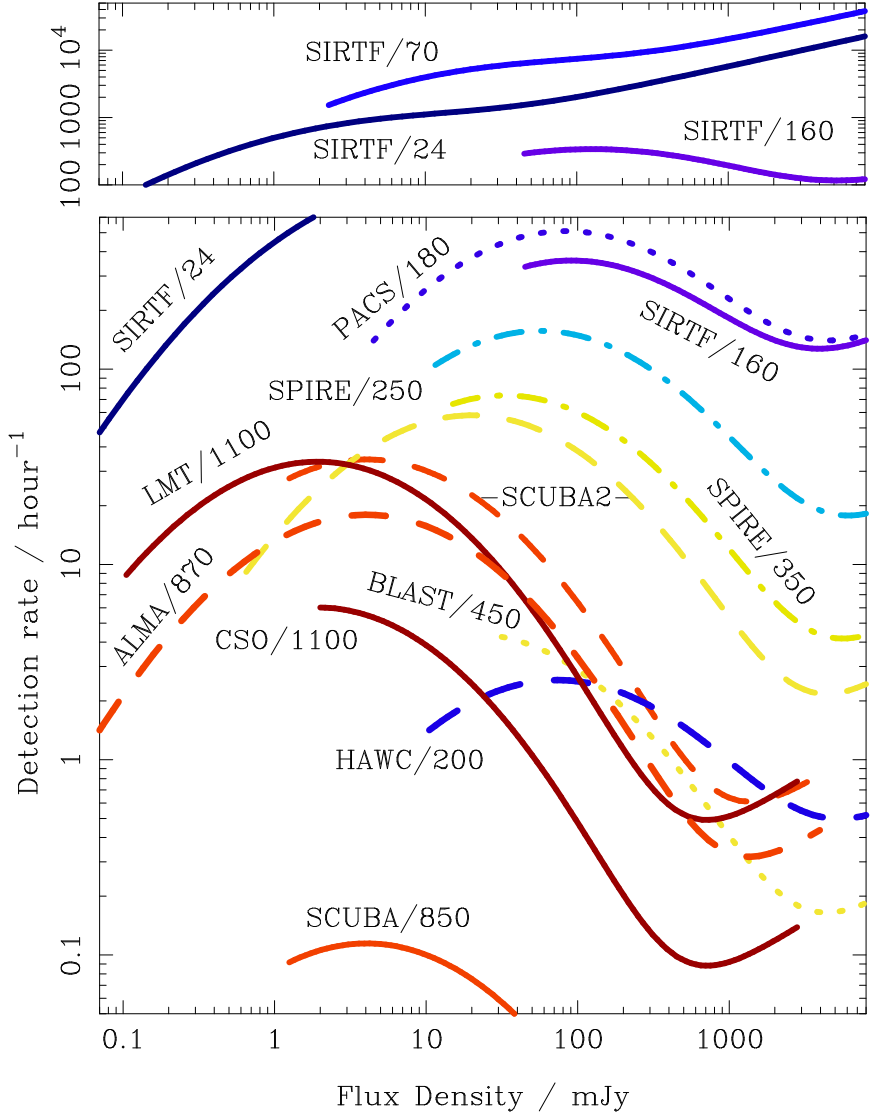

\begin{center}
\hskip 15pt 
\epsfig{file=ratetop.cps, width=2.75cm, angle=-90}
\end{center} 
\begin{center} 
\epsfig{file=rate.cps, width=12.0cm, angle=-90}
\end{center}
\caption{The detection rates expected in a variety of 
forthcoming mm, submm and far-IR surveys. 
The names and wavelengths in microns of the relevant instruments 
are listed: see Tables\,\ref{table:ground} and \ref{table:space}. 
The higher- and lower-peaking SCUBA-II curves correspond to wavelengths
of 450 and 850\,$\mu$m respectively. References 
to instrument performance for these calculations (Blain and Longair, 1996) 
can be 
found in Table\,\ref{table:instrument}. The surface density of galaxies 
assumed follows the models of Blain et al.\ (1999b): 
see Figs.\,\ref{fig:count1} and \ref{fig:count2}. 
Curves stop on the
right if the surface density is expected to fall below a single galaxy
on the whole sky. Curves stop at the left when a relatively optimistic 
definition of the 5$\sigma$ confusion noise level for detection is reached 
(see Fig.\,\ref{fig:confcont}). 
} 
\label{fig:rate}
\end{figure} 

\subsection{Future capabilities and progress} 

The enhanced capabilities of this array of 
new facilities is illustrated in Fig.\,\ref{fig:rate}. The rate at which 
galaxies can be detected is likely to grow dramatically from a
few per day at present to many hundreds per hour. Note that the 
various instruments 
operate at different wavelengths, and so each is  
most sensitive to galaxies at different redshifts and with different 
luminosities. 
However, sample sizes are certain to increase dramatically, 
especially when the $10^4$--10$^5$ galaxies that will be detected in
the {\it Planck Surveyor} all-sky survey are taken into account. 

Multiwavelength 
follow-up observations of all these new submm galaxies are likely to 
remain a time-consuming challenge. However, 
the likely availability of 30-m-aperture ground-based optical/near-IR 
telescopes in the next decades, and the extremely deep imaging capability of 
{\it NGST}, should help us to study a complete sample of submm
galaxies down to luminosities that are only a fraction of $L^*$. 

\begin{table}
\caption{
Wavelengths $\lambda$, sensitivities (as noise equivalent
flux density---NEFD), fields of view (FOV), and confusion
limits due to galaxies and the ISM (in brackets) for existing and 
future ground-based and airborne  
instruments.  
}
\label{table:ground}
\begin{tabular*}{\hsize}{@{\extracolsep{\fill}}*{5}l@{}} 
\hline
Name & $\lambda$ & NEFD & FOV & Confusion\\
& ($\mu$m) & (mJy/$\sqrt{Hz}$) & (arcmin$^2$) & (mJy)\\
\hline
SCUBA & 850 & 80 & 1.7 & 0.12 ($9 \times 10^{-4}$) \\
      & 450 & 160 & 1.7 & 0.053 ($3 \times 10^{-3}$) \\
MAMBO$^{1,2}$ & 1250 & 95 & 1.0 & 0.05 ($7\times10^{-5}$) \\ 
SCUBA-II & 850 & 28 & 64 & 0.12 ($9 \times 10^{-4}$) \\
       & 450 & 90 & 64 & 0.053 ($3 \times 10^{-3}$) \\
HAWC-SOFIA & 200 & 408 & 9.0 & 1.2 (0.30) \\ 
BOLOCAM-CSO &
1100 & 42 & 44 & 0.32 ($2 \times 10^{-3}$) \\
BOLOCAM-LMT &
1100 & 2.8 & 2.5 & $6 \times 10^{-3}$
($4 \times 10^{-5}$)\\
BLAST & 750 & 115 & $\simeq 10$ & 3.9 (0.25) \\ 
	& 450 & 130 & $\simeq 10$ & 6.8 (0.7) \\ 
	& 300 & 150 & $\simeq 10$ & 7.8 (1.1) \\
SMA & 850 & 170 & 0.2 & $< 10^{-7}$ ($< 10^{-6}$) \\
	& 450 & 1700 & 0.05 & $< 10^{-8}$ ($< 10^{-4}$) \\
ALMA
& 870 & 1.9 & 0.050 & $< 10^{-7}$ ($< 10^{-6}$) \\
 & 450 & 11 & 0.013 & $< 10^{-8}$ ($< 10^{-4}$) \\
Extended VLA & 20.5\,cm & 0.40 & 700 & $\sim 0$ ($\sim 0$) \\ 
SKA & 20.5\,cm & $\sim 10^{-2}$ & TBD &  $\sim 0$ ($\sim 0$) \\ 
\hline
\end{tabular*}
An estimate of the speed of a survey down to a chosen depth can be obtained 
by multiplying the FOV by the 
the square of the NEFD value. The approximate extragalactic confusion
noise values are the flux density at which there is one brighter 
source per beam
(Blain et al.,\ 1998; Fig.\,\ref{fig:confcont}). 
This corresponds approximately to the width of peak in the non-Gaussian 
confusion 
noise distribution (see Fig.\,\ref{fig:confhist}). 
The expected ISM confusion noise 
(in brackets) is calculated for a 100-$\mu$m surface
brightness $B_0$=1\,MJy\,sr$^{-1}$ (Helou and Beichman, 1990), 
and scales as $B_0^{1.5}$. Other instruments under development, which 
have not published detailed performance estimates include the 
350-$\mu$m 
SHARC-II camera for the CSO. 
The FOV and NEFD values 
are chosen to provide the correct results for making a fully-sampled 
image of the sky, not measuring the flux from a single galaxy. Updated from 
Table\,1 in Blain (1999b). Relevant references are listed in 
Table\,\ref{table:instrument}. TBD: to be decided.  

$^1$ Note that the FOV of MAMBO is expanded by a factor of 3 for the 
winter of 2001/2002, 
with a 117-bolometer detector array. 
$^2$ A similar device SIMBA is 
being commissioned at SEST. 
\end{table} 
  
\begin{table}
\caption[]{
The equivalent to Table\,\ref{table:ground} for space-borne instrumentation.}  
\label{table:space} 
\begin{tabular*}{\hsize}{@{\extracolsep{\fill}}*{5}l@{}}
\hline
Name & $\lambda$ & NEFD & FOV & Confusion\\
& ($\mu$m) & (mJy/$\sqrt{Hz}$) & (arcmin$^2$) & (mJy)\\
\hline
{\it Herschel}-SPIRE&
500 & 114 & 40 & 2.9 (0.16) \\
                  & 350 & 90 & 40 & 2.6 (0.12) \\
                  & 250 & 84 & 40 & 1.6 (0.24) \\
{\it Herschel}-PACS &
170 & 24 & 6.1 & 0.80 (0.16) \\
                  & 90 & 24 & 6.1 & 0.03 (0.01) \\
{\it SIRTF}-MIPS &
160 & 18 & 2.5 & 6.6 (3.1) \\
                  & 70 & 4.5 & 25 & 0.28 (0.07) \\
                  & 24 & 1.8 & 25 & $6 \times 10^{-4}$
($2 \times 10^{-4}$) \\
{\it SIRTF}-IRAC &
8.0 & 0.15 & 26 & $8 \times 10^{-2}$ ($\sim 10^{-6}$) \\
{\it Planck Surveyor} & 350 & 26 & All-sky & 50 (70) \\
		& 550 & 19 & All-sky & 22 (12) \\
                & 850 & 16 & All-sky & 8.1 (1.6) \\
{\it SPECS} testbed &
250 & 0.17 & 4 & $\sim 10^{-5}$ ($\sim 10^{-3}$) \\
\noalign{\vskip -10pt}
{\it SPIRIT} & & & & \\ 
\hline
\end{tabular*}
Note that the 
values listed for {\it Planck Surveyor} apply to an all-sky survey. 
Another instrument under development, which
has not published detailed performance estimates is
the 50--200-$\mu$m sky survey from the
Japanese {\it ASTRO-F/IRIS} satellite sky survey. 
\end{table}

\begin{table}
\caption[]{
References to instruments listed in 
Tables\,\ref{table:ground} and \ref{table:space}.}
\label{table:instrument} 
\begin{tabular*}{\hsize}{@{\extracolsep{\fill}}*{2}l@{}}
\hline
Name & Information \\ 
\hline
SCUBA & Holland et al.\ (1999) \\ 
MAMBO & Kreysa et al.\ (1998) \\ 
SCUBA-II & www.jach.hawaii.edu/JACpublic/JCMT/\\ 
\noalign{\vskip -10pt}
 & Continuum\_observing/SCUBA-2/home.html \\ 
SOFIA & Davidson et al.\ (1999); sofia.arc.nasa.gov \\
BOLOCAM & Glenn et al.\ (1998); www-lmt.phast.umass.edu/ins/ \\  
\noalign{\vskip -10pt} 
 & continuum/bolocam.html \\
BLAST & Devlin (2001); www.hep.upenn.edu/blast \\
SMA & Ho (2000); sma2.harvard.edu \\ 
ALMA & Wootten (2001); www.alma.nrao.edu \\ 
SKA & www.nfra.nl/skai \\ 
{\it Herschel}-SPIRE & www.ssd.rl.ac.uk/spire \\ 
{\it Herschel}-PACS & pacs.ster.kuleuven.ac.be \\ 
{\it SIRTF}-MIPS \& IRAC  & sirtf.caltech.edu \\ 
{\it Planck Surveyor} & astro.estec.esa.nl/Planck \\ 
{\it SPECS} / {\it SPIRIT} & Mather et al.\ (1998); space.gsfc.nasa.gov/astro/specs \\ 
\hline 
\end{tabular*}\vspace{2pc}
\end{table} 

\section{Summary: key questions and targets for the future} 

The first generation of extragalactic submm-wave surveys have 
provided an important complement to more traditional optical and radio  
searches for distant galaxies, and discovered a cosmologically
significant population of very-luminous, high-redshift dusty galaxies. 

We have found that is very hard to study a complete sample of submm
galaxies at other wavelengths (Smail et al.,\ 2002). The similar experience of 
other groups involved in deep mm/submm surveys 
(Barger et al.,\ 1999a; Eales et al.,\ 2000; 
Carilli et al.,\ 2001; Scott et al.,\ 2002; Webb et al.,\ 2002b) is 
reflected in the relatively few papers describing the individual 
multi-waveband 
properties of the almost 
200 galaxies detected. The most sensitive  
follow-up observations are required in the near-IR, radio and optical 
wavebands to identify and study them (Frayer et al.,\ 2000; 
Ivison et al.,\ 2001), that is very faint 
detection thresholds of order 10\,$\mu$Jy at 
1.4-GHz, $K\simeq23$ and $B\gg26$ respectively.
Much more time has been devoted to multi-waveband follow-up observations
than was spent on the initial submm  
detections. Typical examples of the 
submm population can be detected in imaging-mode SCUBA observations in 
about 10\,h  
of integration. However,
at least 2\,h of near-IR observations at the 10-m Keck 
telescope and about 24\,h 
of integration at the VLA are then 
required in order to find likely counterparts to typical submm galaxies. 
The advantage of the VLA radio observations over those at optical and 
near-IR wavelengths is the very large field of view, which allows many 
galaxies to be detected simultaneously. 
The very brightest optical counterparts to 
submm galaxies can 
be identified spectroscopically in about 7\,h of integration using 
4-m class telescopes (Ivison et al.,\ 1998) and higher-quality spectra 
can be obtained in a comparable time using 8-m class telescopes 
(see the results of a 5-h integration using the UVES spectrograph 
at the European Southern Observatory (ESO) VLT by Vernet and Cimatti, 2001). 
In all cases, identifying a plausible counterpart, where 
possible, is only a first step; finding a redshift for these typically 
faint, red galaxies is much more 
challenging. In this context, the unusual sensitivity of submm 
surveys to the most distant galaxies is almost a drawback, making it 
very hard to detect a complete sample of submm galaxies at other wavelengths. 

Key questions for understanding submm galaxies in the 
future include: 

What are the properties of typical submm galaxies in other 
wavebands, and what is their relationship to other 
high-redshift galaxy samples? 
The submm-selected galaxies appear to a diverse 
mixture of types, including bright merging systems (Ivison et al.,\ 
1998a, 2000a, 2001), optical QSOs  
(Kraiberg Knudsen 
et al.,\ 2001), EROs with $K<20$ (Smail et al.,\ 1999, 2002; 
Gear et al.,\ 2000; Lutz et al.,\ 2001), and much fainter IR-detected galaxies 
(Frayer et al.,\ 2000), which may also turn out to have 
very red colors. It seems that the overlap between the 850-$\mu$m submm 
galaxy population and both the LBGs and faint {\it Chandra} 
X-ray sources is small. Note that some   
of this apparent diversity is sure to be due to the very wide 
redshift distribution of the submm galaxies. 

What is the redshift distribution of the submm galaxies? 
Models of the evolution of submm galaxies that do not grossly 
violate basic observational constraints on the source counts and 
cosmic background 
radiation are easy to generate. However, it is 
vital to predict a plausible redshift distribution, 
with only a small fraction at redshifts less than unity, and 
a probable median redshift of at least 
2--3. It is easy to generate a redshift distribution
that is biased too high. The observational determination of a redshift 
distribution for a well-defined sample of submm galaxies remains 
a crucial goal. This will be easy with ALMA. In the meantime, concerted 
and time-consuming campaigns of optical and near-IR spectroscopy 
will pay off gradually, while observations of cm-wave megamasers 
and the development of wide-band mm- and cm-wave 
spectrometers may offer alternative routes. 
The forthcoming (sub)mm interferometers CARMA and SMA, and developments 
of the IRAM PdBI will also provide accurate 
positions and some CO redshifts for submm galaxies. 

What are the details of the astrophysics responsible for the 
luminosity of the submm galaxies? 
This is very important, as the submm galaxies appear to be signposts to 
some of the 
most luminous and violent phases of galaxy evolution, and could be 
associated with the formation of the bulk of galactic bulges, 
elliptical galaxies and supermassive black holes (Lilly et al.,\ 1999; 
Dunlop, 2001). Whether these galaxies are formed in a single event, or as 
a series of lesser bursts, is a key question for our understanding of  
the process of galaxy formation and 
evolution. Detailed comparisons of the luminosity 
derived from dust continuum emission, the dynamical mass inferred from 
molecular line profiles, the evolved stellar mass inferred
from near-IR observations, and the spatial extent of the activity 
from various 
high-resolution observations will all be important for 
disentangling the complex astrophysics of these systems. 

When can submm instruments be used to resolve and 
study high-redshift galaxies in detail? This is already practical 
given enough observing time at the OVRO MMA and the 
IRAM PdBI. The CARMA and SMA 
interferometers will soon 
have important roles to play in these studies. In about 10 years, 
ALMA will provide the first real chance to detect and study galaxies 
rapidly and  
in great detail using submm observations alone. Luminosities, redshifts, 
dynamical masses and metallicities could all be determined without needing to 
resort to radio, optical and near-IR observations as a matter of 
course. However, because ALMA has a  
relatively small field of view, the most efficient survey strategy may 
be to detect large numbers of galaxies using wide-field mm/submm 
cameras 
like BOLOCAM, SCUBA-II and their successors on 
single-antenna 10--50-m aperture 
survey telescopes, and the {\it Herschel} and {\it Planck Surveyor} 
space missions, and then use ALMA to provide  
detailed images and spectra of all the detected galaxies. 

What is the fundamental limit to making submm observations of distant 
galaxies? 
Submm observations rely on the presence of metals, in the form of 
molecular gas or 
dust grains in order to detect galaxies. 
While submm radiation is able to travel unattenuated  
across the Universe from prior to the epoch of reionization, 
it is possible that a large fraction of pre-reionization `first-light' 
sources are 
insufficiently dusty and metal rich to be detectable as continuum sources. 
Low-metallicity galaxies should be detectable by fine-structure C and O 
far-IR line emission, however.   
It would be 
tremendously exciting to see the birth of the first metal-enriched 
dusty systems with ALMA, and so perhaps to 
determine directly the redshift limit for 
submm surveys. Of course, even if this were possible, 
ALMA would still have a long and fruitful career studying the detailed  
astrophysics of galaxies out to and beyond redshift 5, while the search 
for the most primitive galaxies in the second and third decades of the century 
is taken up by 
space-based mid-IR interferometers and the SKA radio telescope
(see Fig.\,\ref{fig:Svz_highz}). 

Submm observations of the distant Universe are a new tool for 
probing the earliest and most dramatic stages of the evolution of 
galaxies. Over the years to come, the capabilities of submm-wave 
observatories, and our understanding of the Universe in this new window, 
should continue to advance dramatically. 

\section*{Acknowledgments}   

This work is heavily based on results obtained from the SCUBA Lens 
Survey. The following have all been involved with aspects of 
the SCUBA lens survey: Lee Armus, 
Amy Barger, Jocelyn Bezecourt, Leo Blitz, 
Len Cowie, 
John Davies, 
Alastair Edge, Aaron Evans, Andy Fabian, Allon Jameson, Tom Kerr, 
Jean-Francois Le\,Borgne,  
Malcolm Longair, Leo Metcalfe, Glenn Morrison, 
Frazer Owen, Naveen Reddy, 
Nick Scoville, Genevieve Soucail, Jack Welch, Mel Wright 
and Min Yun. 
We thank the staff of the JCMT for operating and the UK ATC for 
providing SCUBA. 

We thank Omar Almaini, Vicki Barnard, Frank Bertoldi, 
Jamie Bock, Chris Carilli, Helmut Dannerbauer, Darren Dowell, 
Steve Eales, Jason Glenn, Sunil Golwala,  
Dean Hines, Kate Isaak, 
Kirsten Kraiberg Knudsen, Attila Kovacs, Andrew Lange, Simon Lilly, 
Ole M\"oller, Priya Natarajan, Max Pettini,  
Tom 
Phillips, Kate Quirk, Enrico Ramirez-Ruiz, Nial Tanvir, 
Neil Trentham, 
Paul van der Werf, 
the editor Marc Kamionkowski, an anonymous referee and 
Roberta Bernstein for useful conversations and comments on the manuscript. 

AWB was supported in Cambridge by the Raymond and Beverly Sackler 
Foundation as part of the Foundation's Deep Sky Initiative Program at the 
IoA. IRS is supported by the Leverhulme Trust 
and the Royal Society. JPK is supported by CNRS. 
Full references and acknowledgement to the instruments and telescopes 
used in this research can be found in Smail et al.\ (2002). 
This research has
made use of the NASA/IPAC Extragalactic Database (NED) which is operated by
the Jet Propulsion Laboratory, California Institute of Technology, under
contract with the National Aeronautics and Space Administration.

\end{document}